\documentclass[12pt]{article}
\input amssym.def
\input amssym.tex

\newcommand{\g}{\mbox{\boldmath$g$}}

\newcommand{\rF}{{\rm F}}
\newcommand{\rG}{{\rm G}}
\newcommand{\rH}{{\rm H}}

\newcommand{\rM}{{\rm M}}

\newcommand{\ga}{{\goth a}}
\newcommand{\gb}{{\goth b}}
\newcommand{\gc}{{\goth c}}
\newcommand{\gd}{{\goth d}}
\newcommand{\goe}{{\goth e}}
\newcommand{\gf}{{\goth f}}

\newcommand{\ba}{{\bf a}}
\newcommand{\bb}{{\bf b}}

\newcommand{\bq}{{\bf q}}
\newcommand{\bp}{{\bf p}}
\newcommand{\br}{{\bf r}}
\newcommand{\bs}{{\bf s}}
\newcommand{\bu}{{\bf u}}
\newcommand{\bv}{{\bf v}}

\newcommand{\bx}{{\bf x}}
\newcommand{\by}{{\bf y}}
\newcommand{\bz}{{\bf z}}

\newcommand{\mbA}{\mbox{\boldmath$A$}}
\newcommand{\mbB}{\mbox{\boldmath$B$}}
\newcommand{\mbC}{\mbox{\boldmath$C$}}

\newcommand{\mbM}{\mbox{\boldmath$M$}}

\newcommand{\mba}{\mbox{\boldmath$a$}}
\newcommand{\mbb}{\mbox{\boldmath$b$}}

\newcommand{\mbu}{\mbox{\boldmath$u$}}
\newcommand{\mbv}{\mbox{\boldmath$v$}}

\newcommand{\cF}{{\cal F}}
\newcommand{\cG}{{\cal G}}

\newcommand{\cM}{{\cal M}}

\newcommand{\cR}{{\cal R}}

\newcommand{\cT}{{\cal T}}
\newcommand{\cV}{{\cal V}}

\newcommand{\wa}{\widetilde{a}}
\newcommand{\wb}{\widetilde{b}}

\newcommand{\wL}{\widetilde{L}}
\newcommand{\wiP}{\widetilde{P}}
\newcommand{\wQ}{\widetilde{Q}}
\newcommand{\wT}{\widetilde{T}}
\newcommand{\wU}{\widetilde{U}}
\newcommand{\wV}{\widetilde{V}}
\newcommand{\wW}{\widetilde{W}}

\newtheorem{theorem}{Theorem}[section]
\newtheorem{proposition}[theorem]{Proposition}

\begin{document}

\begin{center} 
{\LARGE\bf
Miura transformations \\
\vspace{0.05cm}
for Toda--type integrable systems,\\
\vspace{0.05cm}
with applications to the problem \\
\vspace{0.25cm}
of integrable discretizations} \\

\vspace{1.0cm}

{\large\bf Yuri B. SURIS}
\vspace{1.0cm}

{\it Fachbereich Mathematik, Sekr. MA 8-5, Technische Universit\"at Berlin,\\
Str. des 17. Juni 136, 10623 Berlin, Germany\\
E-mail address:} suris @ sfb288.math.tu-berlin.de 
\end{center}
\vspace{2cm}

{\bf Abstract.} We study lattice Miura transformations
for the Toda and Volterra lattices, relativistic Toda
and Volterra lattices, and their modifications. In particular, we give
three successive modifications for the Toda lattice, two for the Volterra
lattice and for the relativistic Toda lattice, and one for the relativistic
Volterra lattice. We discuss Poisson properties of the Miura transformations,
their permutability properties, and their role as localizing changes
of variables in the theory of integrable discretizations.
\newpage
\tableofcontents

\setcounter{equation}{0}
\section{Introduction}

\subsection{The original Miura transformation}

This paper is devoted to lattice Miura transformations. Let us start with
recalling the basic facts about the original Miura transformation, as applied
to the Korteweg--de Vries (KdV) equation.

The KdV equation,
\begin{equation}\label{KdV}
u_t=6uu_x-u_{xxx}\;,
\end{equation}
is a bi--Hamiltonian system, i.e. admits two Hamiltonian formulations with
respect to two compatible local Poisson brackets:
\begin{equation}\label{KdV PB1}
\{\cF,\cG\}_1=\int\frac{\delta\cF}{\delta u} J_1\frac{\delta\cG}{\delta u}dx\;,
\qquad J_1=\partial\;,
\end{equation}
and 
\begin{equation}\label{KdV PB2}
\{\cF,\cG\}_2=\int\frac{\delta\cF}{\delta u} J_2\frac{\delta\cG}{\delta u}dx\;,
\qquad J_2=-\partial^3+2u\partial+2\partial u\;.
\end{equation}
In other words, KdV may be represented in the Hamiltonian form in two
different ways:
\begin{eqnarray}
u_t & = & J_1(3u^2-u_{xx})=
J_1\frac{\delta}{\delta u}\Big(\frac{1}{2}\,u_x^2+u^3\Big)\;,
\\
 & = & J_2(u)=J_2\frac{\delta}{\delta u}\Big(\frac{1}{2}\,u^2\Big)\;.
\end{eqnarray}
The Miura transformation with parameter (also known as the Gardner 
transformation) reads:
\begin{equation}\label{Gardner trans}
u=w+\epsilon w_x+\epsilon^2w^2\;.
\end{equation}
Let us list its most important properties. The Miura transformation is
actually a {\it differential substitution} and therefore, generally speaking,
{\it noninvertible}. Therefore nothing guarantees {\it \'a priori} that
there exists a differential equation for $w$ which is pushed to (\ref{KdV})
by this transformation. Nevertheless, this is the case, and the corresponding
equation is the {\it modified Korteweg--de Vries equation} MKdV$(\epsilon)$:
\begin{equation}\label{MKdV}
w_t=6w(1+\epsilon^2w)w_x-w_{xxx}\;.
\end{equation}

Nothing guaranties {\it \'a priori} also that there exist local Poisson 
brackets in the $w$--space pushed by (\ref{Gardner trans}) to either of
the brackets (\ref{KdV PB1}), (\ref{KdV PB2}). Indeed, this is not the case.
However, this is the case for a certain {\it linear combination} of these
brackets. Namely, 
\[
\{\cdot,\cdot\}_1+\epsilon^2\{\cdot,\cdot\}_2
\]
is the push--forward under (\ref{Gardner trans}) of the bracket
\begin{equation}\label{MKdV PB}
\{\cF,\cG\}=\int\frac{\delta\cF}{\delta w} J_1\frac{\delta\cG}{\delta w}dx
\end{equation}
with the same $J_1=\partial$ as before. Correspondingly, the equation
MKdV$(\epsilon)$ is Hamiltonian with respect to the latter bracket, i.e.
it may be presented as
\begin{equation}
w_t=J_1(3w^2+2\epsilon^2w^3-w_{xx})=
J_1\frac{\delta}{\delta w}\Big(\frac{1}{2}\,w_x^2+u^3+\frac{1}{2}\,\epsilon^2
w^4\Big)\;.
\end{equation}

One can invert (\ref{Gardner trans}) {\it formally}, arriving at the formal
power series
\begin{equation}\label{Gardner formal inv}
w=u-\epsilon u_x+\epsilon^2(u_{xx}-u^2)+\ldots\;.
\end{equation}
Since $w$ is trivially a density of a conservation law for MKdV$(\epsilon)$,
we get an infinite series of conservation laws densities for KdV, as 
coefficients of the above power series. The coefficients by odd powers
of $\epsilon$ deliver trivial conservtion laws (as $u_x$ by the first power 
of $\epsilon$), but the coefficients by even powers give nontrivial ones
(as $u$, $u^2-u_{xx}\sim u^2$, $\ldots$).

Finally, it should be mentioned that the origin of the Miura transformation
(\ref{Gardner trans}) lies in the factorization of Lax operators. Indeed, it
is well known that the Schr\"odinger operator 
\[
\partial^2-u
\]
is a Lax operator for KdV. Now the formula (\ref{Gardner trans}) is equivalent
to the following factorization:
\begin{equation}\label{KdV fact}
1-4\epsilon^2(\partial^2-u)=(1+2\epsilon\partial+2\epsilon^2w)
(1-2\epsilon\partial+2\epsilon^2w)\;.
\end{equation}

\subsection{Lattice Miura transformations}

The main aim of this paper is to collect a huge bulk of variable 
transformations for lattice systems, which can be considered as analogs
of the Miura transformation, and to point out a remarkable similarity
of their properties with the above mentioned ones.

We consider in this paper integrable lattice systems with local
interactions:
\begin{equation}\label{loc syst}
\dot{x}_k=f_{k\,{\rm mod}\,m}
(x_{k-s},\ldots,x_{k-1},x_k,x_{k+1},\ldots,x_{k+s})
\end{equation}
with a {\it fixed} $s\in{\Bbb N}$. They may be considered either on an
infinite lattice $(k\in{\Bbb Z})$, or on a periodic one 
$(k\in{\Bbb Z}/N{\Bbb Z})$. The number $m\in {\Bbb N}$ is called the
{\it number of fields}. It is important that the number $s$, measuring the
locality of interaction, is one and the same for all $k$ (and hence is 
independent on the total number $N$ of particles, or lattice sites, in the 
periodic case).

As two simple examples of such systems, consider the Toda lattice:
\begin{equation}\label{TL introd}
\dot{b}_k=a_k-a_{k-1}\;,\qquad \dot{a}_k=a_k(b_{k+1}-b_k)\;,
\end{equation}
and the relativistic Toda lattice:
\begin{equation}\label{RTL introd}
\dot{b}_k=(1+\alpha b_k)(a_k-a_{k-1})\;,\qquad
\dot{a}_k=a_k(b_{k+1}+\alpha a_{k+1}-b_k-\alpha a_{k-1})\;.
\end{equation}
Both these systems are two--field, i.e. they have the form (\ref{loc syst}) 
with $m=2$ (to see this, set $b_k=x_{2k-1}\,$, $a_k=x_{2k}\,$. For the Toda
lattice we have $s=1$, while for the relativistic Toda lattice $s=2$. 
These two will be among our basic examples considered in detail in the 
main text. Other examples include the Volterra lattice and the relativistic
Volterra lattice, and also {\it modifications} of these systems.

The notion of modification is intimately related to the notion of {\it Miura
transformations}. We do not give here a general definition of a Miura 
transformation, but instead briefly summarize the properties of the Miura 
transformations studied in this paper. It turns out that for all our systems 
(\ref{loc syst}) the corresponding Miura transformations appear in pairs.
One of these maps always has the form
\begin{equation}\label{Miura 1}
\cM_1(\epsilon):\qquad
x_k=y_k+\epsilon\Phi_{k\,{\rm mod}\,m}(y_{k-s},\ldots,y_{k-1},y_k;\epsilon)\;,
\end{equation}
while the second one always has the form
\begin{equation}\label{Miura 2}
\cM_2(\epsilon):\qquad
x_k=y_k+\epsilon\Psi_{k\,{\rm mod}\,m}(y_k,y_{k+1},\ldots,y_{k+s};\epsilon)\;.
\end{equation}
The dependence on the (small) parameter $\epsilon$ is analytic in some
neighborhood of zero. Moreover, there holds the following property:
\[
\Psi_{k\,{\rm mod}\,m}(x_k,x_{k+1},\ldots,x_{k+s};0)-
\Phi_{k\,{\rm mod}\,m}(x_k,x_{k+1},\ldots,x_{k+s};0)=
\]
\begin{equation}\label{Miura approx}
=f_{k\,{\rm mod}\,m}
(x_{k-s},\ldots,x_{k-1},x_k,x_{k+1},\ldots,x_{k+s})\;.
\end{equation}
For instance, the Miura transformations for the Toda lattice (\ref{TL introd})
read:
\begin{equation}\label{MTL Miuras introd}
\rM_1(\epsilon):\;\left\{\begin{array}{l} 
b_k=p_k+\epsilon q_{k-1}\;,\\ \\ a_k=q_k(1+\epsilon p_k)\;, \end{array}\right.
\qquad
\rM_2(\epsilon):\;\left\{\begin{array}{l} 
b_k=p_k+\epsilon q_k\;,\\ \\ a_k=q_k(1+\epsilon p_{k+1})\;.\end{array}\right.
\end{equation}

So, the Miura maps are always given by {\it local formulas}. It is easy to 
understand that in the case of infinite lattices such maps are 
{\it noninvertible}. Of course, in the periodic case the inverse maps always 
exist, but are given in general by {\it nonlocal formulas}. The (small) 
parameter $\epsilon$ is called a {\it modification parameter}. The remarkable 
properties of the Miura maps may be summarized as follows.

First, nothing guarantees {\it a priori} the existence of a local system of 
the type (\ref{loc syst}) in the variables $y_k$ which is pushed to the 
original system (\ref{loc syst}) by a map of the type (\ref{Miura 1}) or 
(\ref{Miura 2}). As we shall be mostly working with the periodic case, we
express the above statement also in this way: nothing guarantees {\it a priori}
that the pull--back of the differential equations of motion (\ref{loc syst}) 
under the change of variables (\ref{Miura 1}) or (\ref{Miura 2}) will be given 
again by local formulas. However, this is the case for the Miura maps,
and can be accepted as their (somewhat informal) definition. Of course,
the very existence of such Miura maps is rather mystifying. The above
mentioned local systems in $y_k$ form one--parameter families of integrable 
deformations, or {\it modifications}, of lattice systems. So, the modified
system (\ref{loc syst}) has the form
\begin{equation}\label{mod syst}
\dot{y}_k=F_{k\,{\rm mod}\,m}
(y_{k-s},\ldots,y_{k-1},y_k,y_{k+1},\ldots,y_{k+s};\epsilon)\;,
\end{equation}
where 
\begin{equation}\label{}
F_{k\,{\rm mod}\,m}(x_{k-s},\ldots,x_{k-1},x_k,x_{k+1},\ldots,x_{k+s};0)\!=
\!f_{k\,{\rm mod}\,m}(x_{k-s},\ldots,x_{k-1},x_k,x_{k+1},\ldots,x_{k+s}).
\end{equation}
For instance, the equations of the modified Toda lattice read:
\begin{equation}\label{MTL introd}
\dot{p}_k=(1+\epsilon p_k)(q_k-q_{k-1})\;,\qquad \dot{q}_k=q_k(p_{k+1}-p_k)\;.
\end{equation}

Further, integrable lattice systems (\ref{loc syst}) often admit a 
{\it Hamiltonian formulation}, the corresponding invariant Poisson bracket
$\{\cdot,\cdot\}$ being given by {\it local formulas}. That means that
\[
\{x_j,x_k\}=0
\]
for $|j-k|$ large enough. Nothing guarantees {\it a priori} that the 
pull--backs of these brackets under the Miura maps (\ref{Miura 1}),
(\ref{Miura 2}) are also given by local formulas. Indeed, as a rule 
these pull--backs are non--local. However, in the multi--Hamiltonian case,
when (\ref{loc syst}) is Hamiltonian with respect to several compatible
local Poisson brackets, it turns out that pull--backs of certain 
{\it linear combinations} of these brackets are local again. These
pull--backs serve then as invariant local Poisson brackets in the Hamiltonian
formulation of the modified systems (\ref{mod syst}). A general observation
is: the number of invariant local Poisson brackets for a modified system 
is, as a rule, by one less than for the original one. For instance, the Toda 
lattice (\ref{TL introd}) is a tri--Hamiltonian system, and the modified Toda
lattice (\ref{MTL introd}) is a bi--Hamiltonian system.

The integrals of motion for the modified systems are always easily 
obtainable as compositions of the Miura maps with the integrals of 
motion for the original systems. Interestingly, just {\it one} integral
of the modified system often allows to find the whole series of integrals 
for the original one. To this end one has to invert the Miura maps 
{\it formally}, as power series in $\epsilon$. For example, the formal
inversion of $M_1(\epsilon)$ for the Toda lattice gives:
\[
\left\{\begin{array}{l}
p_k=b_k-\epsilon a_{k-1}+\epsilon^2 b_{k-1}a_{k-1}-\epsilon^3(b_{k-1}^2a_{k-1}
+a_{k-1}a_{k-2})+\ldots\\ \\
q_k=a_k-\epsilon b_ka_k+\epsilon^2(b_k^2a_k+a_ka_{k-1})-\ldots
\end{array}\right.
\]
Now it is easy to see that $\log(1+\epsilon p_k)$ is a density of a 
conservation law for the modified Toda lattice. The expansion of this 
density in powers of $\epsilon$ reads:
\begin{equation}
\log(1+\epsilon p_k)=\epsilon b_k-\epsilon^2\Big(\frac{1}{2}\,b_k^2+a_{k-1}\Big)
+\epsilon^3\Big(\frac{1}{3}\,b_k^3+b_ka_{k-1}+b_{k-1}a_{k-1}\Big)-\ldots \;.
\end{equation}
The coefficient by $\epsilon^n$ here is nothing but the density of the $n$th 
conservation law for the Toda lattice. This is, probably, the most direct way 
to demonstrate the existence of an infinite series of integrals for this system.

Finally, we point out that the Miura maps turn out to provide {\it localizing
changes of variables} for integrable discretizations. We refer the reader
to the review [S] for a general construction of integrable discretizations
for lattice systems, based on Lax representations and their $r$--matrix
interpretation. We point out here that this construction leads, as a 
rule, to discretizations governed by nonlocal equations. For instance,
equations of motion of the discrete time Toda lattice include continued 
fractions:
\begin{equation}\label{dTL introd}
\wb_k=b_k+h\left(\frac{a_k}{\beta_k}-\frac{a_{k-1}}{\beta_{k-1}}\right)\;, 
\qquad \wa_k=a_k\;\frac{\beta_{k+1}}{\beta_k}\;,
\end{equation}
where
\[
\beta_k=1+hb_k-\frac{h^2a_{k-1}}{1+hb_{k-1}-
\displaystyle\frac{h^2a_{k-2}}{1+hb_{k-2}-\;
\raisebox{-3mm}{$\ddots$}}}\;.
\]
One way to repair this drawback is based on the notion of 
{\it localizing changes of variables} and will be one of our main themes here.
It turns out that the change of variables $\cM_1(h)$,
\begin{equation}\label{loc map}
x_k=\bx_k+h\Phi_{k\,{\rm mod}\,m}(\bx_{k-s},\ldots,\bx_{k-1},\bx_k;h)\;,
\end{equation}
conjugates our discretizations with the maps described by the {\it local} 
formulas:
\begin{equation}\label{loc discr}
\widetilde{\bx}_k+h\Phi_{k\,{\rm mod}\,m}
(\widetilde{\bx}_{k-s},\ldots,\widetilde{\bx}_{k-1},\widetilde{\bx}_k;h)=
\bx_k+h\Psi_{k\,{\rm mod}\,m}(\bx_k,\bx_{k+1},\ldots,\bx_{k+s};h)\;. 
\end{equation}
So, the local discretizations belong to the modified hierarchies, with the
value of the modification parameter equal to the time step of the 
discretization. For instance, for the Toda lattice case the Miura map
\[
M_1(h):\quad \left\{\begin{array}{l} b_k=\bb_k+h\ba_{k-1}\;,\\ \\
a_k=\ba_k(1+h\bb_k)\;,\end{array}\right.
\]
conjugates the discrete time Toda lattice (\ref{dTL introd}) with the 
following map belonging to the modified Toda lattice hierarchy:
\begin{equation}\label{dTL loc introd}
\widetilde{\bb}_k+h\widetilde{\ba}_{k-1}=\bb_k+h\ba_k\;,\qquad
\widetilde{\ba}_k(1+h\widetilde{\bb}_k)=\ba_k(1+h\bb_{k+1})\;.
\end{equation}

Such implicit local discretizations are much more 
satisfying from the esthetical point of view and are much better suited 
for the purposes of numerical simulation.  (If, for instance, one uses the 
Newton's iterative method to solve (\ref{loc discr}) for $\widetilde{\bx}$, 
then one has to solve only linear systems whose matrices are {\it triangular} 
and have a band structure, i.e. only $s$ nonzero diagonals). 
 
In the main body of this paper we provide Miura transformations and
formulate their properties for a wide set of Toda--like systems. After the
results are found and formulated, it is rather straightforward to prove them,
therefore all proofs are omitted. For the notions and notations concerning
the Lax representations, the reader should consult [S] (including them here
would unneccessary increase the length of the paper, which is already much
longer than the author would like it). Many of the results about the
nonrelativistic Toda--like already appeared in the literature, while the
most of the results concerning the relativistic Toda--like systems seem
to be original. The paper ends with some bibliographical remarks.

\newpage
\part{Nonrelativistic systems}
\setcounter{equation}{0}
\section{Toda lattice}\label{Sect Toda}

\subsection{Equations of motion}

The system known as the {\it Toda lattice}, abbreviated as TL,
lives on the space $\cT={\Bbb R}^{2N}(a,b)$ and is described by the
following equations of motion:
\begin{equation}\label{TL}
\dot{b}_k=a_k-a_{k-1}\;,\qquad \dot{a}_k=a_k(b_{k+1}-b_k)\;.
\end{equation}

\subsection{Tri--Hamiltonian structure}

We adopt once and forever the following conventions: the 
Poisson brackets are defined by writing down {\it all nonvanishing} 
brackets between the coordinate functions; the indices in the 
corresponding formulas are taken (mod $N$).
\begin{proposition}\label{linear PB for Toda}
{\rm a)} The relations 
\begin{equation}\label{TL l br}
\{b_k,a_k\}_1=-a_k\;, \qquad \{a_k,b_{k+1}\}_1=-a_k 
\end{equation}
define a Poisson bracket on $\cT$. The system {\rm TL} is a Hamiltonian 
system on $\Big(\cT,\{\cdot,\cdot\}_1\Big)$ with the Hamilton function
\begin{equation}\label{TL H2}
\rH_2(a,b)=\frac{1}{2}\sum_{k=1}^N b_k^2+\sum_{k=1}^{N}a_k\;.
\end{equation}

{\rm b)} The relations
\begin{equation}\label{TL q br}
\begin{array}{cclcccl}
\{b_k,a_k\}_2 & = & -b_ka_k\;, & \quad & 
\{a_k,b_{k+1}\}_2 & = & -a_kb_{k+1}\;, \\ \\
\{b_k,b_{k+1}\}_2 & = & -a_k\;, & \quad & 
\{a_k,a_{k+1}\}_2 & = & -a_ka_{k+1}      
\end{array}
\end{equation}
define a Poisson bracket on $\cT$ compatible with the linear bracket
$\{\cdot,\cdot\}_1$. The system {\rm TL} is a Hamiltonian system on 
$\Big(\cT,\{\cdot,\cdot\}_2\Big)$ with the Hamilton function 
\begin{equation}\label{TL H1}
\rH_1(a,b)=\sum_{k=1}^N b_k\;.
\end{equation}

{\rm c)} The relations
\begin{equation}\label{TL c br}
\begin{array}{cclcccl}
\{b_k,a_k\}_3     & = & -a_k(b_k^2+a_k)\;,    & \quad &
\{a_k,b_{k+1}\}_3 & = & -a_k(b_{k+1}^2+a_k)\;, \\ \\
\{b_k,b_{k+1}\}_3 & = & -a_k(b_k+b_{k+1})\;, & \quad &
\{a_k,a_{k+1}\}_3 & = & -2a_ka_{k+1}b_{k+1}\;,   \\ \\
\{b_k,a_{k+1}\}_3 & = & -a_ka_{k+1}\;,         & \quad & 
\{a_k,b_{k+2}\}_3 & = & -a_ka_{k+1}
\end{array}
\end{equation}
define a Poisson bracket on $\cT$ compatible with $\{\cdot,\cdot\}_1$ and
$\{\cdot,\cdot\}_2$. The system {\rm TL} is a Hamiltonian system on 
$\Big({\cal T},\{\cdot,\cdot\}_3\Big)$ with the Hamilton function 
\[
\rH_0(a,b)=\frac{1}{2}\sum_{k=1}^N\log(a_k)\;.
\]
\end{proposition}

\subsection{Lax representation}

Lax matrix $T:\cT\mapsto\g$:
\begin{equation}\label{TL T}
T(a,b,\lambda)=\lambda^{-1}\sum_{k=1}^N a_kE_{k,k+1}+\sum_{k=1}^N b_kE_{kk}
+\lambda\sum_{k=1}^N E_{k+1,k}\;.
\end{equation}
Lax representation for TL:
\begin{equation}\label{TL Lax}
\dot{T}=[T,B]=[A,T]\;,
\end{equation}
where
\begin{equation}\label{TL B}
B=\pi_+(T)=\sum_{k=1}^N b_kE_{kk}+\lambda\sum_{k=1}^N E_{k+1,k}\;,
\end{equation}
\begin{equation}\label{TL A}
A=\pi_-(T)=\lambda^{-1}\sum_{k=1}^N a_kE_{k,k+1}\;.
\end{equation}

\subsection{Discretization}

Lax representation for the map dTL:
\begin{equation}\label{dTL Lax}
\wT=\mbB^{-1}T\mbB=\mbA T\mbA^{-1} 
\end{equation}
with
\begin{equation}\label{dTL B}
\mbB=\Pi_+(I+hT)=\sum_{k=1}^N\beta_kE_{kk}+h\lambda\sum_{k=1}^{N}E_{k+1,k}\;,
\end{equation}
\begin{equation}\label{dTL A}
\mbA=\Pi_-(I+hT)=I+h\lambda^{-1}\sum_{k=1}^{N}\alpha_kE_{k,k+1}\;.
\end{equation}

\setcounter{equation}{0}
\section{Modified Toda lattice}\label{Sect MTL}

\subsection{Equations of motion}

The phase space of the {\it modified Toda lattice} ${\rm MTL}(\epsilon)$
will be denoted by ${\cal MT}={\Bbb R}^{2N}(q,r)$, the parameter $\epsilon$
will be called the {\it modification parameter}. Equations of motion of
${\rm MTL}(\epsilon)$ read:
\begin{equation}\label{MTL}
\dot{p}_k=(1+\epsilon p_k)(q_k-q_{k-1})\;,\qquad
\dot{q}_k=q_k(p_{k+1}-p_k)\;.
\end{equation}

\subsection{Bi--Hamiltonian structure}
\begin{proposition}\label{MTL first Ham str}
{\rm a)} The relations
\begin{equation}\label{MTL br 1}
\{p_k,q_k\}_{12}=-q_k(1+\epsilon p_k)\;,\qquad 
\{q_k,p_{k+1}\}_{12}=-q_k(1+\epsilon p_{k+1})
\end{equation}
define a Poisson bracket on ${\cal MT}$. The system ${\rm MTL}(\epsilon)$ 
is Hamiltonian with respect to this bracket, with the Hamilton function
\begin{equation}\label{MTL G1}
\rG_1(q,p)=\epsilon^{-1}\sum_{k=1}^N p_k+\sum_{k=1}^N q_k\;.
\end{equation}

{\rm b)} The relations
\begin{eqnarray}
\{p_k,q_k\}_{23} & = & -q_k(p_k+\epsilon q_k)(1+\epsilon p_k)\;,\nonumber\\
\{q_k,p_{k+1}\}_{23} & = & -q_k(p_{k+1}+\epsilon q_k)(1+\epsilon p_{k+1})\;,
\nonumber\\ 
\{p_k,p_{k+1}\}_{23} & = & -q_k(1+\epsilon p_k)(1+\epsilon p_{k+1})\;,
\nonumber\\
\{q_k,q_{k+1}\}_{23} & = & -q_kq_{k+1}(1+\epsilon p_{k+1})
\label{MTL br 2}
\end{eqnarray}
define a Poisson bracket on ${\cal MT}$ compatible with 
{\rm(\ref{MTL br 1})}. The system ${\rm MTL}(\epsilon)$ is Hamiltonian 
with respect to this bracket, with the Hamilton function
\begin{equation}\label{MTL G0}
\rG_0(q,p)=\epsilon^{-1}\sum_{k=1}^N\log(1+\epsilon p_k)\;.
\end{equation}
\end{proposition}
Notice that the function $\rG_1(q,p)$ is singular in $\epsilon$; it may be
done regular, and, moreover, an $O(\epsilon)$--perturbation of $\rH_2(q,p)$
from (\ref{TL H2}), by subtracting
\[
\epsilon^{-2}\sum_{k=1}^N\log(1+\epsilon p_k)\;,
\]
which is a Casimir function of the bracket $\{\cdot,\cdot\}_{12}$.

\subsection{Miura relations}
\begin{theorem}\label{Miuras for MTL}
Define the Miura maps $\rM_{1,2}(\epsilon):\;{\cal MT}(q,p)\mapsto\cT(a,b)$ by:
\begin{equation}\label{MTL Miuras}
\rM_1(\epsilon):\;\left\{\begin{array}{l} 
b_k=p_k+\epsilon q_{k-1}\;,\\ \\ a_k=q_k(1+\epsilon p_k)\;, \end{array}\right.
\qquad
\rM_2(\epsilon):\;\left\{\begin{array}{l} 
b_k=p_k+\epsilon q_k\;,\\ \\ a_k=q_k(1+\epsilon p_{k+1})\;.\end{array}\right.
\end{equation}
Both maps $\rM_{1,2}(\epsilon)$ are Poisson, if ${\cal MT}(q,p)$ is 
equipped with the bracket $\{\cdot,\cdot\}_{12}$, and $\cT(a,b)$ is equipped 
with $\{\cdot,\cdot\}_1+\epsilon\{\cdot,\cdot\}_2$, and also if 
${\cal MT}(q,p)$ is equipped with the bracket $\{\cdot,\cdot\}_{23}$, and 
$\cT(a,b)$ is equipped with $\{\cdot,\cdot\}_2+\epsilon\{\cdot,\cdot\}_3$.
The pull--back of the flow {\rm TL} under either of the Miura maps 
$\rM_{1,2}(\epsilon)$ coincides with ${\rm MTL}(\epsilon)$.
\end{theorem}

\subsection{Lax representation}

Lax matrix $(P,Q):\;{\cal MT}\mapsto\g\otimes\g$:
\begin{eqnarray}
P(q,p,\lambda) & = & 
\sum_{k=1}^N(1+\epsilon p_k)E_{kk}+\epsilon\lambda\sum_{k=1}^N E_{k+1,k}\;,
\label{MTL P}\\  \nonumber\\
Q(q,p,\lambda) & = &
I+\epsilon\lambda^{-1}\sum_{k=1}^N q_k E_{k,k+1}\;.\label{MTL Q}
\end{eqnarray}
Notice that the formulas for the Miura map $\rM_1(\epsilon)$ are equivalent
to the factorization
\begin{equation}\label{TL to MTL fact 1}
I+\epsilon T(a,b,\lambda)=P(q,p,\lambda)Q(q,p,\lambda)\;,
\end{equation}
while the formulas for the Miura map $\rM_2(\epsilon)$ are equivalent
to the factorization
\begin{equation}\label{TL to MTL fact 2}
I+\epsilon T(a,b,\lambda)=Q(q,p,\lambda)P(q,p,\lambda)\;.
\end{equation}
\vspace{2mm}

Lax representation for MTL$(\epsilon)$:
\begin{equation}\label{MTL Lax triads}
\left\{\begin{array}{l}
\dot{P} \;=\; PB_2-B_1P\;=\;A_1P-PA_2\;,\\ \\
\dot{Q} \;=\; QB_1-B_2Q\;=\;A_2Q-QA_1\;,
\end{array}\right.
\end{equation}
where
\begin{eqnarray}
B_1\;=\;\pi_+\Big((PQ-I)/\epsilon\Big) & = & 
\sum_{k=1}^N(p_k+\epsilon q_{k-1})E_{kk}+
\lambda\sum_{k=1}^N E_{k+1,k}\;,\label{MTL B1}\\ \nonumber\\
B_2\;=\;\pi_+\Big((QP-I)/\epsilon\Big) & = & 
\sum_{k=1}^N(p_k+\epsilon q_k)E_{kk}+
\lambda\sum_{k=1}^N E_{k+1,k}\;,\label{MTL B2}\\ \nonumber\\
A_1\;=\;\pi_-\Big((PQ-I)/\epsilon\Big) & = & 
\lambda^{-1}\sum_{k=1}^N q_k(1+\epsilon p_k)E_{k,k+1}\;,
\label{MTL A1}\\    \nonumber\\
A_2\;=\;\pi_-\Big((QP-I)/\epsilon\Big) & = & 
\lambda^{-1}\sum_{k=1}^N q_k(1+\epsilon p_{k+1})E_{k,k+1}\;.
\label{MTL A2} 
\end{eqnarray}
The formulas (\ref{MTL B1}), (\ref{MTL B2}) have to be compared with
(\ref{TL B}) and the definition of the Miura maps $\rM_{1,2}1(\epsilon)$, 
while the formulas (\ref{MTL A1}), (\ref{MTL A2}) have to be
compared with (\ref{TL A}).

\subsection{Discretization}

Lax representation for the map ${\rm dMTL}(\epsilon)$:
\begin{equation}\label{dMTL Lax in g+g}
\wiP=\mbB_1^{-1} P \mbB_2=\mbA_1 P \mbA_2^{-1}\,, \qquad 
\wQ=\mbB_2^{-1} Q \mbB_1=\mbA_2 Q \mbA_1^{-1}
\end{equation}
with
\begin{eqnarray}
\mbB_1\;=\;\Pi_+\Big(I+\frac{h}{\epsilon}\,(PQ-I)\Big) & = &
\sum_{k=1}^N\beta_k^{(1)}E_{kk}+h\lambda\sum_{k=1}^{N}E_{k+1,k}\;,
\label{dMTL B1}\\
\mbB_2\;=\;\Pi_+\Big(I+\frac{h}{\epsilon}\,(QP-I)\Big) & = &
\sum_{k=1}^N\beta_k^{(2)}E_{kk}+h\lambda\sum_{k=1}^{N}E_{k+1,k}\;,
\label{dMTL B2}\\
\mbA_1\;=\;\Pi_-\Big(I+\frac{h}{\epsilon}\,(PQ-I)\Big) & = &
I+h\lambda^{-1}\sum_{k=1}^N\alpha_k^{(1)}E_{k,k+1}\;,
\label{dMTL A1}\\
\mbA_2\;=\;\Pi_-\Big(I+\frac{h}{\epsilon}\,(QP-I)\Big) & = &
I+h\lambda^{-1}\sum_{k=1}^N\alpha_k^{(2)}E_{k,k+1}\;.
\label{dMTL A2}
\end{eqnarray}

\subsection{Application: localizing change of variables for dTL}

Consider again the discretization dTL of the Toda lattice. Compare the formula
\[
I+hT=\mbB\mbA\;,
\]
following from (\ref{dTL B}), (\ref{dTL A}), with (\ref{TL to MTL fact 1}).
This comparison shows immediately that the Miura map $\rM_1(h)$ plays the 
role of the localizing change of variables for dTL. Indeed, define the 
change of variables $\cT(\ba,\bb)\mapsto\cT(a,b)$ by the formulas:
\begin{equation}\label{dTL loc map}
\rM_1(h):\quad\left\{\begin{array}{l}
b_k=\bb_k+h\ba_{k-1}\;,\\ \\
a_k=\ba_k(1+h\bb_k)\;.\end{array}\right.
\end{equation}
Then we find:
\begin{eqnarray*}
\mbB=P(\ba,\bb,\lambda) & = & \sum_{k=1}^N (1+h\bb_k)E_{kk}+
h\lambda\sum_{k=1}^N E_{k+1,k}\;,\\
\mbA=Q(\ba,\bb,\lambda) & = & I+h\lambda^{-1}\sum_{k=1}^N \ba_kE_{k,k+1}\;,
\end{eqnarray*}
so that we immediately get local expressions for the entries of the factors 
$\mbB$, $\mbA\,$: 
\begin{equation}\label{dTL loc beta}
\beta_k=1+h\bb_k\;, \qquad \alpha_k=\ba_k\;.
\end{equation}
\begin{theorem}\label{local dTL}
The change of variables {\rm(\ref{dTL loc map})} conjugates 
the map {\rm dTL} with the map on $\cT(\ba,\bb)$ described by the 
following local equations of motion:
\begin{equation}\label{dTL loc}
\widetilde{\bb}_k+h\widetilde{\ba}_{k-1}=\bb_k+h\ba_k\;,\qquad
\widetilde{\ba}_k(1+h\widetilde{\bb}_k)=\ba_k(1+h\bb_{k+1})\;.
\end{equation}
\end{theorem}
So, the local form of dTL (\ref{dTL loc}) actually belongs to the hierarchy
${\rm MTL}(h)$.
\vspace{2mm}

{\bf Corollary.} {\it The local form of {\rm dTL (\ref{dTL loc})} is Poisson 
with respect to the following brackets on $\cT(\ba,\bb)$:
\begin{equation}\label{dTL loc br 1}
\{\bb_k,\ba_k\}=-\ba_k(1+h\bb_k)\;,\qquad 
\{\ba_k,\bb_{k+1}\}=-\ba_k(1+h\bb_{k+1})\;,
\end{equation}
which is the pull--back of the bracket
\begin{equation}\label{dTL loc m1 br}
\{\cdot,\cdot\}_1+h\{\cdot,\cdot\}_2 
\end{equation}
on $\cT(a,b)$ under the change of variables {\rm(\ref{dTL loc map})}, and
\begin{eqnarray}
\{\bb_k,\ba_k\} & = & -\ba_k(\bb_k+h\ba_k)(1+h\bb_k)\;,\nonumber\\
\{\ba_k,\bb_{k+1}\} & = & -\ba_k(\bb_{k+1}+h\ba_k)(1+h\bb_{k+1})\;,
\nonumber\\ 
\{\bb_k,\bb_{k+1}\} & = & -\ba_k(1+h\bb_k)(1+h\bb_{k+1})\;,\nonumber\\
\{\ba_k,\ba_{k+1}\} & = & -\ba_k\ba_{k+1}(1+h\bb_{k+1})\;,
\label{dTL loc br 2}
\end{eqnarray}
which is the pull--back of the bracket
\begin{equation}\label{dTL loc m2 br}
\{\cdot,\cdot\}_2+h\{\cdot,\cdot\}_3 
\end{equation}
on $\cT(a,b)$ under the change of variables {\rm(\ref{dTL loc map})}.}

\setcounter{equation}{0}
\section{Double modified Toda lattice}\label{Sect M2TL}

\subsection{Equations of motion}

The phase space of the {\it double modified Toda lattice} 
${\rm M^2TL}(\epsilon,\delta)$ will be denoted by
$\cM^2\cT={\Bbb R}^{2N}(r,s)$. The parameters $\epsilon$, $\delta$ will be
called the {\it modification parameters}. The equations of motion read:
\begin{equation}\label{M2TL}
\dot{r}_k=(1+\epsilon r_k)(1+\delta r_k)(s_k-s_{k-1})\;,\qquad
\dot{s}_k=s_k(1+\epsilon\delta s_k)(r_{k+1}-r_k)\;.
\end{equation}
Obviously, the parameters $\epsilon$, $\delta$ enter into the system 
${\rm M^2TL}(\epsilon,\delta)$ symmetrically. 

\subsection{Hamiltonian structure}
\begin{proposition}\label{M2TL Ham str}
The relations
\begin{eqnarray}
\{r_k,s_k\}_{123} & = & 
-s_k(1+\epsilon\delta s_k)(1+\epsilon r_k)(1+\delta r_k)\;, \nonumber\\
\label{M2TL br}\\
\{s_k,r_{k+1}\}_{123} & = & -s_k(1+\epsilon\delta s_k)(1+\epsilon r_{k+1})
(1+\delta r_{k+1}) \nonumber
\end{eqnarray}
define a Poisson bracket on $\cM^2\cT$. 
The system ${\rm M^2TL}(\epsilon,\delta)$ is Hamiltonian with respect to this 
bracket, with the Hamilton function
\begin{equation}\label{M2TL H}
\rF_0(r,s)=(\epsilon\delta)^{-1}\sum_{k=1}^N\log
(1+\delta r_k)+(\epsilon\delta)^{-1}\sum_{k=1}^N\log(1+\epsilon\delta s_k)\;.
\end{equation}
\end{proposition}
The Hamilton function $\rF_0(r,s)$ is given in the form in which the 
$\delta\to 0$ limit is transparent. By adding 
\[
(2\epsilon\delta)^{-1}\sum_{k=1}^N\log\frac{1+\epsilon r_k}{1+\delta r_k}\;,
\]
which is a Casimir function for the bracket $\{\cdot,\cot\}_{123}$, we find
an equivalent Hamilton function symmetric in $\epsilon$, $\delta$.
 
\subsection{Miura relations}
\begin{theorem}\label{Miuras M2TL to MTL}
Define the Miura maps 
$\rM_{1,2}(\epsilon;\delta):\;\cM^2\cT(r,s)\mapsto\cM\cT(q,p)$ by:
\begin{equation}\label{M2TL Miuras}
\rM_1(\epsilon;\delta):\;\left\{\begin{array}{l} 
p_k=r_k+\delta s_{k-1}(1+\epsilon r_k)\;,\\ \\ q_k=s_k(1+\delta r_k)\;, 
\end{array}\right.  \qquad
\rM_2(\epsilon;\delta):\;\left\{\begin{array}{l} 
p_k=r_k+\delta s_k(1+\epsilon r_k)\;,\\ \\ q_k=s_k(1+\delta r_{k+1})\;.
\end{array}\right.
\end{equation}
Both maps $\rM_{1,2}(\epsilon;\delta)$ are Poisson, if $\cM^2\cT(y,z)$ is 
equipped with the bracket $\{\cdot,\cdot\}_{123}$, and $\cM\cT(q,r)$ is 
equipped with $\{\cdot,\cdot\}_{12}+\delta\{\cdot,\cdot\}_{23}$. The pull--back 
of the flow {\rm MTL}$(\epsilon)$ under each one of the Miura maps 
$\rM_{1,2}(\epsilon;\delta)$ coincides with ${\rm M^2TL}(\epsilon,\delta)$.
\end{theorem}

Note that introducing the Miura maps $\rM_{1,2}(\epsilon;\delta)$ we actually 
break the symmetry of ${\rm M^2TL}(\epsilon,\delta)$ in the modification 
parameters. This symmetry may be explained with the help of the following
statement concerning the permutability of the Miura maps.

\begin{theorem}\label{M2TL Miuras permutability}
The following diagram is commutative for all $i=1,2$, $j=1,2$:
\begin{center}
\unitlength1cm
\begin{picture}(9,6.5)
\put(3.5,1.1){\vector(1,0){2}}
\put(3.5,5.1){\vector(1,0){2}}
\put(2,4.1){\vector(0,-1){2}}
\put(7,4.1){\vector(0,-1){2}}
\put(1,0.6){\makebox(2,1){$\cM\cT$}} 
\put(1,4.6){\makebox(2,1){$\cM^2\cT$}}
\put(6,4.6){\makebox(2,1){$\cM\cT$}}
\put(6,0.6){\makebox(2,1){$\cT$}}
\put(0,2.6){\makebox(2,1){$\rM_i(\epsilon;\delta)$}}
\put(7,2.6){\makebox(2,1){$\rM_i(\delta)$}}
\put(3.8,-0.2){\makebox(1.4,1.4){$\rM_j(\epsilon)$}}
\put(3.8,5.0){\makebox(1.4,1.4){$\rM_j(\delta;\epsilon)$}}
\end{picture}
\end{center}
\end{theorem}
In particular, the compositions $\rM_j(\epsilon)\circ\rM_j(\epsilon;\delta)$ 
for $j=1,2$ have to be symmetric in $\epsilon$, $\delta$. Indeed, we have:
\begin{equation}\label{TL to M2TL map 1}
\rM_1(\epsilon)\circ \rM_1(\epsilon;\delta):\;\left\{\begin{array}{l}
b_k=r_k+(\epsilon+\delta)s_{k-1}+\epsilon\delta s_{k-1}(r_{k-1}+r_k)\;, \\ \\
a_k=s_k(1+\epsilon r_k)(1+\delta r_k)(1+\epsilon\delta s_{k-1})\;,
\end{array}\right.
\end{equation}
\begin{equation}\label{TL to M2TL map 2}
\rM_2(\epsilon)\circ \rM_2(\epsilon;\delta):\;\left\{\begin{array}{l}
b_k=r_k+(\epsilon+\delta)s_k+\epsilon\delta s_k(r_k+r_{k+1})\;, \\ \\
a_k=s_k(1+\epsilon r_{k+1})(1+\delta r_{k+1})(1+\epsilon\delta s_{k+1})\;.
\end{array}\right.
\end{equation}
This explains also the symmetry of the Poisson bracket $\{\cdot,\cdot\}_{123}$
(\ref{M2TL br}), since the above compositions are Poisson maps between 
$\cM^2\cT(r,s)$ carrying this bracket, and $\cT(a,b)$ carrying the bracket
\[
\{\cdot,\cdot\}_1+(\epsilon+\delta)\{\cdot,\cdot\}_2+\epsilon\delta
\{\cdot,\cdot\}_3\;.
\]

\subsection{Lax representation and discretization}

Unfortunately, we do not know any interpretation of the Miura maps
(\ref{M2TL Miuras}) in terms of factorization of Lax matrices. 
The only way we can define the discretization ${\rm dM^2TL}(\epsilon,\delta)$ 
is to pull--back the map ${\rm dMTL}(\epsilon)$ under either of the 
Miura transformations $\rM_{1,2}(\epsilon;\delta)$, or to pull--back the 
map ${\rm dMTL}(\delta)$ under either of the Miura transformations 
$\rM_{1,2}(\delta;\epsilon)$.

\subsection{Application: localizing change of variables for 
${\rm dMTL}(\epsilon)$}

It turns out that the Miura map $\rM_1(\epsilon;h)$ plays the role of the
localizing change of variables for ${\rm dMTL}(\epsilon)$. To demonstrate this,
consider the following change of variables $\cM\cT(\bq,\bp)\mapsto
\cM\cT(q,p)$:
\begin{equation}\label{dMTL loc map}
\rM_1(\epsilon;h):\; \left\{\begin{array}{l}
p_k=\bp_k+h\bq_{k-1}(1+\epsilon\bp_k)\;,\\ \\ q_k=\bq_k(1+h\bp_k)\;.
\end{array}\right.
\end{equation}
Then the following expressions may be found for the entries of the factors 
$\mbB_j$, $\mbA_j$ $(j=1,2)$:
\begin{eqnarray}
\beta_k^{(1)}=(1+h\bp_k)(1+h\epsilon\bq_{k-1})\;, & \quad & 
\alpha_k^{(1)}=\bq_k(1+\epsilon\bp_k)\;, \label{dMTL beta in loc map}\\
\nonumber\\
\beta_k^{(2)}=(1+h\bp_k)(1+h\epsilon\bq_k)\;, & \quad &
\alpha_k^{(2)}=\bq_k(1+\epsilon\bp_{k+1})\;.\label{dMTL alpha in loc map}
\end{eqnarray}
\begin{theorem}
The change of variables {\rm(\ref{dMTL loc map})} conjugates the map 
${\rm dMTL}(\epsilon)$ with the map on $\cM\cT(\bq,\bp)$ described
by the following local equations of motion:
\begin{equation}\label{dMTL loc}
\widetilde{\bp}_k+h\widetilde{\bq}_{k-1}(1+\epsilon\widetilde{\bp}_k)
=\bp_k+h\bq_k(1+\epsilon\bp_k)\;,\qquad
\widetilde{\bq}_k(1+h\widetilde{\bp}_k)=\bq_k(1+h\bp_{k+1})\;.
\end{equation}
\end{theorem}
So, the local form of ${\rm dMTL}(\epsilon)$ (\ref{dMTL loc}) actualy
belongs to the hierarchy ${\rm M^2TL}(\epsilon;h)$.
\vspace{2mm}

{\bf Corollary.} {\it The local form of ${\rm dMTL}(\epsilon)$ 
{\rm (\ref{dMTL loc})} is a Poisson map with respect to the following 
bracket on ${\cal MT}(\bq,\br)$:
\begin{eqnarray}
\{\bp_k,\bq_k\} & = & -\bq_k(1+h\epsilon\bq_k)(1+\epsilon\bp_k)(1+h\bp_k)\;,
\nonumber\\   \label{dMTL loc br}\\
\{\bq_k,\bp_{k+1}\} & = & -\bq_k(1+h\epsilon\bq_k)(1+\epsilon\bp_{k+1})
(1+h\bp_{k+1})\;,  \nonumber
\end{eqnarray}
which is the pull--back of the bracket
\begin{equation}\label{MTL mixed br}
\{\cdot,\cdot\}_{12}+h\{\cdot,\cdot\}_{23}
\end{equation}
on ${\cal MT}(q,p)$ under the change of variables {\rm(\ref{dMTL loc map})}.}
\vspace{2mm}

There is an interesting question on the relation between the local forms
of dTL and dMTL$(\epsilon)$. In other words, how do the Miura maps 
$\rM_{1,2}(\epsilon)$ look when seen through the localizing changes of 
variables? The answer is given by Theorem \ref{M2TL Miuras permutability}, 
which in the present case may be reformulated as follows:
\begin{theorem}
{\rm a)} The following diagram is commutative for $j=1,2$:
\begin{center}
\unitlength1cm
\begin{picture}(9,6.5)
\put(3.5,1.1){\vector(1,0){2}}
\put(3.5,5.1){\vector(1,0){2}}
\put(2,4.1){\vector(0,-1){2}}
\put(7,4.1){\vector(0,-1){2}}
\put(1,0.6){\makebox(2,1){$\cM\cT(q,p)$}} 
\put(1,4.6){\makebox(2,1){$\cM\cT(\bq,\bp)$}}
\put(6,4.6){\makebox(2,1){$\cT(\ba,\bb)$}}
\put(6,0.6){\makebox(2,1){$\cT(a,b)$}}
\put(0,2.6){\makebox(2,1){$\rM_1(\epsilon;h)$}}
\put(7,2.6){\makebox(2,1){$\rM_1(h)$}}
\put(3.8,-0.2){\makebox(1.4,1.4){$\rM_j(\epsilon)$}}
\put(3.8,5.0){\makebox(1.4,1.4){$\rM_j(h;\epsilon)$}}
\end{picture}
\end{center}
where the maps $\rM_{1,2}(h;\epsilon)$ are given by
\begin{eqnarray}
\rM_1(h;\epsilon): & \; & \left\{\begin{array}{l} 
\bb_k=\bp_k+\epsilon\bq_{k-1}(1+h\bp_k)\;,\\ \\ 
\ba_k=\bq_k(1+\epsilon\bp_k)\;, 
\end{array}\right. \label{MTL Miura 1 in loc map}\\ 
\nonumber\\  \nonumber\\
\rM_2(h;\epsilon): & \; & \left\{\begin{array}{l} 
\bb_k=\bp_k+\epsilon\bq_k(1+h\bp_k)\;,\\ \\ 
\ba_k=\bq_k(1+\epsilon\bp_{k+1})\;.
\end{array}\right.  \label{MTL Miura 2 in loc map}
\end{eqnarray}

{\rm b)} Either of the Miura maps $\rM_{1,2}(h;\epsilon)$ conjugates the local 
form of ${\rm dMTL}(\epsilon)$ {\rm(\ref{dMTL loc})} with the local form of
{\rm dTL} {\rm(\ref{dTL loc})}.
\end{theorem}

\setcounter{equation}{0}
\section{Triple modified Toda lattice}\label{Sect M3TL}

\subsection{Equations of motion}

The phase space of the {\it triple modified Toda lattice} 
${\rm M^3TL}(\epsilon,\delta,\gamma)$ will be denoted by 
$\cM^3\cT={\Bbb R}^{2N}(\mba,\mbb)$. Three parameters
$\epsilon$, $\delta$, $\gamma$, all with equal rights, will be called the
{\it modification parameters}. Equations of motion of this system read:
\begin{equation}\label{M3TL}
\left\{\begin{array}{l}
\dot{\mbb}_k = (1+\epsilon\mbb_k)(1+\delta\mbb_k)(1+\gamma\mbb_k)
\left(\displaystyle\frac{\mba_k}{1-\epsilon\delta\gamma\mbb_k\mba_k}-
\displaystyle\frac{\mba_{k-1}}{1-\epsilon\delta\gamma\mbb_k\mba_{k-1}}\right)\;, 
\\ \\
\dot{\mba}_k = \mba_k(1+\epsilon\delta\mba_k)(1+\epsilon\gamma\mba_k)
(1+\delta\gamma\mba_k)\left(\displaystyle
\frac{\mbb_{k+1}}{1-\epsilon\delta\gamma\mbb_{k+1}\mba_k}-
\displaystyle\frac{\mbb_k}{1-\epsilon\delta\gamma\mbb_k\mba_k}\right)\;. 
\end{array}\right.
\end{equation}

\subsection{Miura relations}

Actually, the available information about ${\rm M^3TL}(\epsilon,\delta,\gamma)$
is not very rich. Even a local Hamiltonian structure is not known (and
presumably does not exist). All we know is the Miura relation to
${\rm M^2TL}(\epsilon,\delta)$.
\begin{theorem}\label{Miuras for M3TL}
Define the Miura maps $\rM_{1,2}(\epsilon,\delta;\gamma):\;
\cM^3\cT(\mba,\mbb)\mapsto\cM^2\cT(r,s)$ by:
\begin{eqnarray}
\rM_1(\epsilon,\delta;\gamma): & \quad &
\left\{\begin{array}{l}
r_k=\displaystyle
\frac{\mbb_k+\gamma\mba_{k-1}+\gamma(\epsilon+\delta)\mbb_k\mba_{k-1}}
{1-\epsilon\delta\gamma\mbb_k\mba_{k-1}}\;,\\ \\
s_k=\displaystyle
\frac{\mba_k(1+\gamma\mbb_k)}{1-\epsilon\delta\gamma\mbb_k\mba_k}\;, 
\end{array}\right.
\label{M3TL Miura 1}\\ \nonumber\\ \nonumber\\
\rM_2(\epsilon,\delta;\gamma): & \quad &
\left\{\begin{array}{l}
r_k=\displaystyle\frac{\mbb_k+\gamma\mba_k+\gamma(\epsilon+\delta)\mbb_k\mba_k}
{1-\epsilon\delta\gamma\mbb_k\mba_k}\;,\\ \\
s_k=\displaystyle\frac
{\mba_k(1+\gamma\mbb_{k+1})}{1-\epsilon\delta\gamma\mbb_{k+1}\mba_k}\;. 
\end{array}\right.
\label{M3TL Miura 2}
\end{eqnarray}
Than the pull--back of the flow ${\rm M^2TL}(\epsilon,\delta)$ under either
of the Miura maps $\rM_{1,2}(\epsilon,\delta;\gamma)$ 
coincides with ${\rm M^3TL}(\epsilon,\delta,\gamma)$.
\end{theorem}
For these Miura maps again a permutability statement holds:
\begin{theorem}\label{M3TL Miuras permutability}
The following diagram is commutative for all $i=1,2$, $j=1,2$:
\begin{center}
\unitlength1cm
\begin{picture}(9,6.5)
\put(3.5,1.1){\vector(1,0){2}}
\put(3.5,5.1){\vector(1,0){2}}
\put(2,4.1){\vector(0,-1){2}}
\put(7,4.1){\vector(0,-1){2}}
\put(1,0.6){\makebox(2,1){$\cM^2\cT$}} 
\put(1,4.6){\makebox(2,1){$\cM^3\cT$}}
\put(6,4.6){\makebox(2,1){$\cM^2\cT$}}
\put(6,0.6){\makebox(2,1){$\cM\cT$}}
\put(0,2.6){\makebox(2,1){$\rM_i(\epsilon,\delta;\gamma)$}}
\put(7,2.6){\makebox(2,1){$\rM_i(\epsilon;\gamma)$}}
\put(3.8,-0.2){\makebox(1.4,1.4){$\rM_j(\epsilon;\delta)$}}
\put(3.8,5.0){\makebox(1.4,1.4){$\rM_j(\epsilon,\gamma;\delta)$}}
\end{picture}
\end{center}
\end{theorem}

It is of some interest to consider the compositions of the Miura
maps $\rM_j(\epsilon)\circ \rM_j(\epsilon;\delta) \circ 
\rM_j(\epsilon,\delta;\gamma):\;\cM^3\cT(\mba,\mbb)\mapsto\cT(a,b)$ 
bringing ${\rm M^3TL}(\epsilon,\delta,\gamma)$ directly into TL.
These compositions depend symmetrically on all three
modification parameters $\epsilon$, $\delta$, $\gamma$. We give here only
the formulas for the case $j=1$:
\begin{eqnarray*}
a_k & = & \frac{\mba_k(1+\epsilon\mbb_k)(1+\delta\mbb_k)(1+\gamma\mbb_k)
(1+\epsilon\delta\mba_{k-1})(1+\epsilon\gamma\mba_{k-1})
(1+\delta\gamma\mba_{k-1})}
{(1-\epsilon\delta\gamma\mbb_k\mba_k)(1-\epsilon\delta\gamma\mbb_k\mba_{k-1})^2
(1-\epsilon\delta\gamma\mbb_{k-1}\mba_{k-1})}\;,\\ \\
b_k & = & \mbb_k+\frac{\epsilon\delta\gamma\mba_{k-1}}{1-\epsilon\delta\gamma
\mbb_{k-1}\mba_{k-1}}\left(\mbb_k\mbb_{k-1}+\frac{\mbb_k^2+\mba_{k-1}}
{1-\epsilon\delta\gamma\mbb_k\mba_{k-1}}+\frac{\mbb_{k-1}^2+\mba_{k-2}}
{1-\epsilon\delta\gamma\mbb_{k-1}\mba_{k-2}}\right)\\
 & & +\,\frac{(\epsilon\delta+\epsilon\gamma+\delta\gamma)\mba_{k-1}}
{1-\epsilon\delta\gamma\mbb_{k-1}\mba_{k-1}}\left(\frac{\mbb_k}
{1-\epsilon\delta\gamma\mbb_k\mba_{k-1}}+\frac{\mbb_{k-1}}
{1-\epsilon\delta\gamma\mbb_{k-1}\mba_{k-2}}\right)\\
 & & +\,\frac{(\epsilon+\delta+\gamma)\mba_{k-1}}
{1-\epsilon\delta\gamma\mbb_{k-1}\mba_{k-1}}\left(\frac{1}
{1-\epsilon\delta\gamma\mbb_k\mba_{k-1}}+\frac{1}
{1-\epsilon\delta\gamma\mbb_{k-1}\mba_{k-2}}-1\right)\;.
\end{eqnarray*}

\subsection{Application: localizing change of variables for 
${\rm dM^2TL}(\epsilon,\delta)$}
It turns out that the Miura map $\rM_1(\epsilon,\delta;h)$ plays the role
of the localizing change of variables for the map
${\rm dM^2TL}(\epsilon,\delta)$. Indeed, consider the following change 
of variables $\cM^2\cT(\br,\bs)\mapsto\cM^2\cT(r,s)$:
\begin{equation}\label{dM2TL loc map}
\rM_1(\epsilon,\delta;h):\,
\left\{\begin{array}{l}
r_k=\displaystyle\frac{\br_k+h\bs_{k-1}+h(\epsilon+\delta)\br_k\bs_{k-1}}
{1-h\epsilon\delta\br_k\bs_{k-1}}\;,\\ \\
s_k=\displaystyle\frac{\bs_k(1+h\br_k)}{1-h\epsilon\delta\br_k\bs_k}\;.
\end{array}\right.
\end{equation}
It is useful to notice that the first formula in 
(\ref{dM2TL loc map}) may be equivalently rewritten as
\begin{equation}\label{dM2TL loc aux2}
1+\epsilon r_k=\frac{(1+\epsilon\br_k)(1+h\epsilon\bs_{k-1})}
{1-h\epsilon\delta\br_k\bs_{k-1}}\quad\Leftrightarrow\quad
1+\delta r_k=\frac{(1+\delta\br_k)(1+h\delta\bs_{k-1})}
{1-h\epsilon\delta\br_k\bs_{k-1}}\;,
\end{equation}
or else as
\begin{equation}\label{dM2TL loc aux8}
\frac{1+\epsilon r_k}{1+\delta r_k}=
\frac{(1+\epsilon\br_k)(1+h\epsilon\bs_{k-1})}
{(1+\delta\br_k)(1+h\delta\bs_{k-1})}\;,
\end{equation}
while the second formula in (\ref{dM2TL loc map}) has the following two 
equivalent forms:
\begin{equation}\label{dM2TL loc aux1}
1+\epsilon\delta s_k=\frac{1+\epsilon\delta\bs_k}
{1-h\epsilon\delta\br_k\bs_k}\;,
\end{equation}
and
\begin{equation}\label{dM2TL loc aux7}
\frac{s_k}{1+\epsilon\delta s_k}=\frac{\bs_k}{1+\epsilon\delta\bs_k}\,
(1+h\br_k)\;.
\end{equation}
\begin{theorem} The change of variables {\rm(\ref{dM2TL loc map})}
conjugates the map $\;{\rm dM^2TL}(\epsilon,\delta)$ with the map
on $\cM^2\cT(\br,\bs)$ described by the following
local equations of motion:
\begin{equation}\label{dM2TL loc r}
\frac{1+\epsilon\widetilde{\br}_k}{1+\delta\widetilde{\br}_k}
\cdot\frac{1+h\epsilon\widetilde{\bs}_{k-1}}
{1+h\delta\widetilde{\bs}_{k-1}}=
\frac{1+\epsilon\br_k}{1+\delta\br_k}\cdot\frac{1+h\epsilon\bs_k}
{1+h\delta\bs_k}\;,
\end{equation}
\begin{equation}\label{dM2TL loc s}
\frac{\widetilde{\bs}_k}{1+\epsilon\delta\widetilde{\bs}_k}\,
(1+h\widetilde{\br}_k)=\frac{\bs_k}{1+\epsilon\delta\bs_k}\,
(1+h\br_{k+1})\;.
\end{equation}
\end{theorem}

So, the local form (\ref{dM2TL loc r}), (\ref{dM2TL loc s}) belongs 
actually to the hierarchy ${\rm M^3TL}(\epsilon,\delta,h)$.
We discuss now the relation between the local forms of the maps 
${\rm dMTL}(\epsilon)$ and ${\rm dM^2TL}(\epsilon,\delta)$. We have to 
determine how do the Miura transformations $\rM_{1,2}(\epsilon;\delta)$ 
look through the localizing changes of variables. The answer is given by 
Theorem \ref{M3TL Miuras permutability}, which now takes the following form.
\begin{theorem}
{\rm a)} The following diagram is commutative for $j=1,2$:
\begin{center}
\unitlength1cm
\begin{picture}(9,6.5)
\put(3.5,1.1){\vector(1,0){2}}
\put(3.5,5.1){\vector(1,0){2}}
\put(2,4.1){\vector(0,-1){2}}
\put(7,4.1){\vector(0,-1){2}}
\put(1,0.6){\makebox(2,1){$\cM^2\cT(r,s)$}} 
\put(1,4.6){\makebox(2,1){$\cM^2\cT(\br,\bs)$}}
\put(6,4.6){\makebox(2,1){$\cM\cT(\bq,\bp)$}}
\put(6,0.6){\makebox(2,1){$\cM\cT(q,p)$}}
\put(0,2.6){\makebox(2,1){$\rM_1(\epsilon,\delta;h)$}}
\put(7,2.6){\makebox(2,1){$\rM_1(\epsilon;h)$}}
\put(3.8,-0.2){\makebox(1.4,1.4){$\rM_j(\epsilon;\delta)$}}
\put(3.8,5.0){\makebox(1.4,1.4){$\rM_j(\epsilon,h;\delta)$}}
\end{picture}
\end{center}
where the maps $\rM_{1,2}(\epsilon,h;\delta)
:\,\cM^2\cT(\br,\bs)\mapsto \cM\cT(\bq,\bp)$ are given by
\begin{eqnarray}
\rM_1(\epsilon,h;\delta): & \quad & \left\{\begin{array}{l}
\bp_k=\displaystyle
\frac{\br_k+\delta\bs_{k-1}+\delta(\epsilon+h)\br_k\bs_{k-1}}
{1-h\epsilon\delta\br_k\bs_{k-1}}\;,
\\ \\ 
\bq_k=\displaystyle\frac{\bs_k(1+\delta\br_k)}{1-h\epsilon\delta\br_k\bs_k}\;,
\end{array}\right. 
\label{M2TL Miura 1 in loc map}\\ \nonumber\\\nonumber\\
\rM_2(\epsilon,h;\delta): & \quad & \left\{\begin{array}{l}
\bp_k=\displaystyle\frac{\br_k+\delta\bs_k+\delta(\epsilon+h)\br_k\bs_k}
{1-h\epsilon\delta\br_k\bs_k}\;,\\ \\
\bq_k=\displaystyle
\frac{\bs_k(1+\delta\br_{k+1})}{1-h\epsilon\delta\br_{k+1}\bs_k}\;. 
\end{array}\right.
\label{M2TL Miura 2 in loc map} 
\end{eqnarray}

{\rm b)} Either of the Miura maps $\rM_{1,2}(\epsilon,h;\delta)$ conjugates
the local form of ${\rm dM^2TL}(\epsilon,\delta)$ {\rm (\ref{dM2TL loc r}),
(\ref{dM2TL loc s})} with the local form of ${\rm dMTL}(\epsilon)$ 
{\rm (\ref{dMTL loc})}.
\end{theorem}

\setcounter{equation}{0}
\section{Volterra lattice}
\subsection{Equations of motion}
Actually, the {\it Volterra lattice} VL is a more symmetric (and parameter free)
version of the modified Toda lattice. However, as we shall see later on,
analogous statement does not hold anymore for relativistic generalizations.
The phase space of VL will be denoted through 
$\cV={\Bbb R}^{2N}(u,v)$. The equations of motion read:
\begin{equation}\label{VL}
\dot{u}_k=u_k(v_k-v_{k-1})\;,\qquad \dot{v}_k=v_k(u_{k+1}-u_{k})\;.
\end{equation}

\subsection{Bi--Hamiltonian structure}
\begin{proposition}\label{quadratic PB for VL} 
{\rm a)} The relations
\begin{equation}\label{VL q br}
\{u_k,v_k\}_2=-u_kv_k\;,\qquad \{v_k,u_{k+1}\}_2=-v_ku_{k+1}
\end{equation}
define a Poisson bracket on $\cV$. The flow {\rm VL} is a Hamiltonian system 
on $\Big(\cV,\{\cdot,\cdot\}_2\Big)$ with the Hamilton function
\begin{equation}\label{VL q Ham}
\rH_1(u,v)=\sum_{k=1}^N u_k+\sum_{k=1}^N v_k\;.
\end{equation}

{\rm b)} The relations
\begin{equation}\label{VL c br}
\begin{array}{cclcccl}
\{u_k,v_k\}_3 & = & -u_kv_k(u_k+v_k)\;, & \quad &
\{v_k,u_{k+1}\}_3 & = & -v_ku_{k+1}(v_k+u_{k+1})\;,\\ \\
\{u_k,u_{k+1}\}_3 & = & -u_kv_ku_{k+1}\;, & \quad &
\{v_k,v_{k+1}\}_3 & = & -v_ku_{k+1}v_{k+1}
\end{array}
\end{equation}
define a Poisson bracket on $\cV$ compatible with $\{\cdot,\cdot\}_2$. 
The flow {\rm VL} is a Hamiltonian system on $\Big(\cV,\{\cdot,\cdot\}_3\Big)$ 
with the Hamilton function
\begin{equation}\label{VL c Ham}
\rH_0(u,v)=\sum_{k=1}^N \log(u_k)\;,\quad {\rm or}\quad
\rH_0(u,v)=\sum_{k=1}^N \log(v_k)\;.
\end{equation}
\end{proposition}
(The difference of these two functions is a Casimir of the bracket 
$\{\cdot,\cdot\}_3$.)

\subsection{Miura relations}

Define the Miura maps $M_{1,2}:\cV\mapsto\cT$:
\begin{equation}\label{TL Ms}
M_1:\quad \left\{\begin{array}{l}
b_k=u_k+v_{k-1}\;,\\ \\ a_k=u_kv_k\;,\end{array}\right.\qquad
M_2:\quad \left\{\begin{array}{l}
b_k=u_k+v_k\;,\\ \\ a_k=u_{k+1}v_k\;.\end{array}\right.
\end{equation}
\begin{theorem}
{\rm a)} Both maps $M_{1,2}:\cV\mapsto\cT$ are Poisson, if $\cV$ is equipped 
with the bracket {\rm(\ref{VL q br})}, and $\cT$ is equipped with the bracket 
{\rm(\ref{TL q br})}, and also if $\cV$ is equipped with the bracket 
{\rm(\ref{VL c br})}, and $\cT$ is equipped with the bracket 
{\rm(\ref{TL c br})}. 

{\rm b)} The pull--back of the flow {\rm TL} with respect to either of the
Miura maps $M_{1,2}$ coincides with {\rm VL}.
\end{theorem}

\subsection{Lax representation}

Lax matrix: $(U,V):\cV\mapsto\g\otimes\g$:
\begin{eqnarray}
U(u,v,\lambda) & = & \sum_{k=1}^N u_kE_{kk}+\lambda\sum_{k=1}^N E_{k+1,k}\;,
\label{VL U}\\
V(u,v,\lambda) & = & I+\lambda^{-1}\sum_{k=1}^N v_kE_{k,k+1}\;. 
\label{VL V}
\end{eqnarray}

The origin of the Miura transformations $M_{1,2}$ lies in the following
factorizations of the Toda Lax matrix $T(a,b,\lambda)$: the formulas for
the map $M_1$ are equivalent to
\begin{equation}\label{TL T fact}
T(a,b,\lambda)=U(u,v,\lambda)V(u,v,\lambda)\;,
\end{equation}
while the formulas for the map $M_2$ are equivalent to
\begin{equation}\label{TL T fact-}
T(a,b,\lambda)=V(u,v,\lambda)U(u,v,\lambda)\;.
\end{equation}

Lax representation for VL:
\begin{equation}\label{VL Lax in g+g}
\left\{\begin{array}{l}
\dot{U}=UB_2-B_1U=C_1U-UC_2\;, \\  \\ 
\dot{V}=VB_1-B_2V=C_2V-VC_1\;,  \end{array}\right.
\end{equation}
with 
\begin{eqnarray}
B_1\;=\;\pi_+(UV) & = & \sum_{k=1}^N (u_k+v_{k-1})E_{kk}+
\lambda\sum_{k=1}^NE_{k+1,k}\;,
\label{VL B1}\\
B_2\;=\;\pi_+(VU) & = & \sum_{k=1}^N (u_k+v_k)E_{kk}+
\lambda\sum_{k=1}^NE_{k+1,k}\;,
\label{VL B2}\\
C_1\;=\;\pi_-(UV) & = & \lambda^{-1}\sum_{k=1}^N u_kv_kE_{k,k+1}\;,
\label{VL C1}\\
C_2\;=\;\pi_-(VU) & = & \lambda^{-1}\sum_{k=1}^N u_{k+1}v_kE_{k,k+1}\;.
\label{VL C2}
\end{eqnarray}

\subsection{Discretization}

Lax representation for the map dVL:
\begin{equation}\label{dVL Lax in g+g}
\wU=\mbB_1^{-1}U\mbB_2=\mbC_1U\mbC_2^{-1}\;, \qquad 
\wV=\mbB_2^{-1}U\mbB_1=\mbC_2V\mbC_1^{-1}
\end{equation}
with
\begin{eqnarray}
\mbB_1\;=\;\Pi_+(I+hUV) & = & 
\sum_{k=1}^N\beta_k^{(1)}E_{kk}+h\lambda\sum_{k=1}^{N}E_{k+1,k}\;,
\label{dVL B1}\\
\mbB_2\;=\;\Pi_+(I+hVU) & = & 
\sum_{k=1}^N\beta_k^{(2)}E_{kk}+h\lambda\sum_{k=1}^{N}E_{k+1,k}\;,
\label{dVL B2}\\
\mbC_1\;=\;\Pi_-(I+hUV) & = & 
I+h\lambda^{-1}\sum_{k=1}^{N}\gamma_k^{(1)}E_{k,k+1}\;,
\label{dVL C1}\\
\mbC_2\;=\;\Pi_-(I+hVU) & = & 
I+h\lambda^{-1}\sum_{k=1}^{N}\gamma_k^{(2)}E_{k,k+1}\;.
\label{dVL C2}
\end{eqnarray}

\setcounter{equation}{0}
\section{Modified Volterra lattice}
\label{Sect MVL}

\subsection{Equations of motion}

The phase space of the {\it modified Volterra lattice}
${\rm MVL}(\epsilon)$ will be denoted by 
$\cM\cV={\Bbb R}^{2N}(y,z)$, the {\it modification 
parameter} here is $\epsilon$, and the equations of motion read:
\begin{equation}\label{MVL}
\dot{y}_k=y_k(1+\epsilon y_k)(z_k-z_{k-1})\;,\qquad
\dot{z}_k=z_k(1+\epsilon z_k)(y_{k+1}-y_k)\;.
\end{equation}

\subsection{Hamiltonian structure}
\begin{proposition}
The relations
\begin{equation}\label{MVL br}
\begin{array}{rcl}
\{y_k,z_k\}_{23} & = & -y_kz_k(1+\epsilon y_k)(1+\epsilon z_k)\;,\\ \\
\{z_k,y_{k+1}\}_{23} & = & -z_ky_{k+1}(1+\epsilon z_k)(1+\epsilon y_{k+1})
\end{array}
\end{equation}
define a Poisson bracket on $\cM\cV(y,z)$. The system ${\rm MVL}(\epsilon)$ 
{\rm (\ref{MVL})} is Hamiltonian with respect to this bracket, with the 
Hamilton function
\begin{equation}\label{MVL H}
\rG_0(y,z)=\epsilon^{-1}\sum_{k=1}^N\log(1+\epsilon y_k)+
\epsilon^{-1}\sum_{k=1}^N\log(1+\epsilon z_k)\;.
\end{equation}
\end{proposition}

\subsection{Miura relations}
The modified Volterra lattice is Miura related to the Volterra lattice 
as well as to the modified Toda lattice.
\begin{theorem}\label{MVL Miuras}
{\rm a)} The flow ${\rm MVL}(\epsilon)$ {\rm(\ref{MVL})} is the pull--back 
of the flow {\rm VL (\ref{VL})} under either of the Miura transformations 
$\mbM_{1,2}(\epsilon):\,\cM\cV(y,z)\mapsto\cV(u,v)$ defined by
\begin{equation}\label{MVL Miuras to VL}
\mbM_1(\epsilon)\;:\left\{\begin{array}{l}
u_k=y_k(1+\epsilon z_{k-1})\;,\\ \\ v_k=z_k(1+\epsilon y_k)\;,\end{array}
\right.\qquad
\mbM_2(\epsilon)\;:\left\{\begin{array}{l}
u_k=y_k(1+\epsilon z_k)\;,\\ \\ v_k=z_k(1+\epsilon y_{k+1})\;.\end{array}
\right.
\end{equation}
Both Miura transformations $\mbM_{1,2}(\epsilon)$ are Poisson, if $\cM\cV(y,z)$ 
is equipped with the bracket $\{\cdot,\cdot\}_{23}$ and $\cV(u,v)$ is equipped 
with the bracket $\{\cdot,\cdot\}_2+\epsilon\{\cdot,\cdot\}_3$.

{\rm b)} The flow ${\rm MVL}(\epsilon)$ {\rm(\ref{MVL})} is the pull--back 
of the flow ${\rm MTL}(\epsilon)$ {\rm(\ref{MTL})} under either of the Miura 
transformations $M_{1,2}(\epsilon):\,\cM\cV(y,z)\mapsto\cM\cT(q,p)$ 
defined by 
\begin{equation}\label{MVL Miuras to MTL}
M_1(\epsilon)\;:\left\{\begin{array}{l}
p_k=y_k+z_{k-1}+\epsilon y_kz_{k-1}\;,\\ \\ q_k=y_kz_k\;,\end{array}
\right.\qquad
M_2(\epsilon)\;:\left\{\begin{array}{l}
p_k=y_k+z_k+\epsilon y_kz_k\;,\\ \\ q_k=y_{k+1}z_k\;.\end{array}
\right.
\end{equation}
Both Miura transformations $M_{1,2}(\epsilon)$ are Poisson, if 
$\cM\cV(y,z)$ and $\cM\cT(q,p)$ carry the corresponding brackets 
$\{\cdot,\cdot\}_{23}$.

{\rm c)} The following diagram is commutative for all $i=1,2$, $j=1,2$:
\begin{center}
\unitlength1cm
\begin{picture}(9,6.5)
\put(3.5,1.1){\vector(1,0){2}}
\put(3.5,5.1){\vector(1,0){2}}
\put(2,4.1){\vector(0,-1){2}}
\put(7,4.1){\vector(0,-1){2}}
\put(1,0.6){\makebox(2,1){$\cV$}} 
\put(1,4.6){\makebox(2,1){$\cM\cV$}}
\put(6,4.6){\makebox(2,1){$\cM\cT$}}
\put(6,0.6){\makebox(2,1){$\cT$}}
\put(0,2.6){\makebox(2,1){$\mbM_i(\epsilon)$}}
\put(7,2.6){\makebox(2,1){$\rM_i(\epsilon)$}}
\put(3.8,-0.2){\makebox(1.4,1.4){$M_j$}}
\put(3.8,5.0){\makebox(1.4,1.4){$M_j(\epsilon)$}}
\end{picture}
\end{center}
\end{theorem}

\subsection{Discretization}
 
The map $\cM\cV\mapsto\cM\cV$ denoted by ${\rm dMVL}(\epsilon)$ is by 
definition conjugated with dVL by means of $\mbM_{1,2}(\epsilon)$, 
and conjugated with ${\rm dMTL}(\epsilon)$ by means of $M_{1,2}(\epsilon)$. 

\subsection{Application: localizing change of variables for dVL}

The Miura map $\mbM_1(h)$ turns out to play the role of the localizing
change of variables for the map dVL. Indeed, consider the following change 
of variables $\cV(\bu,\bv)\mapsto\cV(u,v)$:
\begin{equation}\label{dVL loc map}
\mbM_1(h):\quad \left\{\begin{array}{l}
u_k=\bu_k(1+h\bv_{k-1})\;, \\ \\ v_k=\bv_k(1+h\bu_k)\;.
\end{array}\right.
\end{equation}
Then the entries of the factors $\mbB_j$, $\mbC_j$ 
(see (\ref{dVL B1})--(\ref{dVL C2})) admit local expressions in the 
coordinates $(\bu,\bv)$:
\begin{eqnarray}
\beta_k^{(1)}=(1+h\bu_k)(1+h\bv_{k-1})\;, & \quad &
\gamma_k^{(1)}=\bu_k\bv_k\;,\label{dVL loc beta}\\ \nonumber\\
\beta_k^{(2)}=(1+h\bv_k)(1+h\bu_k)\;, & \quad &
\gamma_k^{(2)}=\bu_{k+1}\bv_k\;.  \label{dVL loc gamma}
\end{eqnarray}

\begin{theorem}\label{local dVL}
The change of variables {\rm(\ref{dVL loc map})} conjugates 
the map {\rm dVL} with the map on $\cV(\bu,\bv)$ described by the 
following local equations of motion:
\begin{equation}\label{dVL loc}
\widetilde{\bu}_k(1+h\widetilde{\bv}_{k-1})=\bu_k(1+h\bv_k)\;,\qquad
\widetilde{\bv}_k(1+h\widetilde{\bu}_k)=\bv_k(1+h\bu_{k+1})\;.
\end{equation}
\end{theorem}
So, the local form of dVL (\ref{dVL loc}) belongs actually to the 
hierarchy ${\rm dMVL}(h)$.
\vspace{1.5mm}

{\bf Corollary.} {\it The local form of {\rm dVL (\ref{dVL loc})} is a 
Poisson map with respect to the following  bracket on $\cV(\bu,\bv)$:
\begin{equation}\label{dVL loc PB loc}
\begin{array}{rcl}
\{\bu_k,\bv_k\} & = & -\bu_k\bv_k(1+h\bu_k)(1+h\bv_k)\;, \\ \\
\{\bv_k,\bu_{k+1}\} & = & -\bv_k\bu_{k+1}(1+h\bv_k)(1+h\bu_{k+1})\;,
\end{array}
\end{equation}
which is the pull--back of the bracket
\begin{equation}\label{dVL loc PB}
\{\cdot,\cdot\}_2+h\{\cdot,\cdot\}_3
\end{equation}
on $\cV(u,v)$ under the change of variables {\rm(\ref{dVL loc map})}.}
\vspace{2mm}

Now we translate the Miura relations existing between dTL and dVL, 
as members of the TL and VL hierarchies, into the localizing variables.
This is achieved by the following reformulation of the part c)
of Theorem \ref{MVL Miuras} 
\begin{theorem} 
\begin{itemize}
\item[\rm a)] The following diagram is commutative for $j=1,2$:
\begin{center}
\unitlength1cm
\begin{picture}(9,6.5)
\put(3.5,1.1){\vector(1,0){2}}
\put(3.5,5.1){\vector(1,0){2}}
\put(2,4.1){\vector(0,-1){2}}
\put(7,4.1){\vector(0,-1){2}}
\put(1,0.6){\makebox(2,1){$\cV(u,v)$}} 
\put(1,4.6){\makebox(2,1){$\cV(\bu,\bv)$}}
\put(6,4.6){\makebox(2,1){$\cT(\ba,\bb)$}}
\put(6,0.6){\makebox(2,1){$\cT(a,b)$}}
\put(0,2.6){\makebox(2,1){$\mbM_1(h)$}}
\put(7,2.6){\makebox(2,1){$\rM_1(h)$}}
\put(3.8,-0.2){\makebox(1.4,1.4){$M_j$}}
\put(3.8,5.0){\makebox(1.4,1.4){$M_j(h)$}}
\end{picture}
\end{center}
where the maps $M_{1,2}(h):\,\cV(\bu,\bv)\mapsto\cT(\ba,\bb)$ are given
by the formulas
\begin{equation}\label{loc dTL M1}
M_1(h)\,: \quad \left\{\begin{array}{l}
\bb_k=\bu_k+\bv_{k-1}+h\bu_k\bv_{k-1}\;, \\ \\
\ba_k=\bu_k\bv_k\;,\end{array}\right.
\end{equation}
and
\begin{equation}\label{loc dTL M2}
M_2(h)\,: \quad \left\{\begin{array}{l}
\bb_k=\bu_k+\bv_k+h\bu_k\bv_k\;, \\ \\
\ba_k=\bu_{k+1}\bv_k\;.\end{array}\right.
\end{equation}
\item[\rm b)] Both maps $M_{1,2}(h)$ are Poisson, if $\cV(\bu,\bv)$ is 
equipped with the bracket {\rm(\ref{dVL loc PB loc})}, and $\cT(\ba,\bb)$ is 
equipped with the bracket {\rm(\ref{dTL loc br 2})}.
\item[\rm c)] The local form of {\rm dVL (\ref{dVL loc})} is conjugated 
with the local form of {\rm dTL (\ref{dTL loc})} by either of the maps 
$M_{1,2}(h)$. 
\end{itemize}
\end{theorem}

\setcounter{equation}{0}
\section{Double modified Volterra lattice}

\subsection{Equations of motion}

The phase space of the {\it double modified Volterra lattice}
${\rm M^2VL}(\epsilon,\delta)$ will be denoted by 
$\cM^2\cV={\Bbb R}^{2N}(\mbu,\mbv)$, the modification parameters by $\epsilon$,
$\delta$. The equations of motion read:
\begin{eqnarray}\label{M2VL}
\dot{\mbu}_k & = & \mbu_k(1+\epsilon\mbu_k)(1+\delta\mbu_k)\left(
\frac{\mbv_k}{1-\epsilon\delta\mbv_k\mbu_k}-
\frac{\mbv_{k-1}}{1-\epsilon\delta\mbu_k\mbv_{k-1}}
\right)\;,\nonumber\\ \\
\dot{\mbv}_k & = & \mbv_k(1+\epsilon\mbv_k)(1+\delta\mbv_k)\left(
\frac{\mbu_{k+1}}{1-\epsilon\delta\mbu_{k+1}\mbv_k}-
\frac{\mbu_k}{1-\epsilon\delta\mbv_k\mbu_k}
\right)\;.\nonumber
\end{eqnarray}

\subsection{Miura relations}

\begin{theorem}\label{M2VL Miuras}
{\rm a)} Define the Miura maps 
$\mbM_{1,2}(\epsilon;\delta):\cM^2\cV(\mbu,\mbv)\mapsto\cM\cV(y,z)$ by:
\begin{eqnarray}
\mbM_1(\epsilon;\delta): & \quad & \left\{\begin{array}{l}
y_k=\displaystyle\frac{\mbu_k(1+\delta\mbv_{k-1})}
{1-\epsilon\delta\mbu_k\mbv_{k-1}}\;,\\ \\
z_k=\displaystyle\frac{\mbv_k(1+\delta\mbu_k)}
{1-\epsilon\delta\mbv_k\mbu_k}\;,
\end{array}\right. \\ \nonumber\\  \nonumber\\
\mbM_2(\epsilon;\delta): & \quad & \left\{\begin{array}{l}
y_k=\displaystyle\frac{\mbu_k(1+\delta\mbv_k)}
{1-\epsilon\delta\mbv_k\mbu_k}\;,\\ \\
z_k=\displaystyle\frac{\mbv_k(1+\delta\mbu_{k+1})}
{1-\epsilon\delta\mbu_{k+1}\mbv_k}\;.
\end{array}\right.
\end{eqnarray}
Then the pull--back of the flow ${\rm MVL}(\epsilon)$ under either of 
the Miura maps $\mbM_{1,2}(\epsilon;\delta)$ coincides with 
${\rm M^2VL}(\epsilon,\delta)$.

{\rm b)} Define the Miura maps 
$M_{1,2}(\epsilon,\delta):\cM^2\cV(\mbu,\mbv)\mapsto\cM^2\cT(r,s)$ by:
\begin{eqnarray}
M_1(\epsilon,\delta): & \quad & \left\{\begin{array}{l}
r_k=\displaystyle\frac{\mbu_k+\mbv_{k-1}+(\epsilon+\delta)\mbu_k\mbv_{k-1}}
{1-\epsilon\delta\mbu_k\mbv_{k-1}}\;,\\ \\
s_k=\displaystyle\frac{\mbu_k\mbv_k}
{1-\epsilon\delta\mbu_k\mbv_k}\;,
\end{array}\right. \\ \nonumber\\  \nonumber\\
M_2(\epsilon,\delta): & \quad & \left\{\begin{array}{l}
r_k=\displaystyle\frac{\mbu_k+\mbv_k+(\epsilon+\delta)\mbu_k\mbv_k}
{1-\epsilon\delta\mbu_k\mbv_k}\;,\\ \\
s_k=\displaystyle\frac{\mbu_{k+1}\mbv_k}
{1-\epsilon\delta\mbu_{k+1}\mbv_k}\;.
\end{array}\right.
\end{eqnarray}
Then the pull--back of the flow ${\rm M^2TL}(\epsilon,\delta)$ under 
either of the Miura maps $M_{1,2}(\epsilon,\delta)$ coincides with 
${\rm M^2VL}(\epsilon,\delta)$.

{\rm c)} The following diagram is commutative for all $i=1,2$, $j=1,2$:
\begin{center}
\unitlength1cm
\begin{picture}(9,6.5)
\put(3.5,1.1){\vector(1,0){2}}
\put(3.5,5.1){\vector(1,0){2}}
\put(2,4.1){\vector(0,-1){2}}
\put(7,4.1){\vector(0,-1){2}}
\put(1,0.6){\makebox(2,1){$\cM\cV$}} 
\put(1,4.6){\makebox(2,1){$\cM^2\cV$}}
\put(6,4.6){\makebox(2,1){$\cM\cV$}}
\put(6,0.6){\makebox(2,1){$\cV$}}
\put(0,2.6){\makebox(2,1){$\mbM_i(\epsilon;\delta)$}}
\put(7,2.6){\makebox(2,1){$\mbM_i(\delta)$}}
\put(3.8,-0.2){\makebox(1.4,1.4){$\mbM_j(\epsilon)$}}
\put(3.8,5.0){\makebox(1.4,1.4){$\mbM_j(\delta;\epsilon)$}}
\end{picture}
\end{center}

{\rm d)} The following diagram is commutative for all $i=1,2$, $j=1,2$:
\begin{center}
\unitlength1cm
\begin{picture}(9,6.5)
\put(3.5,1.1){\vector(1,0){2}}
\put(3.5,5.1){\vector(1,0){2}}
\put(2,4.1){\vector(0,-1){2}}
\put(7,4.1){\vector(0,-1){2}}
\put(1,0.6){\makebox(2,1){$\cM\cV$}} 
\put(1,4.6){\makebox(2,1){$\cM^2\cV$}}
\put(6,4.6){\makebox(2,1){$\cM^2\cT$}}
\put(6,0.6){\makebox(2,1){$\cM\cT$}}
\put(0,2.6){\makebox(2,1){$\mbM_i(\epsilon;\delta)$}}
\put(7,2.6){\makebox(2,1){$\rM_i(\epsilon;\delta)$}}
\put(3.8,-0.2){\makebox(1.4,1.4){$M_j(\epsilon)$}}
\put(3.8,5.0){\makebox(1.4,1.4){$M_j(\epsilon;\delta)$}}
\end{picture}
\end{center}
\end{theorem}

\subsection{Application: localizing change of variables for 
${\rm dMVL}(\epsilon)$}

It turns out that the Miura map $\mbM_1(\epsilon;h)$ plays the role
of the localizing change of variables for the map ${\rm dMVL}(\epsilon)$.
Indeed, consider the following change of variables $\cM\cV(\by,\bz)\mapsto
\cM\cV(y,z)$:
\begin{equation}\label{dMVL loc map}
\mbM_1(\epsilon;h):\quad
y_k=\frac{\by_k(1+h\bz_{k-1})}{1-h\epsilon\by_k\bz_{k-1}}\;,\qquad
z_k=\frac{\bz_k(1+h\by_k)}{1-h\epsilon\bz_k\by_k}\;.
\end{equation}
\begin{theorem}
The change of variables {\rm(\ref{dMVL loc map})} conjugates the map 
${\rm dMVL}(\epsilon)$ with the map on $\cM\cV(\by,\bz)$ described by the
following local equations of motion:
\begin{eqnarray}\label{dMVL loc}
\frac{\widetilde{\by}_k}{1+\epsilon\widetilde{\by}_k}\,
(1+h\widetilde{\bz}_{k-1}) & = & \frac{\by_k}{1+\epsilon\by_k}\,
(1+h\bz_k)\;,\nonumber\\ \\
\frac{\widetilde{\bz}_k}{1+\epsilon\widetilde{\bz}_k}\,
(1+h\widetilde{\by}_k) & = & \frac{\bz_k}{1+\epsilon\bz_k}\,
(1+h\by_{k+1})\;.\nonumber
\end{eqnarray}
\end{theorem}
We see that the local form of the map ${\rm dMVL}(\epsilon)$ 
(\ref{dMVL loc}) belongs actually to the hierarchy 
${\rm M^2VL}(\epsilon,h)$.

Finally, we give the translations of the Miura relations of 
dMVL$(\epsilon)$ to both dVL and dMTL$(\epsilon)$ into the 
language of localizing variables. This is achieved by reformulating the
parts c) and d) of Theorem \ref{M2VL Miuras}.
\begin{theorem} 
{\rm a)} The following diagram is commutative for $j=1,2$:
\begin{center}
\unitlength1cm
\begin{picture}(9,6.5)
\put(3.5,1.1){\vector(1,0){2}}
\put(3.5,5.1){\vector(1,0){2}}
\put(2,4.1){\vector(0,-1){2}}
\put(7,4.1){\vector(0,-1){2}}
\put(1,0.6){\makebox(2,1){$\cM\cV(y,z)$}} 
\put(1,4.6){\makebox(2,1){$\cM\cV(\by,\bz)$}}
\put(6,4.6){\makebox(2,1){$\cV(\bu,\bv)$}}
\put(6,0.6){\makebox(2,1){$\cV(u,v)$}}
\put(0,2.6){\makebox(2,1){$\mbM_1(\epsilon;h)$}}
\put(7,2.6){\makebox(2,1){$\mbM_1(h)$}}
\put(3.8,-0.2){\makebox(1.4,1.4){$\mbM_{1,2}(\epsilon)$}}
\put(3.8,5.0){\makebox(1.4,1.4){$\mbM_{1,2}(h;\epsilon)$}}
\end{picture}
\end{center}
Here the maps $\mbM_{1,2}(h;\epsilon):\,\cM\cV(\by,\bz)\mapsto\cV(\bu,\bv)$ 
are given by the formulas
\begin{equation}\label{Miura 1 MVL VL loc}
\mbM_1(h;\epsilon)\,: \quad \left\{\begin{array}{l}
\bu_k=\displaystyle\frac{\by_k(1+\epsilon\bz_{k-1})}
{1-h\epsilon\by_k\bz_{k-1}}\;,\\ \\
\bv_k=\displaystyle\frac{\bz_k(1+\epsilon\by_k)}{1-h\epsilon\bz_k\by_k}\;,
\end{array}\right.
\end{equation}
and
\begin{equation}\label{Miura 2 MVL VL loc}
\mbM_2(h;\epsilon)\,: \quad \left\{\begin{array}{l}
\bu_k=\displaystyle\frac{\by_k(1+\epsilon\bz_k)}
{1-h\epsilon\bz_k\by_k}\;,\\ \\
\bv_k=\displaystyle\frac{\bz_k(1+\epsilon\by_{k+1})}
{1-h\epsilon\by_{k+1}\bz_k}\;, 
\end{array}\right.
\end{equation} 
Either of the maps $\mbM_{1,2}(h;\epsilon)$ conjugates the local form of 
{\rm dMVL$(\epsilon)$ (\ref{dMVL loc})} with the local form of {\rm dVL 
(\ref{dVL loc})}. 

{\rm b)} The following diagram is commutative for $j=1,2$:
\begin{center}
\unitlength1cm
\begin{picture}(9,6.5)
\put(3.5,1.1){\vector(1,0){2}}
\put(3.5,5.1){\vector(1,0){2}}
\put(2,4.1){\vector(0,-1){2}}
\put(7,4.1){\vector(0,-1){2}}
\put(1,0.6){\makebox(2,1){$\cM\cV(y,z)$}} 
\put(1,4.6){\makebox(2,1){$\cM\cV(\by,\bz)$}}
\put(6,4.6){\makebox(2,1){$\cM\cT(\bq,\bp)$}}
\put(6,0.6){\makebox(2,1){$\cM\cT(q,p)$}}
\put(0,2.6){\makebox(2,1){$\mbM_1(\epsilon;h)$}}
\put(7,2.6){\makebox(2,1){$\rM_1(\epsilon;h)$}}
\put(3.8,-0.2){\makebox(1.4,1.4){$M_{1,2}(\epsilon)$}}
\put(3.8,5.0){\makebox(1.4,1.4){$M_{1,2}(\epsilon,h)$}}
\end{picture}
\end{center}
Here the maps $M_{1,2}(\epsilon,h):\,\cM\cV(\by,\bz)\mapsto\cM\cT(\bq,\br)$ 
are given by the formulas
\begin{eqnarray}
M_1(\epsilon,h)\,: & \quad & \left\{\begin{array}{l}
\bp_k=\displaystyle\frac{\by_k+\bz_{k-1}+(\epsilon+h)\by_k\bz_{k-1}}
{1-h\epsilon\by_k\bz_{k-1}}\;,\\ \\
\bq_k=\displaystyle\frac{\by_k\bz_k}{1-h\epsilon\by_k\bz_k}\;,
\end{array}\right. \label{Miura 1 MVL MTL loc}\\ \nonumber\\
M_2(\epsilon,h)\,: & \quad & \left\{\begin{array}{l}
\bp_k=\displaystyle\frac{\by_k+\bz_k+(\epsilon+h)\by_k\bz_k}
{1-h\epsilon\by_k\bz_k}\;,\\ \\
\bq_k=\displaystyle\frac{\by_{k+1}\bz_k}{1-h\epsilon\by_{k+1}\bz_k}\;,
\end{array}\right.  \label{Miura 2 MVL MTL loc}
\end{eqnarray}
Either of the maps $M_{1,2}(\epsilon,h)$ conjugates the local form of 
{\rm dMVL$(\epsilon)$ (\ref{dMVL loc})} with the local form of 
{\rm dMTL$(\epsilon)$ (\ref{dMTL loc})}. 
\end{theorem}

\newpage
\part{Relativistic systems}
\setcounter{equation}{0}
\section{Relativistic Toda lattice}
\label{Sect RTL param}
\subsection{Equations of motion}
We now turn to a tower of modifications connected with the {\it relativistic
Toda lattice} hierarchy, or ${\rm RTL}(\alpha)$. Here $\alpha$ is a (small)
parameter playing the role of the inverse speed of light. A typical phenomenon
encountered for these relativistic systems is the {\it splitting} of some
notions and results. For example, in the ${\rm RTL}(\alpha)$ hierarchy there
exist actually {\it two} simple flows approximating the usual TL as 
$\alpha\to 0$. The phase space of this hierarchy will be denoted by
$\cR\cT={\Bbb R}^{2N}(a,b)$.

The flow ${\rm RTL}_+(\alpha)$ is a polynomial perturbation of TL: 
\begin{equation}\label{RTL+ param}
\dot{b}_k=(1+\alpha b_k)(a_k-a_{k-1})\;, \qquad \dot{a}_k=
a_k(b_{k+1}-b_k+\alpha a_{k+1}-\alpha a_{k-1})\;.
\end{equation}
The flow ${\rm RTL}_-(\alpha)$ is a rational perturbation of TL:
\begin{equation}\label{RTL- param}
\dot{b}_k=\frac{a_k}{1+\alpha b_{k+1}}-\frac{a_{k-1}}{1+\alpha b_{k-1}}\;, 
\qquad \dot{a}_k=
a_k\left(\frac{b_{k+1}}{1+\alpha b_{k+1}}-\frac{b_k}{1+\alpha b_k}\right)\;.
\end{equation}
The first one of these flows resembles the modified Toda lattice ${\rm MTL}
(\alpha)$, however its properties are somewhat different. In particular, it
is a tri--Hamiltonian system, while ${\rm MTL}(\alpha)$ admits only two
local hamiltonian formulations.

\subsection{Tri--Hamiltonian structure}
\begin{proposition} 
{\rm a)} The relations
\begin{equation}\label{RTL param l br}
\begin{array}{lcl}
\{b_k,a_k\}_{1\alpha}=-a_k\;, & \quad & \{a_k,b_{k+1}\}_{1\alpha}=-a_k\;,\\ \\
\{b_k,b_{k+1}\}_{1\alpha}=\alpha a_k & & 
\end{array}
\end{equation}
define a Poisson bracket on $\cR\cT$. The flows ${\rm RTL}_{\pm}(\alpha)$ 
are Hamiltonian systems on $\Big(\cR\cT,\{\cdot,\cdot\}_{1\alpha}\Big)$ with the 
Hamilton functions 
\begin{equation}\label{RTL+ param H2}
\rH_{2\alpha}^{(+)}(a,b)=\sum_{k=1}^N \left(\frac{1}{2}\,b_k^2+a_k\right)
+\alpha\sum_{k=1}^N(b_k+b_{k+1})a_k+
\alpha^2\sum_{k=1}^N\left(\frac{1}{2}\,a_k^2+a_ka_{k+1}\right)\;,
\end{equation}
\begin{equation}\label{RTL- param H2}
\rH_{2\alpha}^{(-)}(a,b)=-\alpha^{-2}\sum_{k=1}^N \log(1+\alpha b_k)\;,
\end{equation}
respectively. The functions $\rH_{2\alpha}^{(+)}$ and $\rH_{2\alpha}^{(-)}$
are in involution in the bracket $\{\cdot,\cdot\}_{1\alpha}$.

{\rm b)} The relations
\begin{equation}\label{RTL param q br}
\begin{array}{lcl}
\{b_k,a_k\}_{2\alpha}=-b_ka_k\;, & \quad &
\{a_k,b_{k+1}\}_{2\alpha}=-a_kb_{k+1}\;, \\ \\
\{b_k,b_{k+1}\}_{2\alpha}=-a_k\;, & \quad & 
\{a_k,a_{k+1}\}_{2\alpha}=-a_ka_{k+1} 
\end{array}
\end{equation}
define a Poisson bracket on $\cR\cT(a,b)$, compatible with 
$\{\cdot,\cdot\}_{1\alpha}$. The flows ${\rm RTL}_{\pm}(\alpha)$ are 
Hamiltonian on $\Big(\cR\cT,\{\cdot,\cdot\}_{2\alpha}\Big)$ with the 
Hamilton functions 
\begin{equation}\label{RTL+ param H1}
\rH_{1\alpha}^{(+)}(a,b)=\sum_{k=1}^N b_k+\alpha \sum_{k=1}^N a_k\;,
\end{equation}
\begin{equation}\label{RTL- param H1}
\rH_{1\alpha}^{(-)}(a,b)=\alpha^{-1}\,\sum_{k=1}^N\log(1+\alpha b_k)\;,
\end{equation}
respectively. The functions $\rH_{1\alpha}^{(+)}$ and $\rH_{1\alpha}^{(-)}$
are in involution in the bracket $\{\cdot,\cdot\}_{2\alpha}$.

{\rm c)} The relations
\begin{equation}\label{RTL param c br}
\begin{array}{l}
\{b_k,a_k\}_{3\alpha}     = -a_k(b_k^2+a_k)-\alpha b_ka_k^2\;, \\ \\ 
\{a_k,b_{k+1}\}_{3\alpha} = -a_k(b_{k+1}^2+a_k)-\alpha a_k^2b_{k+1}\;,  \\ \\
\{b_k,b_{k+1}\}_{3\alpha} = -a_k(b_k+b_{k+1})-\alpha b_ka_kb_{k+1}\;, \\ \\
\{a_k,a_{k+1}\}_{3\alpha} = 
         -2a_kb_{k+1}a_{k+1}-\alpha a_ka_{k+1}(a_k+a_{k+1})\;, \\ \\
\{b_k,a_{k+1}\}_{3\alpha} = -a_ka_{k+1}-\alpha b_ka_ka_{k+1}\;, \\ \\
\{a_k,b_{k+2}\}_{3\alpha} = -a_ka_{k+1}-\alpha a_ka_{k+1}b_{k+2}\;,\\ \\
\{a_k,a_{k+2}\}_{3\alpha} = -\alpha a_ka_{k+1}a_{k+2}
\end{array}
\end{equation}
define a Poisson bracket on $\cR\cT(a,b)$ compatible with 
$\{\cdot,\cdot\}_{1\alpha}$ and $\{\cdot,\cdot\}_{2\alpha}$. The flows 
${\rm RTL}_{\pm}(\alpha)$ are Hamiltonian on 
$\Big(\cR\cT,\{\cdot,\cdot\}_{3\alpha}\Big)$ 
with the Hamilton functions 
\begin{eqnarray}
\rH_{0\alpha}^{(+)}(a,b) & = & \frac{1}{2}\,\sum_{k=1}^N\log(a_k)\;, 
\label{RTL+ param H0}\\ \nonumber\\
\rH_{0\alpha}^{(-)}(a,b) & = & \frac{1}{2}\,\sum_{k=1}^N\log(a_k)-
\sum_{k=1}^N\log(1+\alpha b_k)\;,\label{RTL- param H0}
\end{eqnarray}
respectively. The functions $\rH_{0\alpha}^{(+)}$ and $\rH_{0\alpha}^{(-)}$
are in involution in the bracket $\{\cdot,\cdot\}_{3\alpha}$.
\end{proposition}

Notice that the function $\rH_{2\alpha}^{(-)}(a,b)$, singular in $\alpha$,
becomes regular, and, moreover, an $O(\alpha)$--perturbation of
$\rH_2(a,b)$ from (\ref{TL H2}), upon adding
\[
\alpha^{-1}\sum_{k=1}^N b_k+\sum_{k=1}^N a_k\;,
\]
which is a Casimir function of the bracket $\{\cdot,\cdot\}_{1\alpha}$.

Notice also that the bracket $\{\cdot,\cdot\}_{2\alpha}$ is actually
independent on $\alpha$ in coordinates $(a,b)$ and {\it literally coincides} 
with the invariant quadratic bracket (\ref{TL q br}) of the TL hierarchy. 

\subsection{Lax representation}
Lax matrix $(L,W^{-1}):\cR\cT\mapsto \g\otimes\g$:
\begin{eqnarray}
L(a,b,\lambda) & = & \sum_{k=1}^N (1+\alpha b_k)E_{kk}+
\alpha\lambda\sum_{k=1}^N E_{k+1,k}\;,
\label{RTL param L}\\
W(a,b,\lambda) & = & I-\alpha\lambda^{-1}\sum_{k=1}^N a_kE_{k,k+1}\;.
\label{RTL param U}
\end{eqnarray}

Lax representation of ${\rm RTL}_+(\alpha)$:
\begin{equation}\label{RTL+ param triads}
\dot{L}=LA_2-A_1L\;, \qquad \dot{W}=WA_2-A_1W\;,
\end{equation}
with 
\begin{eqnarray}
A_1=\pi_+\Big((LW^{-1}-I)/\alpha\Big) & = & 
\sum_{k=1}^N(b_k+\alpha a_{k-1})E_{kk}+\lambda\sum_{k=1}^N E_{k+1,k}\;, 
\label{RTL+ param A1}\\
A_2=\pi_+\Big((W^{-1}L-I)/\alpha\Big) & = & 
\sum_{k=1}^N(b_k+\alpha a_k)E_{kk}+\lambda\sum_{k=1}^N E_{k+1,k}\;. 
\label{RTL+ param A2}
\end{eqnarray}

Lax representation of ${\rm RTL}_-(\alpha)$:
\begin{equation}
\dot{L}=C_1L-LC_2\;, \qquad \dot{W}=C_1W-WC_2\;,
\end{equation}
with 
\begin{eqnarray}
C_1=\pi_-\Big((I-WL^{-1})/\alpha\Big) & = & \lambda^{-1}\sum_{k=1}^N 
\frac{a_k}{1+\alpha b_{k+1}}\,E_{k,k+1}\;, \label{RTL- param C}\\
C_2=\pi_-\Big((I-L^{-1}W)/\alpha\Big) & = & \lambda^{-1}\sum_{k=1}^N 
\frac{a_k}{1+\alpha b_k}\,E_{k,k+1}\;. \label{RTL- param D}
\end{eqnarray}

\subsection{Discretization}
Lax representation of the map ${\rm dRTL}_+(\alpha)$:
\begin{equation}\label{dRTL+ param Lax}
\wL=\mbA_1^{-1}L\mbA_2\;,\qquad \wW=\mbA_1^{-1}W\mbA_2\;,
\end{equation}
where
\begin{eqnarray}
\mbA_1=\Pi_+\Big(I+\frac{h}{\alpha}\,(LW^{-1}-I)\Big) & = & 
\sum_{k=1}^N \ga_k E_{kk}+h\lambda\sum_{k=1}^N E_{k+1,k}\;,
\label{dRTL+ param A1}\\
\mbA_2=\Pi_+\Big(I+\frac{h}{\alpha}\,(W^{-1}L-I)\Big) & = & 
\sum_{k=1}^N \gb_k E_{kk}+h\lambda\sum_{k=1}^N E_{k+1,k}\;.
\label{dRTL+ param A2}
\end{eqnarray}
Lax representation of the map ${\rm dRTL}_-(\alpha)$:
\begin{equation}\label{dRTL- param Lax}
\wL=\mbC_1L\mbC_2^{-1}\;,\qquad \wW=\mbC_1W\mbC_2^{-1}\;,
\end{equation}
where
\begin{eqnarray}
\mbC_1=\Pi_-\Big(I+\frac{h}{\alpha}\,(I-WL^{-1})\Big) & = & 
I+h\lambda^{-1}\sum_{k=1}^N \gc_k E_{k,k+1}\;, \label{dRTL- param C1}\\
\mbC_2=\Pi_-\Big(I+\frac{h}{\alpha}\,(I-L^{-1}W)\Big) & = & 
I+h\lambda^{-1}\sum_{k=1}^N \gd_k E_{k,k+1}\;.  \label{dRTL- param C2}
\end{eqnarray}

\setcounter{equation}{0}
\section{Modified relativistic Toda lattice I}
\label{Sect modified RTL}
As a next example of the ``relativistic splitting'', let us mention the 
existence of {\it two} different ways to modify the ${\rm RTL}(\alpha)$
hierarchy. In other words, there exist two different {\it modified relativistic
Toda hierarchies}. In this section we consider one of them.
\subsection{Equations of motion}
The phase space of the ${\rm MRTL}^{(+)}(\alpha;\epsilon)$ hierarchy will
be denoted by ${\cal MRT}={\Bbb R}^{2N}(q,p)$. As usual, in this relativistic
hierarchy there exist two simplest flows.

Equations of motion of ${\rm MRTL}^{(+)}_+(\alpha;\epsilon)$:
\begin{equation}\label{MRTL++}
\begin{array}{l}
\dot{p}_k=(1+\alpha p_k)(1+\epsilon p_k)(q_k-q_{k-1})\;,\\ \\
\dot{q}_k=q_k(1+\epsilon\alpha q_k)\Big(p_{k+1}+\alpha q_{k+1}+
\epsilon\alpha p_{k+1}q_{k+1}-p_k-\alpha q_{k-1}-\epsilon\alpha p_kq_{k-1}
\Big)\;.
\end{array}
\end{equation}
Equations of motion of ${\rm MRTL}^{(+)}_-(\alpha;\epsilon)$:
\begin{equation}\label{MRTL+-}
\begin{array}{l}
\dot{p}_k=(1+\epsilon p_k)\left(\displaystyle\frac
{q_k}{(1+\alpha p_{k+1})(1+\epsilon\alpha q_k)}-
\displaystyle\frac{q_{k-1}}{(1+\alpha p_{k-1})(1+\epsilon\alpha q_{k-1})}
\right)\;,\\ \\
\dot{q}_k=q_k\left(\displaystyle\frac
{p_{k+1}}{1+\alpha p_{k+1}}-\displaystyle\frac{p_k}{1+\alpha p_k}\right)\;.
\end{array}
\end{equation}

Obviously, the equations of ${\rm MRTL}^{(+)}_+(\alpha;\epsilon)$
are similar to (and slightly more complicated than)
the equations of motion of the double modification 
${\rm M^2TL}(\alpha,\epsilon)$ of the Toda lattice. However, these systems
have somewhat different properties. Most obvious are the non--symmetric roles
played in ${\rm MRTL}^{(+)}_+(\alpha;\epsilon)$ by the parameters $\alpha$ and
$\epsilon$. More serious differences will become apparent later on, in 
particular, the system ${\rm MRTL}^{(+)}_+(\alpha;\epsilon)$ possesses two 
local invariant Poisson brackets, while for ${\rm M^2TL}(\alpha,\epsilon)$ 
only one local invariant Poisson bracket is known.

\subsection{Bi--Hamiltonian structure}
\begin{proposition}\label{MRTL first Ham str}
{\rm a)} The relations
\begin{eqnarray}\label{MRTL+ br 1}
\{p_k,q_k\}_{12\alpha} & = & -q_k(1+\epsilon p_k)\;,\nonumber\\ \nonumber\\
\{q_k,p_{k+1}\}_{12\alpha} & = & -q_k(1+\epsilon p_{k+1})\;,\\ \nonumber\\
\{p_k,p_{k+1}\}_{12\alpha} & = & \alpha\,\displaystyle\frac{q_k}{1+\epsilon
\alpha q_k}\,(1+\epsilon p_k)(1+\epsilon p_{k+1}) \nonumber
\end{eqnarray}
define a Poisson bracket on ${\cal MRT}(q,p)$. 
The systems ${\rm MRTL}^{(+)}_{\pm}(\alpha;\epsilon)$ are Hamiltonian 
with respect to the bracket {\rm (\ref{MRTL+ br 1})}, with the Hamilton 
functions
\begin{equation}\label{MRTL++ H2}
\rG_{1\alpha}^{(+)}(q,p)=\epsilon^{-1}\sum_{k=1}^N p_k
+(1+\epsilon^{-1}\alpha)\sum_{k=1}^N q_k 
+\alpha\sum_{k=1}^N (p_k+p_{k+1})q_k+
\alpha^2\sum_{k=1}^N q_kq_{k+1}(1+\epsilon p_{k+1})\;,
\end{equation}
\begin{equation}\label{MRTL+- H2}
\rG_{1\alpha}^{(-)}(q,p)=(\epsilon-\alpha)^{-1}\alpha^{-1}
\sum_{k=1}^N \log\Big((1+\alpha p_k)(1+\epsilon\alpha q_k)\Big)\;,
\end{equation}  
respectively. The functions $\rG_{1\alpha}^{(+)}$ and $\rG_{1\alpha}^{(-)}$
are in involution in the bracket $\{\cdot,\cdot\}_{12\alpha}$.

{\rm b)} The relations
\begin{eqnarray}
\{p_k,q_k\}_{23\alpha} & = & -q_k(p_k+\epsilon q_k+\epsilon\alpha p_kq_k)
(1+\epsilon p_k)\;,\nonumber\\ \nonumber\\
\{q_k,p_{k+1}\}_{23\alpha} & = & -q_k(p_{k+1}+\epsilon q_k+
\epsilon\alpha p_{k+1}q_k)(1+\epsilon p_{k+1})\;,
\nonumber\\ \label{MRTL br 2}\\
\{p_k,p_{k+1}\}_{23\alpha} & = & -\displaystyle\frac{q_k}
{1+\epsilon\alpha q_k}\,
(1+\epsilon p_k)(1+\epsilon p_{k+1})\;, \nonumber\\  \nonumber\\
\{q_k,q_{k+1}\}_{23\alpha} & = & -q_kq_{k+1}(1+\epsilon p_{k+1})
(1+\epsilon\alpha q_k)(1+\epsilon\alpha q_{k+1})\nonumber
\end{eqnarray}
define a Poisson bracket on ${\cal MRT}$ compatible with 
{\rm(\ref{MRTL+ br 1})}. The systems
${\rm MRTL}_{\pm}^{(+)}(\alpha;\epsilon)$ are Hamiltonian 
with respect to this bracket with the Hamilton function 
\begin{eqnarray}
\rG_{0\alpha}^{(+)}(q,p) & = & \epsilon^{-1}\sum_{k=1}^N 
\log(1+\epsilon p_k)+
\epsilon^{-1}\sum_{k=1}^N \log(1+\epsilon \alpha q_k)\;, 
\label{MRTL++ H1}\\ \nonumber\\
\rG_{0\alpha}^{(-)}(q,p) & = & (\epsilon-\alpha)^{-1}\sum_{k=1}^N 
\log\frac{1+\epsilon p_k}{1+\alpha p_k}\;,\label{MRTL+- H1}
\end{eqnarray}
respectively. The functions $\rG_{0\alpha}^{(+)}$ and $\rG_{0\alpha}^{(-)}$
are in involution in the bracket $\{\cdot,\cdot\}_{23\alpha}$.
\end{proposition}

Notice that the function $\rG_{1\alpha}^{(+)}(q,p)$ becomes regular in
$\epsilon$ upon subtracting
\[
\epsilon^{-2}\sum_{k=1}^N \log(1+\epsilon p_k)+
\epsilon^{-2}\sum_{k=1}^N \log(1+\epsilon\alpha q_k)\;,
\] 
which is a Casimir of the bracket $\{\cdot,\cdot\}_{12\alpha}$.

\subsection{Miura relations}
The relation between ${\rm MRTL}^{(+)}(\alpha;\epsilon)$ and 
${\rm RTL}(\alpha)$ hierarchies 
is established in the following statement.
\begin{proposition}\label{Miuras for MRTL}
Define the Miura maps $\rM_{1,2}^{(+)}(\alpha;\epsilon):\;
{\cal MRT}(q,p)\mapsto\cR\cT(a,b)$ by:
\begin{eqnarray}
 \rM_1^{(+)}(\alpha;\epsilon): & & \left\{\begin{array}{l} 
b_k=p_k+\epsilon q_{k-1}+\epsilon\alpha p_kq_{k-1}\;,\\ \\ 
a_k=q_k(1+\epsilon p_k)(1+\epsilon\alpha q_{k-1})\;, 
\end{array}\right. 
\label{MRTL+ Miura 1} \\ \nonumber\\  \nonumber\\
  \rM_2^{(+)}(\alpha;\epsilon): & & \left\{\begin{array}{l} 
b_k=p_k+\epsilon q_k+\epsilon\alpha p_kq_k\;,\\ \\ 
a_k=q_k(1+\epsilon p_{k+1})(1+\epsilon\alpha q_{k+1})\;.\end{array}\right. 
\label{MRTL+ Miura 2}
\end{eqnarray}
Both maps $\rM_{1,2}^{(+)}(\alpha;\epsilon)$ are Poisson, if 
${\cal MRT}(q,p)$ is equipped with the bracket 
$\{\cdot,\cdot\}_{12\alpha}$, and $\cR\cT(a,b)$ is 
equipped with $\{\cdot,\cdot\}_{1\alpha}+\epsilon\{\cdot,\cdot\}_{2\alpha}$, 
and also if ${\cal MRT}(q,p)$ is equipped with the bracket 
$\{\cdot,\cdot\}_{23\alpha}$, and $\cR\cT(a,b)$ is equipped with 
$\{\cdot,\cdot\}_{2\alpha}+\epsilon\{\cdot,\cdot\}_{3\alpha}$.
The pull--back of the flows ${\rm RTL}_{\pm}(\alpha)$ under either of 
the Miura maps $\rM_{1,2}^{(+)}(\alpha;\epsilon)$ coincides with 
${\rm MRTL}_{\pm}^{(+)}(\alpha;\epsilon)$.
\end{proposition}
In the rest of this section we shall sometimes denote by 
$a_k^{(1)}=a_k^{(1)}(q,r)$, $b_k^{(1)}=b_k^{(1)}(q,r)$ the functions given 
by the formulas (\ref{MRTL+ Miura 1}), and by $a_k^{(2)}=a_k^{(2)}(q,r)$, 
$b_k^{(2)}=b_k^{(2)}(q,r)$ the functions given by the formulas 
(\ref{MRTL+ Miura 2}).

\subsection{Lax representations}
The next interesting issue of the ``relativistic splitting'' is the
following: the ${\rm MRTL}^{(+)}$ hierarchy possesses two different
Lax representations.

The first Lax matrix $(P_1,W_1^{-1},Q_1):\cM\cR\cT\mapsto\g\otimes\g\otimes\g$:
\begin{eqnarray}
P_1(q,p,\lambda) & = & \sum_{k=1}^N (1+\epsilon p_k)(1+\epsilon\alpha q_{k-1})
E_{kk}+\epsilon\lambda\sum_{k=1}^N E_{k+1,k}\;, \label{MRTL+ P1}\\
\nonumber\\
Q_1(q,p,\lambda) & = & I+(\epsilon-\alpha)\lambda^{-1}\sum_{k=1}^N 
q_k E_{k,k+1}\;,  \label{MRTL+ Q1}\\  \nonumber\\
W_1(q,p,\lambda) & = & I -\alpha\lambda^{-1}\sum_{k=1}^N q_k(1+\epsilon p_k)
(1+\epsilon\alpha q_{k-1})E_{k,k+1}\;.\label{MRTL+ W1}
\end{eqnarray}
Notice that the formulas (\ref{MRTL+ Miura 1}) for the Miura map 
$\rM_1^{(+)}(\alpha;\epsilon)$ are equivalent to the factorization
\begin{equation}\label{MRTL fact 1}
\Big(1-\frac{\epsilon}{\alpha}\Big)W(a^{(1)},b^{(1)},\lambda)+
\frac{\epsilon}{\alpha}\,L(a^{(1)},b^{(1)},\lambda)=
P_1(q,p,\lambda)Q_1(q,p,\lambda)\;,
\end{equation}
along with the formula $W(a^{(1)},b^{(1)},\lambda)=W_1(q,p,\lambda)$.
\vspace{2mm}

The second Lax matrix $(P_2,Q_2,W_2^{-1}):\cM\cR\cT\mapsto\g\otimes\g\otimes\g$:
\begin{eqnarray}
P_2(q,p,\lambda) & = & \sum_{k=1}^N (1+\epsilon p_k)(1+\epsilon\alpha q_k)
E_{kk}+\epsilon\lambda\sum_{k=1}^N E_{k+1,k}\;, \label{MRTL P2}\\
\nonumber\\
Q_2(q,p,\lambda) & = & I+(\epsilon-\alpha)\lambda^{-1}\sum_{k=1}^N 
q_k E_{k,k+1}\;,  \label{MRTL Q2}\\ \nonumber\\
W_2(q,p,\lambda) & = & I -\alpha\lambda^{-1}\sum_{k=1}^N q_k(1+\epsilon p_{k+1})
(1+\epsilon\alpha q_{k+1})E_{k,k+1}\;.
\end{eqnarray}
The formulas (\ref{MRTL+ Miura 2}) for the Miura map 
$\rM_2^{(+)}(\alpha;\epsilon)$ are equivalent to the factorization
\begin{equation}\label{MRTL fact 2}
\Big(1-\frac{\epsilon}{\alpha}\Big)W(a^{(2)},b^{(2)},\lambda)+
\frac{\epsilon}{\alpha}\,L(a^{(2)},b^{(2)},\lambda)=
Q_2(q,p,\lambda)P_2(q,p,\lambda)\;,
\end{equation}
along with the formula $W(a^{(2)},b^{(2)},\lambda)=W_2(q,p,\lambda)$.
\vspace{2mm}

The relation between these two Lax representation is established via the
following formulas:
\begin{equation}\label{MRTL+ Lax rel}
Q_1=Q_2\;,\qquad W_1^{-1}P_1=P_2W_2^{-1}\;.
\end{equation}
\vspace{2mm}

The first Lax representation of ${\rm MRTL}_+^{(+)}(\alpha;\epsilon)$:
\begin{eqnarray}
\dot{P}_1 & = & P_1A_3-A_1P_1\;,\label{MRTL++ Lax triads P1}\\ \nonumber\\
\dot{W}_1 & = & W_1A_2-A_1W_1\;,\label{MRTL++ Lax triads W1}\\ \nonumber\\
\dot{Q}_1 & = & Q_1A_2-A_3Q_1\;,\label{MRTL++ Lax triads Q1}
\end{eqnarray}
where
\begin{eqnarray*}
A_1=\pi_+\Big((P_1Q_1W_1^{-1}-I)/\epsilon\Big) & = & 
\sum_{k=1}^N\Big(b_k^{(1)}+\alpha a_{k-1}^{(1)}\Big)E_{kk}+
\lambda\sum_{k=1}^N E_{k+1,k}\;,   \label{MRTL+ A1}\\ 
A_2=\pi_+\Big((W_1^{-1}P_1Q_1-I)/\epsilon\Big) & = & 
\sum_{k=1}^N\Big(b_k^{(1)}+\alpha a_k^{(1)}\Big)E_{kk}+
\lambda\sum_{k=1}^N E_{k+1,k}\;, \label{MRTL+ A2}\\ 
A_3=\pi_+\Big((Q_1W_1^{-1}P_1-I)/\epsilon\Big) & = & 
\sum_{k=1}^N\Big(b_k^{(2)}+\alpha a_{k-1}^{(2)}\Big)E_{kk}+
\lambda\sum_{k=1}^N E_{k+1,k}\;.  \label{MRTL+ A3} 
\end{eqnarray*}
The second Lax representation of ${\rm MRTL}_+^{(+)}(\alpha;\epsilon)$: 
\begin{eqnarray}
\dot{P}_2 & = & P_2B_3-B_1P_2\;,\label{MRTL++ Lax triads P2}\\ \nonumber\\
\dot{Q}_2 & = & Q_2B_1-B_2Q_2\;,\label{MRTL++ Lax triads Q2}\\ \nonumber\\
\dot{W}_2 & = & W_2B_3-B_2W_2\;,\label{MRTL++ Lax triads W2}
\end{eqnarray}
where
\begin{eqnarray*}
B_1=\pi_+\Big((P_2W_2^{-1}Q_2-I)/\epsilon\Big) & = &
\sum_{k=1}^N\Big(b_k^{(1)}+\alpha a_k^{(1)}\Big)E_{kk}+
\lambda\sum_{k=1}^N E_{k+1,k}\;,\\
B_2=\pi_+\Big((Q_2P_2W_2^{-1}-I)/\epsilon\Big) & = & 
\sum_{k=1}^N\Big(b_k^{(2)}+\alpha a_{k-1}^{(2)}\Big)E_{kk}+
\lambda\sum_{k=1}^N E_{k+1,k}\;,  \\ 
B_3=\pi_+\Big((W_2^{-1}Q_2P_2-I)/\epsilon\Big) & = & 
\sum_{k=1}^N\Big(b_k^{(2)}+\alpha a_k^{(2)}\Big)E_{kk}+
\lambda\sum_{k=1}^N E_{k+1,k}\;. 
\end{eqnarray*}
The relations (\ref{MRTL+ Lax rel}) assure the identities
\[
A_2=B_1\;,\qquad A_3=B_2\;.
\]
Let us mention the following expressions for the entries of the matrices
$A_j$, $B_j$ above:
\begin{eqnarray*}
b_k^{(1)}+\alpha a_{k-1}^{(1)}\!\! & = & \!\!
p_k+(\epsilon+\alpha)q_{k-1}+\epsilon\alpha
(q_{k-1}p_{k-1}+p_kq_{k-1}+\alpha q_{k-1}q_{k-2}+
\epsilon\alpha q_{k-1}p_{k-1}q_{k-2}),\\
b_k^{(1)}+\alpha a_k^{(1)}\!\! & = & \!\!
p_k+\alpha q_k+\epsilon q_{k-1}+\epsilon\alpha
(p_kq_{k-1}+q_kp_k+\alpha q_kq_{k-1}+\epsilon\alpha q_kp_kq_{k-1})\;,\\ 
b_k^{(2)}+\alpha a_{k-1}^{(2)}\!\! & = & \!\! 
p_k+\alpha q_{k-1}+\epsilon q_k+\epsilon\alpha
(p_kq_{k-1}+q_kp_k+\alpha q_kq_{k-1}+\epsilon\alpha q_kp_kq_{k-1})\;.\\
b_k^{(2)}+\alpha a_k^{(2)}\!\! & = & \!\!
p_k+(\epsilon+\alpha)q_k+\epsilon\alpha
(q_kp_k+p_{k+1}q_k+\alpha q_{k+1}q_k+
\epsilon\alpha q_{k+1}p_{k+1}q_k)\;.
\end{eqnarray*}
\vspace{2mm}

The first Lax representation of ${\rm MRTL}_-^{(+)}(\alpha;\epsilon)$:
\begin{eqnarray}
\dot{P}_1 & = & C_1P_1-P_1C_3\;,\label{MRTL+- Lax triads P1}\\ \nonumber\\
\dot{W}_1 & = & C_1W_1-W_1C_2\;,\label{MRTL+- Lax triads W1}\\ \nonumber\\
\dot{Q}_1 & = & C_3Q_1-Q_1C_2\;,\label{MRTL+- Lax triads Q1}
\end{eqnarray}
with 
\begin{eqnarray*}
C_1 & = & 
\lambda^{-1}\sum_{k=1}^N 
\frac{a_k^{(1)}}
{1+\alpha b_{k+1}^{(1)}}\,E_{k,k+1}=
\lambda^{-1}\sum_{k=1}^N 
\frac{q_k(1+\epsilon p_k)(1+\epsilon\alpha q_{k-1})}
{(1+\alpha p_{k+1})(1+\epsilon\alpha q_k)}\,E_{k,k+1}\;, \\
C_2 & = & 
\lambda^{-1}\sum_{k=1}^N 
\frac{a_k^{(1)}}{1+\alpha b_k^{(1)}}\,E_{k,k+1}= \lambda^{-1}\sum_{k=1}^N 
\frac{q_k(1+\epsilon p_k)}{1+\alpha p_k}\,E_{k,k+1}\;,\\
C_3 & = & 
\lambda^{-1}\sum_{k=1}^N 
\frac{a_k^{(2)}}{1+\alpha b_{k+1}^{(2)}}\,E_{k,k+1}= \lambda^{-1}\sum_{k=1}^N 
\frac{q_k(1+\epsilon p_{k+1})}{1+\alpha p_{k+1}}\,E_{k,k+1}\;.
\end{eqnarray*}
The second Lax representation of ${\rm MRTL}_-^{(+)}(\alpha;\epsilon)$:
\begin{eqnarray}
\dot{P}_2 & = & D_1P_2-P_2D_3\;,\label{MRTL+- Lax triads Q2}\\ \nonumber\\
\dot{Q}_2 & = & D_2Q_2-Q_2D_1\;,\label{MRTL+- Lax triads W2}\\ \nonumber\\
\dot{W}_2 & = & D_2W_2-W_2D_3\;,\label{MRTL+- Lax triads P2}
\end{eqnarray}
with 
\begin{eqnarray*}
D_1 & = & 
\lambda^{-1}\sum_{k=1}^N 
\frac{a_k^{(1)}}{1+\alpha b_k^{(1)}}\,E_{k,k+1}= \lambda^{-1}\sum_{k=1}^N 
\frac{q_k(1+\epsilon p_k)}{1+\alpha p_k}\,E_{k,k+1}\;,\\
D_2 & = & 
\lambda^{-1}\sum_{k=1}^N 
\frac{a_k^{(2)}}{1+\alpha b_{k+1}^{(2)}}\,E_{k,k+1}=
\lambda^{-1}\sum_{k=1}^N 
\frac{q_k(1+\epsilon p_{k+1})}
{1+\alpha p_{k+1}}\,E_{k,k+1}\;, \\
D_3 & = & 
\lambda^{-1}\sum_{k=1}^N 
\frac{a_k^{(2)}}{1+\alpha b_k^{(2)}}\,E_{k,k+1}= \lambda^{-1}\sum_{k=1}^N 
\frac{q_k(1+\epsilon p_{k+1})(1+\epsilon\alpha q_{k+1})}
{(1+\alpha p_k)(1+\epsilon\alpha q_k)}\,E_{k,k+1}\;.
\end{eqnarray*}

\subsection{Discretization}
We discuss here only the discretization of the flow 
${\rm MRTL}_+^{(+)}(\alpha;\epsilon)$.

The first Lax representation of ${\rm dMRTL}_+^{(+)}(\alpha;\epsilon)$:
\begin{equation}\label{dMRTL++ 1 Lax}
\wiP_1=\mbA_1^{-1}P_1\mbA_3\;,\qquad \wW_1=\mbA_1^{-1}W_1\mbA_2\;,\qquad
\wQ_1=\mbA_3^{-1}Q_1\mbA_2
\end{equation}
with
\begin{eqnarray}
\mbA_1=\Pi_+\Big(I+\frac{h}{\epsilon}\,(P_1Q_1W_1^{-1}-I)\Big) & = & 
\sum_{k=1}^N \ga_kE_{kk}+h\lambda\sum_{k=1}^NE_{k+1,k}\;,\label{dMRTL+ A1}\\
\mbA_2=\Pi_+\Big(I+\frac{h}{\epsilon}\,(W_1^{-1}P_1Q_1-I)\Big) & = & 
\sum_{k=1}^N \gb_kE_{kk}+h\lambda\sum_{k=1}^NE_{k+1,k}\;,\label{dMRTL+ A2}\\
\mbA_3=\Pi_+\Big(I+\frac{h}{\epsilon}\,(Q_1W_1^{-1}P_1-I)\Big) & = & 
\sum_{k=1}^N \goe_kE_{kk}+h\lambda\sum_{k=1}^N E_{k+1,k}\;.\label{dMRTL+ A3}
\end{eqnarray}
The second Lax representation of ${\rm dMRTL}_+^{(+)}(\alpha;\epsilon)$:
\begin{equation}\label{dMRTL++ 2 Lax}
\wiP_2=\mbB_1^{-1}P_2\mbB_3\;,\qquad \wQ_2=\mbB_2^{-1}Q_2\mbB_1\;,\qquad
\wW_2=\mbB_2^{-1}W_2\mbB_3
\end{equation}
with
\begin{eqnarray}
\mbB_1=\Pi_+\Big(I+\frac{h}{\epsilon}\,(P_2W_2^{-1}Q_2-I)\Big) & = & 
\sum_{k=1}^N \gb_kE_{kk}+h\lambda\sum_{k=1}^N E_{k+1,k}\;, \label{dMRTL+ B1}\\
\mbB_2=\Pi_+\Big(I+\frac{h}{\epsilon}\,(Q_2P_2W_2^{-1}-I)\Big) & = & 
\sum_{k=1}^N \goe_kE_{kk}+h\lambda\sum_{k=1}^NE_{k+1,k}\;, \label{dMRTL+ B2}\\
\mbB_3=\Pi_+\Big(I+\frac{h}{\epsilon}\,(W_2^{-1}Q_2P_2-I)\Big) & = & 
\sum_{k=1}^N \gf_kE_{kk}+h\lambda\sum_{k=1}^NE_{k+1,k}\;.  \label{dMRTL+ B3}
\end{eqnarray}

\subsection{Application: localizing change of variables for 
${\rm dRTL}_+(\alpha)$}
The role of the localizing change of variables for the map
${\rm dRTL}_+(\alpha)$ is played by the Miura transformation
$\rM_1^{(+)}(\alpha;h)$. Indeed, we have:
\[
\Pi_+\bigg(I+\frac{h}{\alpha}\,\Big(LW^{-1}-I\Big)\bigg)=
\Pi_+\bigg(\Big(1-\frac{h}{\alpha}\Big)W+\frac{h}{\alpha}\,L\bigg)\;.
\]
Compare this with (\ref{MRTL fact 1}). We see that if we define the 
change of variables $\cR\cT(\ba,\bb)\mapsto\cR\cT(a,b)$ by the formulas
\begin{equation}\label{dRTL+ param loc map}
\rM_1^{(+)}(\alpha;h)\,:\quad\left\{\begin{array}{l}
b_k=\bb_k+h\ba_{k-1}+h\alpha\bb_k\ba_{k-1}\;,\\ \\
a_k=\ba_k(1+h\bb_k)(1+h\alpha \ba_{k-1})\;,
\end{array}\right.
\end{equation}
then
\[
\Pi_+\bigg(I+\frac{h}{\alpha}\,\Big(LW^{-1}-I\Big)\bigg)=
P_1(\ba,\bb,\lambda)=\sum_{k=1}^N(1+h\bb_k)(1+h\alpha\ba_{k-1})E_{kk}+
h\lambda\sum_{k=1}^N E_{k+1,k}\;.
\] 
So, we get the following local expressions 
for the entries of the factors (\ref{dRTL+ param A1}), (\ref{dRTL+ param A2}):
\[
\ga_k=(1+h\bb_k)(1+h\alpha\ba_{k-1})\;,\qquad
\gb_k=(1+h\bb_k)(1+h\alpha\ba_k)\;.
\]
\begin{theorem}\label{dRTL+ param in localization variables}
The change of variables {\rm(\ref{dRTL+ param loc map})} conjugates 
${\rm dRTL}_+(\alpha)$  with the map on $\cR\cT(\ba,\bb)$ described by 
the following local equations of motion:
\begin{equation}\label{dRTL+ param loc}
\begin{array}{l}
\widetilde{\bb}_k+h\widetilde{\ba}_{k-1}(1+\alpha\widetilde{\bb}_k)=
\bb_k+h\ba_k(1+\alpha\bb_k)\;,\\ \\
\widetilde{\ba}_k(1+h\widetilde{\bb}_k)(1+h\alpha\widetilde{\ba}_{k-1})
=\ba_k(1+h\bb_{k+1})(1+h\alpha\ba_{k+1})\;.
\end{array}
\end{equation}
\end{theorem}
So, the local form of ${\rm dRTL}_+(\alpha)$, i.e. the map 
(\ref{dRTL+ param loc}), belongs actually to the hierarchy 
${\rm MRTL}^{(+)}(\alpha;h)$.
\vspace{1.5mm}

{\bf Corollary.} {\it The local form of ${\rm dRTL}_+(\alpha)$ 
{\rm(\ref{dRTL+ param loc})} is a Poisson map with respect to the following
Poisson bracket on $\cR\cT(\ba,\bb)$:
\begin{eqnarray}
\{\bb_k,\ba_k\} & = & -\ba_k(1+h\bb_k)\;, \nonumber\\
\{\ba_k,\bb_{k+1}\} & = & -\ba_k(1+h\bb_{k+1})\;,
\label{dRTL+ param loc PB1}\\
\{\bb_k,\bb_{k+1}\} & = & \alpha\,\displaystyle\frac
{\ba_k}{1+h\alpha\ba_k}\,(1+h\bb_k)(1+h\bb_{k+1})\;,
\nonumber
\end{eqnarray}
which is the pull--back of the bracket 
\begin{equation}\label{dRTL+ param loc PB1 target}
\{\cdot,\cdot\}_{1\alpha}+h\{\cdot,\cdot\}_{2\alpha}
\end{equation}
on $\cR\cT(a,b)$ under the change of variables {\rm(\ref{dRTL+ param loc map})},
and also with respect to the following Poisson bracket on $\cR\cT(\ba,\bb)$,
compatible with {\rm(\ref{dRTL+ param loc PB1})}:
\begin{eqnarray}
\{\bb_k,\ba_k\} & = & -\ba_k(\bb_k+h\ba_k+h\alpha\bb_k\ba_k)(1+h\bb_k) \;,
\nonumber\\ \nonumber\\
\{\ba_k,\bb_{k+1}\} & = & -\ba_k(\bb_{k+1}+h\ba_k+h\alpha\bb_{k+1}\ba_k)
(1+h\bb_{k+1})\;,\nonumber\\
\label{dRTL+ param loc PB2}\\
\{\bb_k,\bb_{k+1}\} & = & -\displaystyle\frac
{\ba_k}{1+h\alpha\ba_k}\,(1+h\bb_k)(1+h\bb_{k+1})\;, \nonumber\\ \nonumber\\
\{\ba_k,\ba_{k+1}\} & = & -\ba_k\ba_{k+1}(1+h\alpha\ba_k)(1+h\bb_{k+1})
(1+h\alpha\ba_{k+1})\;,\nonumber
\end{eqnarray}
which is the pull--back of the bracket
\begin{equation}\label{dRTL+ param loc PB2 target}
\{\cdot,\cdot\}_{2\alpha}+h\{\cdot,\cdot\}_{3\alpha}
\end{equation}
on $\cR\cT(a,b)$ under the change of variables {\rm(\ref{dRTL+ param loc map})}.}

\setcounter{equation}{0}
\section{Modified relativistic Toda lattice II}

We introduce now a {\it modified relativistic Toda hierarchy} 
${\rm MRTL}^{(-)}(\alpha;\epsilon)$ different from 
the one considered in the previous section. The presentation will be parallel
to that of the previous section, and there will be several similar objects 
denoted by the same letters. This should not lead to a confusion.

\subsection{Equations of motion}
In particular, we shall use the notation ${\cal MRT}={\Bbb R}^{2N}(q,p)$
for the phase space of the ${\rm MRTL}^{(-)}(\alpha;\epsilon)$ hierarchy.

Equations of motion of ${\rm MRTL}^{(-)}_+(\alpha;\epsilon)$:
\begin{equation}\label{MRTL-+}
\begin{array}{l}
\dot{p}_k=\Big(1+(\epsilon+\alpha)p_k\Big)(q_k-q_{k-1})\;,\\ \\
\dot{q}_k=q_k\left(p_{k+1}+\alpha q_{k+1}+
\displaystyle\frac{\epsilon\alpha p_{k+1}q_{k+1}}{1+\alpha p_{k+1}}
-p_k-\alpha q_{k-1}-\displaystyle\frac{\epsilon\alpha p_kq_{k-1}}
{1+\alpha p_k}\right).
\end{array}
\end{equation}
Equations of motion of ${\rm MRTL}^{(-)}_-(\alpha;\epsilon)$:
\begin{equation}\label{MRTL--}
\begin{array}{l}
\dot{p}_k=\!\Big(1+(\epsilon+\alpha)p_k\Big)\!
\left(\displaystyle\frac
{q_k}{(1+\alpha p_k)(1+\alpha p_{k+1})+\epsilon\alpha q_k}-
\displaystyle\frac{q_{k-1}}
{(1+\alpha p_{k-1})(1+\alpha p_k)+\epsilon\alpha q_{k-1}}\right)\!,
\\ \\
\dot{q}_k=\displaystyle\frac{q_k(p_{k+1}-p_k)}
{(1+\alpha p_k)(1+\alpha p_{k+1})+\epsilon\alpha q_k}\,.
\end{array}
\end{equation}

\subsection{Bi--Hamiltonian structure}
\begin{proposition}
{\rm a)} The relations
\begin{eqnarray}
\{p_k,q_k\}_{12\alpha} & = & -q_k\Big(1+(\epsilon+\alpha)p_k\Big)\;, 
\nonumber\\
\{q_k,p_{k+1}\}_{12\alpha} & = & 
-q_k\Big(1+(\epsilon+\alpha)p_{k+1}\Big)\;,
\label{MRTL- br 1}\\
\{q_k,q_{k+1}\}_{12\alpha} & = & -\alpha q_kq_{k+1}
\left(1+\displaystyle\frac
{\epsilon p_{k+1}}{1+\alpha p_{k+1}}\right)\,.
\nonumber
\end{eqnarray}
define a Poisson bracket on ${\cal MRT}(q,p)$. 
The systems ${\rm MRTL}^{(-)}_{\pm}(\alpha;\epsilon)$ are Hamiltonian 
with respect to the bracket {\rm (\ref{MRTL+ br 1})}, with the Hamilton 
functions
\begin{equation}\label{MRTL-+ H2}
\rG_{1\alpha}^{(+)}(q,p)=(\epsilon+\alpha)^{-1}\sum_{k=1}^N p_k
+\sum_{k=1}^N q_k \;,
\end{equation}
\begin{equation}\label{MRTL-- H2}
\rG_{1\alpha}^{(-)}(q,p)=(\epsilon\alpha)^{-1}\sum_{k=1}^N \log(1+\alpha p_k)
+(\epsilon\alpha)^{-1}\sum_{k=1}^N \log\left(1+\frac{\epsilon\alpha q_k}
{(1+\alpha p_k)(1+\alpha p_{k+1})}\right),
\end{equation}  
respectively. The functions $\rG_{1\alpha}^{(+)}$ and $\rG_{1\alpha}^{(-)}$
are in involution in the bracket $\{\cdot,\cdot\}_{12\alpha}$.

{\rm b)} The relations
\begin{eqnarray}
\{p_k,q_k\}_{23\alpha}     & = & -q_k\Big(p_k+(\epsilon+\alpha)q_k\Big)
\Big(1+(\epsilon+\alpha)p_k\Big)\;,\nonumber\\
\{q_k,_{k+1}\}_{23\alpha}  & = & -q_k\Big(p_{k+1}+(\epsilon+\alpha)q_k\Big)
\Big(1+(\epsilon+\alpha)p_{k+1}\Big)\;,\nonumber\\
\{p_k,p_{k+1}\}_{23\alpha} & = & -q_k\Big(1+(\epsilon+\alpha)p_k\Big)
(1+(\epsilon+\alpha)p_{k+1}\Big)\;,\nonumber\\
\{q_k,q_{k+1}\}_{23\alpha} & = & -q_kq_{k+1}\Big(1+2\alpha p_{k+1}
+\alpha(\epsilon+\alpha)(q_k+q_{k+1})\Big)
\left(1+\displaystyle\frac{\epsilon p_{k+1}}{1+\alpha p_{k+1}}\right),
\nonumber\\
\{p_k,q_{k+1}\}_{23\alpha} & = & -\alpha q_k q_{k+1}
\Big(1+(\epsilon+\alpha)p_k\Big)
\left(1+\displaystyle\frac{\epsilon p_{k+1}}{1+\alpha p_{k+1}}\right)\;,
\label{MRTL- br 2}\\
\{q_k,p_{k+2}\}_{23\alpha} & = & -\alpha q_kq_{k+1}
\Big(1+(\epsilon+\alpha)p_{k+2}\Big)
\left(1+\displaystyle\frac{\epsilon p_{k+1}}{1+\alpha p_{k+1}}\right)\;,
\nonumber\\
\{q_k,q_{k+2}\}_{23\alpha} & = & -\alpha^2 q_kq_{k+1}q_{k+2}
\left(1+\displaystyle\frac{\epsilon p_{k+1}}{1+\alpha p_{k+1}}\right)
\left(1+\displaystyle\frac{\epsilon p_{k+2}}{1+\alpha p_{k+2}}\right)\;.
\nonumber
\end{eqnarray}
define a Poisson bracket on ${\cal MRT}$ compatible with 
{\rm(\ref{MRTL- br 1})}. The systems
${\rm MRTL}_{\pm}^{(-)}(\alpha;\epsilon)$ are Hamiltonian 
with respect to this bracket with the Hamilton function 
\begin{equation}\label{MRTL-+ H1}
\rG_{1\alpha}^{(+)}(q,p)=(\epsilon+\alpha)^{-1}\sum_{k=1}^N
\log\Big(1+(\epsilon+\alpha)p_k\Big)\;, 
\end{equation}
\begin{equation}\label{MRTL-- H1}
\rG_{1\alpha}^{(-)}(q,p)=\epsilon^{-1}\sum_{k=1}^N
\log\left(1+\frac{\epsilon p_k}{1+\alpha p_k}\right)
-\epsilon^{-1}\sum_{k=1}^N\log\left(
1+\frac{\epsilon\alpha q_k}{(1+\alpha p_k)(1+\alpha p_{k+1})}\right),
\end{equation}
respectively. The functions $\rG_{0\alpha}^{(+)}$ and $\rG_{0\alpha}^{(-)}$
are in involution in the bracket $\{\cdot,\cdot\}_{23\alpha}$.
\end{proposition}
Notice that the function $\rG_{1\alpha}^{(-)}(q,p)$ becomes regular in 
$\epsilon$ upon subtracting
\[
(\epsilon\alpha)^{-1}\sum_{k=1}^N
\log\Big(1+(\epsilon+\alpha)p_k\Big)\;,
\]
which is a Casimir function of the bracket $\{\cdot,\cdot\}_{12\alpha}$.

\subsection{Miura relations}
The relation between ${\rm MRTL}^{(-)}(\alpha;\epsilon)$ and 
${\rm RTL}(\alpha)$ hierarchies 
is established in the following statement.
\begin{proposition}\label{Miuras for MRTL-}
Define the Miura maps $\rM_{1,2}^{(-)}(\alpha;\epsilon):\;
{\cal MRT}(q,p)\mapsto\cR\cT(a,b)$ by:
\begin{eqnarray}
 \rM_1^{(-)}(\alpha;\epsilon): & & \left\{\begin{array}{l} 
b_k=p_k+\displaystyle\frac{\epsilon q_{k-1}}{1+\alpha p_{k-1}}\;,\\ \\ 
a_k=q_k\left(1+\displaystyle\frac{\epsilon p_k}{1+\alpha p_k}\right)\;, 
\end{array}\right. 
\label{MRTL- Miura 1} \\ \nonumber\\  \nonumber\\
\rM_2^{(-)}(\alpha;\epsilon): & & \left\{\begin{array}{l} 
b_k=p_k+\displaystyle\frac{\epsilon q_k}{1+\alpha p_{k+1}}\;,\\ \\ 
a_k=q_k\left(1+\displaystyle\frac{\epsilon p_{k+1}}{1+\alpha p_{k+1}}\right)\;. 
\end{array}\right. 
\label{MRTL- Miura 2}
\end{eqnarray}
Both maps $\rM_{1,2}^{(-)}(\alpha;\epsilon)$ are Poisson, if 
${\cal MRT}(q,p)$ is equipped with the bracket 
$\{\cdot,\cdot\}_{12\alpha}$, and $\cR\cT(a,b)$ is 
equipped with $\{\cdot,\cdot\}_{1\alpha}+(\epsilon+\alpha)
\{\cdot,\cdot\}_{2\alpha}$, 
and also if ${\cal MRT}(q,p)$ is equipped with the bracket 
$\{\cdot,\cdot\}_{23\alpha}$, and $\cR\cT(a,b)$ is equipped with 
$\{\cdot,\cdot\}_{2\alpha}+(\epsilon+\alpha)\{\cdot,\cdot\}_{3\alpha}$.
The pull--back of the flows ${\rm RTL}_{\pm}(\alpha)$ under either of 
the Miura maps $\rM_{1,2}^{(-)}(\alpha;\epsilon)$ coincides with 
${\rm MRTL}_{\pm}^{(-)}(\alpha;\epsilon)$.
\end{proposition}
In the rest of this section we shall sometimes denote by 
$a_k^{(1)}=a_k^{(1)}(q,r)$, $b_k^{(1)}=b_k^{(1)}(q,r)$ the functions given 
by the formulas (\ref{MRTL- Miura 1}), and by $a_k^{(2)}=a_k^{(2)}(q,r)$, 
$b_k^{(2)}=b_k^{(2)}(q,r)$ the functions given by the formulas 
(\ref{MRTL- Miura 2}).

\subsection{Lax representations}
The ${\rm MRTL}^{(-)}$ hierarchy, like the ${\rm MRTL}^{(+)}$ one, possesses 
two different Lax representations.

The first Lax matrix $(P_1,W_1^{-1},Q_1):\cM\cR\cT\mapsto\g\otimes\g\otimes\g$:
\begin{eqnarray}
P_1(q,p,\lambda) & = & \sum_{k=1}^N \Big(1+(\epsilon+\alpha)p_k\Big)E_{kk}
+(\epsilon+\alpha)\lambda\sum_{k=1}^N E_{k+1,k}\;, \label{MRTL- P1}\\
\nonumber\\
Q_1(q,p,\lambda) & = & I+\epsilon\lambda^{-1}\sum_{k=1}^N 
\displaystyle\frac{q_k}{1+\alpha p_k} E_{k,k+1}\;,  \label{MRTL- Q1}\\  
\nonumber\\
W_1(q,p,\lambda) & = & I -\alpha\lambda^{-1}\sum_{k=1}^N q_k\left(1+
\displaystyle\frac{\epsilon p_k}{1+\alpha p_k}\right)E_{k,k+1}\;.
\label{MRTL- W1}
\end{eqnarray}
The formulas (\ref{MRTL- Miura 1}) for the Miura map 
$\rM_1^{(-)}(\alpha;\epsilon)$ are equivalent to the factorization
\begin{equation}\label{MRTL- fact 1}
\Big(1+\frac{\epsilon}{\alpha}\Big)L(a^{(1)},b^{(1)},\lambda)-
\frac{\epsilon}{\alpha}\,W(a^{(1)},b^{(1)},\lambda)=
P_1(q,p,\lambda)Q_1(q,p,\lambda)\;,
\end{equation}
along with the formula $W(a^{(1)},b^{(1)},\lambda)=W_1(q,p,\lambda)$.
\vspace{2mm}

The second Lax matrix $(P_2,Q_2,W_2^{-1}):\cM\cR\cT\mapsto\g\otimes\g\otimes\g$:
\begin{eqnarray}
P_2(q,p,\lambda) & = & \sum_{k=1}^N \Big(1+(\epsilon+\alpha)p_k\Big)E_{kk}
+(\epsilon+\alpha)\lambda\sum_{k=1}^N E_{k+1,k}\;, \label{MRTL- P2}\\
\nonumber\\
Q_2(q,p,\lambda) & = & I+\epsilon\lambda^{-1}\sum_{k=1}^N 
\displaystyle\frac{q_k}{1+\alpha p_{k+1}} E_{k,k+1}\;,  \label{MRTL- Q2}\\  
\nonumber\\
W_2(q,p,\lambda) & = & I -\alpha\lambda^{-1}\sum_{k=1}^N q_k\left(1+
\displaystyle\frac{\epsilon p_{k+1}}{1+\alpha p_{k+1}}\right)E_{k,k+1}\;.
\label{MRTL- W2}
\end{eqnarray}
The formulas (\ref{MRTL- Miura 2}) for the Miura map 
$\rM_2^{(-)}(\alpha;\epsilon)$ are equivalent to the factorization
\begin{equation}\label{MRTL- fact 2}
\Big(1+\frac{\epsilon}{\alpha}\Big)L(a^{(2)},b^{(2)},\lambda)-
\frac{\epsilon}{\alpha}\,W(a^{(2)},b^{(2)},\lambda)=
Q_2(q,p,\lambda)P_2(q,p,\lambda)\;,
\end{equation}
along with the formula $W(a^{(2)},b^{(2)},\lambda)=W_2(q,p,\lambda)$.
\vspace{2mm}

The relation between these two Lax representation is established via the
following formulas:
\begin{equation}\label{MRTL- Lax rel}
P_1=P_2\;,\qquad Q_1W_1^{-1}=W_2^{-1}Q_2\;.
\end{equation}
\vspace{2mm}

The first Lax representation of ${\rm MRTL}_+^{(-)}(\alpha;\epsilon)$:
\begin{eqnarray}
\dot{P}_1 & = & P_1A_3-A_1P_1\;,\label{MRTL-+ Lax triads P1}\\ \nonumber\\
\dot{W}_1 & = & W_1A_2-A_1W_1\;,\label{MRTL-+ Lax triads W1}\\ \nonumber\\
\dot{Q}_1 & = & Q_1A_2-A_3Q_1\;,\label{MRTL-+ Lax triads Q1}
\end{eqnarray}
where
\[
A_j=\pi_+\Big((\epsilon+\alpha)^{-1}(T_j-I)\Big)\;,\quad j=1,2,3\;,
\]
with
\[
T_1=P_1Q_1W_1^{-1}\;,\quad T_2=W_1^{-1}P_1Q_1\;,\quad T_3=Q_1W_1^{-1}P_1\;.
\]
So, we have the following expressions:
\begin{eqnarray*}
A_1 & = & \sum_{k=1}^N\Big(b_k^{(1)}+\alpha a_{k-1}^{(1)}\Big)E_{kk}+
\lambda\sum_{k=1}^N E_{k+1,k}\\ 
& = & \sum_{k=1}^N\Big(p_k+(\epsilon+\alpha)q_{k-1}\Big)E_{kk}+
\lambda\sum_{k=1}^N E_{k+1,k}\;,\\ 
A_2 & = & \sum_{k=1}^N\Big(b_k^{(1)}+\alpha a_k^{(1)}\Big)E_{kk}+
\lambda\sum_{k=1}^N E_{k+1,k}\\ 
& = & \sum_{k=1}^N\left(p_k+\alpha q_k+\displaystyle\frac
{\epsilon q_{k-1}}{1+\alpha p_{k-1}}+\displaystyle\frac
{\epsilon\alpha q_kp_k}{1+\alpha p_k}\right)E_{kk}+
\lambda\sum_{k=1}^N E_{k+1,k}\;,\\ 
A_3 & = & \sum_{k=1}^N\Big(b_k^{(2)}+\alpha a_k^{(2)}\Big)E_{kk}+
\lambda\sum_{k=1}^N E_{k+1,k}\\
& = & \sum_{k=1}^N\Big(p_k+(\epsilon+\alpha)q_k\Big)E_{kk}+
\lambda\sum_{k=1}^N E_{k+1,k}\;.
\end{eqnarray*}

\vspace{2mm}

The second Lax representation of ${\rm MRTL}_+^{(-)}(\alpha;\epsilon)$: 
\begin{eqnarray}
\dot{P}_2 & = & P_2B_3-B_1P_2\;,\label{MRTL-+ Lax triads P2}\\ \nonumber\\
\dot{Q}_2 & = & Q_2B_1-B_2Q_2\;,\label{MRTL-+ Lax triads Q2}\\ \nonumber\\
\dot{W}_2 & = & W_2B_3-B_2W_2\;,\label{MRTL-+ Lax triads W2}
\end{eqnarray}
where
\[
B_j=\pi_+\Big((\epsilon+\alpha)^{-1}(T_j-I)\Big)\;,\quad j=1,2,3\;,
\]
with
\[
T_1=P_2W_2^{-1}Q_2\;,\quad T_2=Q_2P_2W_2^{-1}\;,\quad T_3=W_2^{-1}Q_2P_2\;.
\]
So, we have now the following expressions:
\begin{eqnarray*}
B_1 & = & \sum_{k=1}^N\Big(b_k^{(1)}+\alpha a_{k-1}^{(1)}\Big)E_{kk}+
\lambda\sum_{k=1}^N E_{k+1,k}\\ 
& = & \sum_{k=1}^N\Big(p_k+(\epsilon+\alpha)q_{k-1}\Big)E_{kk}+
\lambda\sum_{k=1}^N E_{k+1,k}\;,\\ 
B_2 & = & \sum_{k=1}^N\Big(b_k^{(2)}+\alpha a_{k-1}^{(2)}\Big)E_{kk}+
\lambda\sum_{k=1}^N E_{k+1,k}  \\ 
& = & \sum_{k=1}^N\left(p_k+\alpha q_{k-1}+\displaystyle\frac
{\epsilon q_k}{1+\alpha p_{k+1}}+\displaystyle\frac
{\epsilon\alpha q_{k-1}p_k}{1+\alpha p_k}\right)E_{kk}+
\lambda\sum_{k=1}^N E_{k+1,k}\;,\\ 
B_3 & = & \sum_{k=1}^N\Big(b_k^{(2)}+\alpha a_k^{(2)}\Big)E_{kk}+
\lambda\sum_{k=1}^N E_{k+1,k}\\
& = & \sum_{k=1}^N\Big(p_k+(\epsilon+\alpha)q_k\Big)E_{kk}+
\lambda\sum_{k=1}^N E_{k+1,k}\;.
\end{eqnarray*}
The relations (\ref{MRTL- Lax rel}) assure the identities
\[
A_1=B_1\;,\qquad A_3=B_3\;.
\]

Turning to the flow ${\rm MRTL}_-^{(+)}(\alpha;\epsilon)$, we have the first 
Lax representation:
\begin{eqnarray}
\dot{P}_1 & = & C_1P_1-P_1C_3\;,\label{MRTL-- Lax triads P1}\\ \nonumber\\
\dot{W}_1 & = & C_1W_1-W_1C_2\;,\label{MRTL-- Lax triads W1}\\ \nonumber\\
\dot{Q}_1 & = & C_3Q_1-Q_1C_2\;,\label{MRTL-- Lax triads Q1}
\end{eqnarray}
with 
\begin{eqnarray*}
C_1 & = & 
\lambda^{-1}\sum_{k=1}^N \frac{a_k^{(1)}}
{1+\alpha b_{k+1}^{(1)}}\,E_{k,k+1}\\
& = & \lambda^{-1}\sum_{k=1}^N 
\frac{q_k\Big(1+(\epsilon+\alpha)p_k\Big)}
{(1+\alpha p_{k+1})(1+\alpha p_k)+\epsilon\alpha q_k}\,E_{k,k+1}\;, \\
C_2 & = & \lambda^{-1}\sum_{k=1}^N 
\frac{a_k^{(1)}}{1+\alpha b_k^{(1)}}\,E_{k,k+1}\\  
& = & \lambda^{-1}\sum_{k=1}^N 
\frac{q_k\Big(1+(\epsilon+\alpha)p_k\Big)\Big(1+\alpha p_{k-1}\Big)}
{\Big((1+\alpha p_k)(1+\alpha p_{k-1})+\epsilon\alpha q_{k-1}\Big)
\Big(1+\alpha p_k\Big)}\,E_{k,k+1}\,,\\
C_3 & = & \lambda^{-1}\sum_{k=1}^N 
\frac{a_k^{(2)}}{1+\alpha b_k^{(2)}}\,E_{k,k+1}\\
& = &  \lambda^{-1}\sum_{k=1}^N 
\frac{q_k\Big(1+(\epsilon+\alpha)p_{k+1}\Big)}
{(1+\alpha p_{k+1})(1+\alpha p_k)+\epsilon\alpha q_k}\,E_{k,k+1}\;.
\end{eqnarray*}
The second Lax representation of ${\rm MRTL}_-^{(+)}(\alpha;\epsilon)$:
\begin{eqnarray}
\dot{P}_2 & = & D_1P_2-P_2D_3\;,\label{MRTL-- Lax triads Q2}\\ \nonumber\\
\dot{Q}_2 & = & D_2Q_2-Q_2D_1\;,\label{MRTL-- Lax triads W2}\\ \nonumber\\
\dot{W}_2 & = & D_2W_2-W_2D_3\;,\label{MRTL-- Lax triads P2}
\end{eqnarray}
with 
\begin{eqnarray*}
D_1 & = & \lambda^{-1}\sum_{k=1}^N 
\frac{a_k^{(1)}}{1+\alpha b_{k+1}^{(1)}}\,E_{k,k+1}\\
& = & \lambda^{-1}\sum_{k=1}^N 
\frac{q_k\Big(1+(\epsilon+\alpha)p_k\Big)}
{(1+\alpha p_{k+1})(1+\alpha p_k)+\epsilon\alpha q_k}\,E_{k,k+1}\;, \\
D_2 & = & \lambda^{-1}\sum_{k=1}^N 
\frac{a_k^{(2)}}{1+\alpha b_{k+1}^{(2)}}\,E_{k,k+1}\\
& = &  \lambda^{-1}\sum_{k=1}^N 
\frac{q_k\Big(1+(\epsilon+\alpha)p_{k+1}\Big)\Big(1+\alpha p_{k+2}\Big)}
{\Big((1+\alpha p_{k+2})(1+\alpha p_{k+1})+\epsilon\alpha q_{k+1}\Big)
\Big(1+\alpha p_{k+1}\Big)}\,E_{k,k+1}\,,\\
D_3 & = & \lambda^{-1}\sum_{k=1}^N 
\frac{a_k^{(2)}}{1+\alpha b_k^{(2)}}\,E_{k,k+1}\\
& = &  \lambda^{-1}\sum_{k=1}^N 
\frac{q_k\Big(1+(\epsilon+\alpha)p_{k+1}\Big)}
{(1+\alpha p_{k+1})(1+\alpha p_k)+\epsilon\alpha q_k}\,E_{k,k+1}\;.
\end{eqnarray*}

\subsection{Discretization}
We discuss here only the discretization of the flow 
${\rm MRTL}_+^{(+)}(\alpha;\epsilon)$.

The first Lax representation of ${\rm dMRTL}_+^{(-)}(\alpha;\epsilon)$:
\begin{equation}\label{dMRTL-+ 1 Lax}
\wiP_1=\mbA_1^{-1}P_1\mbA_3\;,\qquad \wW_1=\mbA_1^{-1}W_1\mbA_2\;,\qquad
\wQ_1=\mbA_3^{-1}Q_1\mbA_2
\end{equation}
with
\begin{eqnarray*}
\mbA_1=\Pi_+\Big(I+\frac{h}{\epsilon+\alpha}\,(P_1Q_1W_1^{-1}-I)\Big) & = & 
\sum_{k=1}^N \ga_kE_{kk}+h\lambda\sum_{k=1}^NE_{k+1,k}\;,\\
\mbA_2=\Pi_+\Big(I+\frac{h}{\epsilon+\alpha}\,(W_1^{-1}P_1Q_1-I)\Big) & = & 
\sum_{k=1}^N \gb_kE_{kk}+h\lambda\sum_{k=1}^NE_{k+1,k}\;,\\
\mbA_3=\Pi_+\Big(I+\frac{h}{\epsilon+\alpha}\,(Q_1W_1^{-1}P_1-I)\Big) & = & 
\sum_{k=1}^N \goe_kE_{kk}+h\lambda\sum_{k=1}^N E_{k+1,k}\;.
\end{eqnarray*}
The second Lax representation of ${\rm dMRTL}_+^{(+)}(\alpha;\epsilon)$:
\begin{equation}\label{dMRTL-+ 2 Lax}
\wiP_2=\mbB_1^{-1}P_2\mbB_3\;,\qquad \wQ_2=\mbB_2^{-1}Q_2\mbB_1\;,\qquad
\wW_2=\mbB_2^{-1}W_2\mbB_3
\end{equation}
with
\begin{eqnarray*}
\mbB_1=\Pi_+\Big(I+\frac{h}{\epsilon+\alpha}\,(P_2W_2^{-1}Q_2-I)\Big)
 & = & \sum_{k=1}^N \gb_kE_{kk}+h\lambda\sum_{k=1}^N E_{k+1,k}\;,\\
\mbB_2=\Pi_+\Big(I+\frac{h}{\epsilon+\alpha}\,(Q_2P_2W_2^{-1}-I)\Big)
 & = & \sum_{k=1}^N \goe_kE_{kk}+h\lambda\sum_{k=1}^NE_{k+1,k}\;,\\
\mbB_3=\Pi_+\Big(I+\frac{h}{\epsilon+\alpha}\,(W_2^{-1}Q_2P_2-I)\Big)
 & = & \sum_{k=1}^N \gf_kE_{kk}+h\lambda\sum_{k=1}^NE_{k+1,k}\;.
\end{eqnarray*}

\subsection{Application: localizing change of variables for 
${\rm dRTL}_-(\alpha)$}

The Miura transformation $\rM_1^{(-)}(\alpha;h)$ plays the role of the 
localizing change of variables for ${\rm dRTL}_-(\alpha)$. The reason
for this lies in the following identity:
\[
\Pi_-\bigg(I+\frac{h}{\alpha}\Big(I-L^{-1}W\Big)\bigg)=
\Pi_-\bigg(\Big(1+\frac{h}{\alpha}\Big)L-\frac{h}{\alpha}\,W\bigg).
\]
This has to be compared with (\ref{MRTL- fact 1}). This comparison implies
that, if we define the change of variables $\cR\cT(\ba,\bb)\mapsto\cR\cT(a,b)$
by the formulas
\begin{equation}\label{dRTL- param loc map}
\rM_1^{(-)}(\alpha;h)\,:\quad \left\{\begin{array}{l}
b_k=\bb_k+\displaystyle\frac{h\ba_{k-1}}{1+\alpha\bb_{k-1}}\;,\\ \\
a_k=\ba_k\left(1+\displaystyle\frac{h\bb_k}{1+\alpha\bb_k}\right)\;,
\end{array}\right.
\end{equation}
then
\[
\Pi_-\bigg(I+\frac{h}{\alpha}\Big(I-L^{-1}W\Big)\bigg)=Q_1(\ba,\bb,\lambda)=
I+h\lambda^{-1}\sum_{k=1}^N\frac{\ba_k}{1+\alpha\bb_k}\,E_{k,k+1}\;.
\]
So, we get for the entries of the factors (\ref{dRTL- param C1}), 
(\ref{dRTL- param C2}) the following local expressions:
\[
\gc_k=\frac{\ba_k}{1+\alpha\bb_{k+1}}\;,\qquad 
\gd_k=\frac{\ba_k}{1+\alpha\bb_k}\;.
\]
\begin{theorem}\label{dRTL- param in localization variables}
The change of variables {\rm(\ref{dRTL- param loc map})} conjugates 
${\rm dRTL}_-(\alpha)$  with the map on $\cR\cT(\ba,\bb)$ described by 
the following local equations of motion:
\begin{equation}\label{dRTL- param loc}
\begin{array}{l}
\widetilde{\bb}_k+
\displaystyle\frac{h\widetilde{\ba}_{k-1}}
{\raisebox{-1.5mm}{$\,1+\alpha\widetilde{\bb}_{k-1}$}}=
\bb_k+\displaystyle\frac{h\ba_k}{1+\alpha\bb_{k+1}}\;,\\ \\
\widetilde{\ba}_k\left(1+\displaystyle\frac{h\widetilde{\bb}_k}
{\raisebox{-1.5mm}{$\,1+\alpha\widetilde{\bb}_k$}}\right)
=\ba_k\left(1+\displaystyle\frac{h\bb_{k+1}}{1+\alpha\bb_{k+1}}\right)\;.
\end{array}
\end{equation}
\end{theorem}
So, the local form of ${\rm dRTL}_-(\alpha)$ (\ref{dRTL- param loc}) belongs
to the hierarchy ${\rm MRTL}^{(-)}(\alpha;h)$.

{\bf Corollary.} {\it The local form of ${\rm dRTL}_-(\alpha)$ 
{\rm(\ref{dRTL- param loc})} is a Poisson map with respect to
the following Poisson bracket on $\cR\cT(\ba,\bb)$:
\begin{eqnarray}
\{\bb_k,\ba_k\} & = & -\ba_k\Big(1+(h+\alpha)\bb_k\Big)\;, \nonumber\\
\{\ba_k,\bb_{k+1}\} & = & -\ba_k\Big(1+(h+\alpha)\bb_{k+1}\Big)\;,
\label{dRTL- param loc PB1}\\
\{\ba_k,\ba_{k+1}\} & = & -\alpha\ba_k\ba_{k+1}\left(1+\displaystyle\frac
{h\bb_{k+1}}{1+\alpha\bb_{k+1}}\right)\,,
\nonumber
\end{eqnarray}
which is the pull--back of the bracket 
\begin{equation}\label{dRTL- param loc PB1 target}
\{\cdot,\cdot\}_{1\alpha}+(h+\alpha)\{\cdot,\cdot\}_{2\alpha}
\end{equation}
on $\cR\cT(a,b)$ under the change of variables {\rm(\ref{dRTL- param loc map})},
and also with respect to the following Poisson bracket on 
$\cR\cT(\ba,\bb)$ compatible with {\rm(\ref{dRTL- param loc PB1})}:
\begin{eqnarray}
\{\bb_k,\ba_k\}     & = & -\ba_k\Big(\bb_k+(h+\alpha)\ba_k\Big)
\Big(1+(h+\alpha)\bb_k\Big)\;,\nonumber\\
\{\ba_k,\bb_{k+1}\} & = & -\ba_k\Big(\bb_{k+1}+(h+\alpha)\ba_k\Big)
\Big(1+(h+\alpha)\bb_{k+1}\Big)\;,\nonumber\\
\{\bb_k,\bb_{k+1}\} & = & -\ba_k\Big(1+(h+\alpha)\bb_k\Big)
(1+(h+\alpha)\bb_{k+1}\Big)\;,\nonumber\\
\{\ba_k,\ba_{k+1}\} & = & -\ba_k\ba_{k+1}\Big(1+2\alpha\bb_{k+1}
+\alpha(h+\alpha)(\ba_k+\ba_{k+1})\Big)
\left(1+\displaystyle\frac{h\bb_{k+1}}{1+\alpha\bb_{k+1}}\right),
\nonumber\\
\{\bb_k,\ba_{k+1}\} & = & -\alpha\ba_k\ba_{k+1}
\Big(1+(h+\alpha)\bb_k\Big)
\left(1+\displaystyle\frac{h\bb_{k+1}}{1+\alpha\bb_{k+1}}\right)\;,
\label{dRTL- param loc PB2}\\
\{\ba_k,\bb_{k+2}\} & = & -\alpha\ba_k\ba_{k+1}\Big(1+(h+\alpha)\bb_{k+2}\Big)
\left(1+\displaystyle\frac{h\bb_{k+1}}{1+\alpha\bb_{k+1}}\right)\;,\nonumber\\
\{\ba_k,\ba_{k+2}\} & = & -\alpha^2\ba_k\ba_{k+1}\ba_{k+2}
\left(1+\displaystyle\frac{h\bb_{k+1}}{1+\alpha\bb_{k+1}}\right)
\left(1+\displaystyle\frac{h\bb_{k+2}}{1+\alpha\bb_{k+2}}\right)\;,\nonumber
\end{eqnarray}
which is the pull--back of the bracket 
\begin{equation}\label{dRTL- param loc PB2 target}
\{\cdot,\cdot\}_{2\alpha}+(h+\alpha)\{\cdot,\cdot\}_{3\alpha}
\end{equation}
on $\cR\cT(a,b)$ under the change of variables 
{\rm(\ref{dRTL- param loc map})}.}

\setcounter{equation}{0}
\section{Double modified relativistic Toda lattice}

\subsection{Equations of motion}
We shall not elaborate all branches of the tree of successive 
modifications of the relativistic Toda lattice, and restrict ourselves
with one possible modification of one of the modified relativistic Todas,
namely of ${\rm MRTL}^{(+)}(\alpha;\epsilon)$. The phase space of the
corresponding hierarchy ${\rm M^2RTL}^{(+)}(\alpha;\epsilon,\delta)$
will be denoted by $\cM^2\cR\cT={\Bbb R}^{2N}(r,s)$. The equations of motion
of the simplest representative of this hierarchy read:
\begin{eqnarray}
\dot{r}_k & = & 
\frac{(1+\alpha r_k)(1+\epsilon r_k)(1+\delta r_k)(s_k-s_{k-1})}
{1-\epsilon\delta\alpha D(s_k,r_k,s_{k-1})}\;,  \label{M2RTL r}\\ 
\nonumber\\
\dot{s}_k & = & 
s_k(1+\epsilon\alpha s_k)(1+\delta\alpha s_k)(1+\epsilon\delta s_k)
\bigg(\,\frac{r_{k+1}+\alpha s_{k+1}+\alpha(\epsilon+\delta)r_{k+1}s_{k+1}}
{1-\epsilon\delta\alpha D(s_{k+1},r_{k+1},s_k)}- \nonumber\\
 & & \qquad-\frac{r_k+\alpha s_{k-1}+\alpha(\epsilon+\delta)r_ks_{k-1}}
{1-\epsilon\delta\alpha D(s_k,r_k,s_{k-1})}\,\bigg)\,,  \label{M2RTL s}
\end{eqnarray}
where 
\begin{equation}\label{M2RTL D}
D(s_k,r_k,s_{k-1})=r_ks_{k-1}+s_kr_k+\alpha s_ks_{k-1}+\alpha(\epsilon+\delta)
s_kr_ks_{k-1}\;.
\end{equation}

\subsection{Hamiltonian structure}
\begin{proposition}
The relations
\begin{eqnarray}
\{r_k,s_k\}_{123\alpha} & = & -s_k(1+\epsilon\delta s_k)
(1+\epsilon r_k)(1+\delta r_k)\;,
\nonumber\\
\{s_k,r_{k+1}\}_{123\alpha} & = & -s_k(1+\epsilon\delta s_k)
(1+\epsilon r_{k+1})(1+\delta r_{k+1})\;,  \label{M2RTL PB}\\
\{r_k,r_{k+1}\}_{123\alpha} & = & \alpha\,\frac{s_k(1+\epsilon\delta s_k)}
{(1+\epsilon\alpha s_k)(1+\delta\alpha s_k)}\,
(1+\epsilon r_k)(1+\delta r_k)(1+\epsilon r_{k+1})(1+\delta r_{k+1})\;.
\nonumber
\end{eqnarray}
define a Poisson bracket on $\cM^2\cR\cT$. The flow 
${\rm M^2RTL}_+^{(+)}(\alpha;\epsilon,\delta)$ is
Hamiltonian with respect to this bracket with the Hamilton function
\begin{eqnarray}
\rF_{0\alpha}(r,s) & = & (\epsilon\delta)^{-1}\sum_{k=1}^N\log(1+\delta r_k)
+(\epsilon\delta)^{-1}\sum_{k=1}^N\log(1+\epsilon\delta s_k)
+(\epsilon\delta)^{-1}\sum_{k=1}^N\log(1+\delta\alpha s_k)\nonumber\\
& & -(\epsilon\delta)^{-1}\sum_{k=1}^N\log\Big(
1-\epsilon\delta\alpha D(s_k,r_k,s_{k-1})\Big)\;.\label{M2RTL F0}
\end{eqnarray}
\end{proposition}

\subsection{Miura relations}
\begin{theorem}
Define the Miura maps $\rM_{1,2}^{(+)}(\alpha;\epsilon;\delta):\cM^2\cR\cT
\mapsto\cM\cR\cT$ by the formulas:
\begin{equation}\label{M2RTL Miura 1}
\rM_1^{(+)}(\alpha;\epsilon;\delta):\quad
\left\{\begin{array}{l}
p_k=\displaystyle\frac{r_k+\delta s_{k-1}+\delta(\epsilon+\alpha)r_ks_{k-1}}
{1-\epsilon\delta\alpha r_ks_{k-1}}\;,\\ \\
q_k=\displaystyle\frac{s_k(1+\delta r_k)(1+\delta\alpha s_{k-1})}
{1-\epsilon\delta\alpha D(s_k,r_k,s_{k-1})}\;,
\end{array}\right.
\end{equation}
\begin{equation}\label{M2RTL Miura 2}
\rM_2^{(+)}(\alpha;\epsilon;\delta):\quad
\left\{\begin{array}{l}
p_k=\displaystyle\frac{r_k+\delta s_k+\delta(\epsilon+\alpha)r_ks_k}
{1-\epsilon\delta\alpha r_ks_k}\;,\\ \\
q_k=\displaystyle\frac{s_k(1+\delta r_{k+1})(1+\delta\alpha s_{k+1})}
{1-\epsilon\delta\alpha D(s_{k+1},r_{k+1},s_k)}\;.
\end{array}\right.
\end{equation}
Both maps $\rM_{1,2}^{(+)}(\alpha;\epsilon;\delta)$ are Poisson, if 
$\cM^2\cR\cT$ is equipped with the bracket $\{\cdot,\cdot\}_{123\alpha}$,
and $\cM\cR\cT$ is equipped with $\{\cdot,\cdot\}_{12\alpha}+\delta
\{\cdot,\cdot\}_{23\alpha}$. The pull--back of the flow 
${\rm MRTL}_+^{(+)}(\alpha;\epsilon)$ with respect to either of the Miura maps
$\rM_{1,2}^{(+)}(\alpha;\epsilon;\delta)$ coincides with 
${\rm M^2RTL}_+^{(+)}(\alpha;\epsilon,\delta)$.
\end{theorem}
The following statement holds concerning the permutability of the Miura maps.
\begin{theorem}\label{M2RTL Miuras permutability}
The following diagram is commutative for all $i=1,2$, $j=1,2$:
\begin{center}
\unitlength1cm
\begin{picture}(9,6.5)
\put(3.5,1.1){\vector(1,0){2}}
\put(3.5,5.1){\vector(1,0){2}}
\put(2,4.1){\vector(0,-1){2}}
\put(7,4.1){\vector(0,-1){2}}
\put(1,0.6){\makebox(2,1){$\cM\cR\cT$}} 
\put(1,4.6){\makebox(2,1){$\cM^2\cR\cT$}}
\put(6,4.6){\makebox(2,1){$\cM\cR\cT$}}
\put(6,0.6){\makebox(2,1){$\cR\cT$}}
\put(-0.4,2.6){\makebox(2,1){$\rM_i^{(+)}(\alpha;\epsilon;\delta)$}}
\put(7.2,2.6){\makebox(2,1){$\rM_i^{(+)}(\alpha;\delta)$}}
\put(3.8,-0.2){\makebox(1.4,1.4){$\rM_j^{(+)}(\alpha;\epsilon)$}}
\put(3.8,5.0){\makebox(1.4,1.4){$\rM_j^{(+)}(\alpha;\delta;\epsilon)$}}
\end{picture}
\end{center}
\end{theorem}

\subsection{Application: localizing change of variables for 
${\rm dMRTL}_+^{(+)}(\alpha;\epsilon)$}
The map $\rM_1^{(+)}(\alpha;\epsilon;h)$ turns out to play the role
of the localizing change of variables for 
${\rm dMRTL}_+^{(+)}(\alpha;\epsilon)$. Indeed, define the change of variables
$\cM\cR\cT(\bq,\bp)\mapsto\cM\cR\cT(q,p)$ as
\begin{equation}\label{dMRTL loc map}
\rM_1^{(+)}(\alpha;\epsilon;h):\quad
\left\{\begin{array}{l}
p_k=\displaystyle\frac{\bp_k+h\bq_{k-1}+h(\epsilon+\alpha)\bp_k\bq_{k-1}}
{1-h\epsilon\alpha\bp_k\bq_{k-1}}\;,\\ \\
q_k=\displaystyle\frac{\bq_k(1+h\bp_k)(1+h\alpha\bq_{k-1})}
{1-h\epsilon\alpha D(\bq_k,\bp_k,\bq_{k-1})}\;,
\end{array}\right.
\end{equation}
where, as before,
\begin{equation}\label{dMRTL loc D}
D(\bq_k,\bp_k,\bq_{k-1})=\bq_k\bp_k+\bp_k\bq_{k-1}+\alpha\bq_k\bq_{k-1}+
(h+\epsilon)\alpha\bq_k\bp_k\bq_{k-1}\;.
\end{equation}

Let us notice that the expression for $p_k$ is symmetric in $\alpha$, 
$\epsilon$, and may be represented in two equivalent alternative forms:
\begin{equation}\label{dMRTL loc map r alt1}
1+\epsilon p_k=\frac{(1+\epsilon\bp_k)(1+h\epsilon\bq_{k-1})}
{1-h\epsilon\alpha\bp_k\bq_{k-1}}\quad\Leftrightarrow\quad
1+\alpha p_k=\frac{(1+\alpha\bp_k)(1+h\alpha\bq_{k-1})}
{1-h\epsilon\alpha\bp_k\bq_{k-1}}\;,
\end{equation}
or else as
\begin{equation}\label{dMRTL loc map r alt2}
\frac{1+\epsilon p_k}{1+\alpha p_k}=
\frac{(1+\epsilon\bp_k)(1+h\epsilon\bq_{k-1})}
{(1+\alpha\bp_k)(1+h\alpha\bq_{k-1})}\;.
\end{equation}
The expression for $q_k$ in (\ref{dMRTL loc map}) does not enjoy the
$\alpha\leftrightarrow\epsilon$ symmetry anymore; it may be equivalently 
rewritten as 
\begin{equation}\label{dMRTL loc map q alt1}
1+\epsilon\alpha q_k=\frac{(1+\epsilon\alpha\bq_k)
(1-h\epsilon\alpha\bp_k\bq_{k-1})}{1-h\epsilon\alpha 
D(\bq_k,\bp_k,\bq_{k-1})}\;,
\end{equation}
or else as
\begin{equation}\label{dMRTL loc map q alt2}
\frac{q_k}{1+\epsilon\alpha q_k}=\frac{\bq_k(1+h\bp_k)(1+h\alpha\bq_{k-1})}
{(1+\epsilon\alpha\bq_k)(1-h\epsilon\alpha\bp_k\bq_{k-1})}\;.
\end{equation}
It can be proved that under this change of variables the following local
expressions for the factors (\ref{dMRTL+ A1})--(\ref{dMRTL+ A3}) and
(\ref{dMRTL+ B1})--(\ref{dMRTL+ B3}) hold:
\begin{equation}\label{dMRTL a in loc map}
\ga_k=\frac{(1+h\bp_k)(1+h\alpha\bq_{k-1})(1+h\epsilon\bq_{k-1})
(1-h\epsilon\alpha\bp_{k-1}\bq_{k-2})}{(1-h\epsilon\alpha\bp_k\bq_{k-1})
(1-h\epsilon\alpha D(\bq_{k-1},\bp_{k-1},\bq_{k-2}))}\;,
\end{equation}
\begin{equation}\label{dMRTL b in loc map}
\gb_k=\frac{(1+h\bp_k)(1+h\alpha\bq_k)(1+h\epsilon\bq_{k-1})}
{1-h\epsilon\alpha D(\bq_k,\bp_k,\bq_{k-1})}\;,
\end{equation}
\begin{equation}\label{dMRTL e in loc map}
\goe_k=\frac{(1+h\bp_k)(1+h\alpha\bq_{k-1})(1+h\epsilon\bq_k)}
{1-h\epsilon\alpha D(\bq_k,\bp_k,\bq_{k-1})}\;.
\end{equation}
\begin{equation}\label{dMRTL f in loc map}
\gf_k=\frac{(1+h\bp_k)(1+h\alpha\bq_k)(1+h\epsilon\bq_k)
(1-h\epsilon\alpha\bp_{k+1}\bq_{k+1})}{(1-h\epsilon\alpha\bp_k\bq_k)
(1-h\epsilon\alpha D(\bq_{k+1},\bp_{k+1},\bq_k))}\;.
\end{equation}

\begin{theorem}
The change of variables {\rm (\ref{dMRTL loc map})} conjugates
${\rm dMRTL}_+^{(+)}(\alpha;\epsilon)$ with the map on $\cM\cR\cT(\bq,\bp)$ 
described by the following local equations of motion:
\begin{eqnarray}
\frac{(1+\epsilon\widetilde{\bp}_k)(1+h\epsilon\widetilde{\bq}_{k-1})}
{1-h\epsilon\alpha\widetilde{\bp}_k\widetilde{\bq}_{k-1}} & = &
\frac{(1+\epsilon\bp_k)(1+h\epsilon\bq_k)}{1-h\epsilon\alpha\bq_k\bp_k}\;,
\label{dMRTL loc r}\\ \nonumber\\
\frac{\widetilde{\bq}_k(1+h\widetilde{\bp}_k)(1+h\alpha\widetilde{\bq}_{k-1})}
{1-h\epsilon\alpha D(\widetilde{\bq}_k,\widetilde{\bp}_k,\widetilde{\bq}_{k-1})}
 & = & 
\frac{\bq_k(1+h\bp_{k+1})(1+h\alpha\bq_{k+1})}{1-h\epsilon\alpha 
D(\bq_{k+1},\bp_{k+1},\bq_k)}\;,
\label{dMRTL loc q}
\end{eqnarray}
The equivalent form of the above equations of motion:
\begin{eqnarray}
\frac{1+\epsilon\widetilde{\bp}_k}{1+\alpha\widetilde{\bp}_k}\cdot\frac
{1+h\epsilon\widetilde{\bq}_{k-1}}{1+h\alpha\widetilde{\bq}_{k-1}} & = &
\frac{1+\epsilon\bp_k}{1+\alpha\bp_k}\cdot\frac
{1+h\epsilon\bq_k}{1+h\alpha\bq_k}\;,
\label{dMRTL loc r alt}\\  \nonumber\\
\frac{\widetilde{\bq}_k}{1+\epsilon\alpha\widetilde{\bq}_k}\cdot
\frac{(1+h\widetilde{\bp}_k)(1+h\alpha\widetilde{\bq}_{k-1})}
{1-h\epsilon\alpha\widetilde{\bp}_k\widetilde{\bq}_{k-1}} & = &
\frac{\bq_k}{1+\epsilon\alpha\bq_k}\cdot\frac
{(1+h\bp_{k+1})(1+h\alpha\bq_{k+1})}{1-h\epsilon\alpha\bq_{k+1}\bp_{k+1}}\;.
\nonumber\\  \label{dMRTL loc q alt}
\end{eqnarray}
\end{theorem}

We see that the local form of ${\rm dMRTL}_+^{(+)}(\alpha;\epsilon)$ given 
in the previous theorem belongs actually to the hierarchy 
${\rm M^2RTL}^{(+)}(\alpha;\epsilon,h)$.

{\bf Corollary.} {\it The local form of ${\rm dMRTL}_+^{(+)}(\alpha;\epsilon)$
is a Poisson map with respect to the following Poisson bracket on 
$\cM\cR\cT(\bq,\br)$:
\begin{eqnarray}
\{\br_k,\bq_k\} & = & -\bq_k(1+h\epsilon\bq_k)(1+h\br_k)(1+\epsilon\br_k)\;,
\nonumber\\
\{\bq_k,\br_{k+1}\} & = & -\bq_k(1+h\epsilon\bq_k)(1+h\br_{k+1})
(1+\epsilon\br_{k+1})\;,  \label{dMRTL loc PB loc}\\
\{\br_k,\br_{k+1}\} & = & \alpha\,\frac{\bq_k(1+h\epsilon\bq_k)}
{(1+\epsilon\alpha\bq_k)(1+h\alpha\bq_k)}\,
(1+h\br_k)(1+\epsilon\br_k)(1+h\br_{k+1})(1+\epsilon\br_{k+1})\;,\nonumber
\end{eqnarray}
which is the pull--back of the bracket
\begin{equation}\label{M2RTL br}
\{\cdot,\cdot\}_{12\alpha}+h\{\cdot,\cdot\}_{23\alpha}
\end{equation}
on $\cM\cR\cT(q,r)$ under the change of variables {\rm (\ref{dMRTL loc map})}.}

Finally, let us consider the relation between the local forms of the maps
${\rm dRTL}_+(\alpha)$ and ${\rm dMRTL}_+(\alpha;\epsilon)$. In other words, 
how do the Miura maps $\rM_{1,2}^{(+)}(\alpha;\epsilon)$ look when seen through 
the localizing changes of variables? The answer is given by Theorem
\ref{M2RTL Miuras permutability}, which can be reformulated as follows:
\begin{theorem}
The following diagram is commutative for $j=1,2$:
\begin{center}
\unitlength1cm
\begin{picture}(9,6.5)
\put(3.5,1.1){\vector(1,0){2}}
\put(3.5,5.1){\vector(1,0){2}}
\put(2,4.1){\vector(0,-1){2}}
\put(7,4.1){\vector(0,-1){2}}
\put(1,0.6){\makebox(2,1){$\cM\cR\cT(q,p)$}} 
\put(1,4.6){\makebox(2,1){$\cM\cR\cT(\bq,\bp)$}}
\put(6,4.6){\makebox(2,1){$\cR\cT(\ba,\bb)$}}
\put(6,0.6){\makebox(2,1){$\cR\cT(a,b)$}}
\put(-0.4,2.6){\makebox(2,1){$\rM_1^{(+)}(\alpha;\epsilon;h)$}}
\put(7.2,2.6){\makebox(2,1){$\rM_1^{(+)}(\alpha;h)$}}
\put(3.8,-0.2){\makebox(1.4,1.4){$\rM_j^{(+)}(\alpha;\epsilon)$}}
\put(3.8,5.0){\makebox(1.4,1.4){$\rM_j^{(+)}(\alpha;h;\epsilon)$}}
\end{picture}
\end{center}
where the maps $\rM_{1,2}^{(+)}(\alpha;h;\epsilon)$ are given by
\begin{equation}\label{MRTL Miura 1 in loc map}
\rM_1^{(+)}(\alpha;h;\epsilon):\;\left\{\begin{array}{l} 
\bb_k=\displaystyle\frac{\bp_k+\epsilon\bq_{k-1}+
\epsilon(h+\alpha)\bp_k\bq_{k-1}}{1-h\epsilon\alpha\bp_k\bq_{k-1}}\;,\\ \\ 
\ba_k=\displaystyle\frac{\bq_k(1+\epsilon\bp_k)(1+\epsilon\alpha\bq_{k-1})}
{1-h\epsilon\alpha D(\bq_k,\bp_k,\bq_{k-1})}\;, 
\end{array}\right.
\end{equation}
\begin{equation}\label{MRTL Miura 2 in loc map}
\rM_2^{(+)}(\alpha;h;\epsilon):\;\left\{\begin{array}{l} 
\bb_k=\displaystyle\frac{\bp_k+\epsilon\bq_k+
\epsilon(h+\alpha)\bq_k\bp_k}{1-h\epsilon\alpha\bq_k\bp_k}\;,\\ \\ 
\ba_k=\displaystyle\frac{\bq_k(1+\epsilon\bp_{k+1})(1+\epsilon\alpha\bq_{k+1})}
{1-h\epsilon\alpha D(\bq_{k+1},\bp_{k+1},\bq_k)}\;.
\end{array}\right.
\end{equation}
\end{theorem}

\setcounter{equation}{0}
\section{Relativistic Volterra lattice}
\label{Sect rel Volterra}
\subsection{Equations of motion}
The systems related to the flows ${\rm RTL}_{\pm}(\alpha)$ in the same way
as VL is related to TL, are naturally called {\it relativistic Volterra
lattices} ${\rm RVL}_{\pm}(\alpha)$. Their phase space will be denoted by
$\cR\cV={\Bbb R}^{2N}(u,v)$. At this point the next instance of the
``relativistic splitting'' appears: there are two systems playing the role
of  ${\rm RVL}_+(\alpha)$, namely
\begin{equation}\label{RVL1+}
\left\{\begin{array}{l}
\dot{u}_k = u_k(v_k-v_{k-1}+\alpha u_kv_k-\alpha u_{k-1}v_{k-1})\;,\\ \\
\dot{v}_k = v_k(u_{k+1}-u_k+\alpha u_{k+1}v_{k+1}-\alpha u_kv_k)\;,
\end{array}\right.
\end{equation}
and
\begin{equation}\label{RVL2+}
\left\{\begin{array}{l}
\dot{u}_k = u_k(v_k-v_{k-1}+\alpha u_{k+1}v_k-\alpha u_kv_{k-1})\;,\\ \\
\dot{v}_k = v_k(u_{k+1}-u_k+\alpha u_{k+1}v_k-\alpha u_kv_{k-1})\;.
\end{array}\right.
\end{equation}
Similarly, there are two systems playing the role of ${\rm RVL}_-(\alpha)$, 
namely
\begin{equation}\label{RVL1-}
\left\{\begin{array}{l}
\dot{u}_k = u_k\left(\displaystyle\frac{v_k}{1+\alpha(u_{k+1}+v_k)}
-\displaystyle\frac{v_{k-1}}{1+\alpha(u_k+v_{k-1})}\right)\;,\\ \\
\dot{v}_k = v_k\left(\displaystyle\frac{u_{k+1}}{1+\alpha(u_{k+1}+v_k)}
-\displaystyle\frac{u_k}{1+\alpha (u_k+v_{k-1})}\right)\;,
\end{array}\right.
\end{equation}
and
\begin{equation}\label{RVL2-}
\left\{\begin{array}{l}
\dot{u}_k = u_k\left(\displaystyle\frac{v_k}{1+\alpha(u_k+v_k)}
-\displaystyle\frac{v_{k-1}}{1+\alpha(u_{k-1}+v_{k-1})}\right)\;,\\ \\
\dot{v}_k = v_k\left(\displaystyle\frac{u_{k+1}}{1+\alpha(u_{k+1}+v_{k+1})}
-\displaystyle\frac{u_k}{1+\alpha (u_k+v_k)}\right)\;.
\end{array}\right.
\end{equation}
Obviously, (\ref{RVL1+}) and (\ref{RVL2+}) are related by means of a simple
shift
\begin{equation}\label{RVL shift}
\bar{u}_k=v_k\;,\quad \bar{v}_k=u_{k+1}\;,
\end{equation}
and the same holds for (\ref{RVL1-}) and (\ref{RVL2-}). Therefore we restrict
ourselves to considering only one system of each pair. We reserve the notation
${\rm RVL}_+(\alpha)$ for the flow (\ref{RVL1+}), and ${\rm RVL}_-(\alpha)$ 
for the flow (\ref{RVL1-}).

\subsection{Bi--Hamiltonian structure}
\begin{proposition}\label{RVL bi-Hamiltonian}
{\rm a)} The formulas 
\begin{equation}\label{RVL q br}
\{u_k,v_k\}_{2\alpha}=-u_kv_k\;,\qquad \{v_k,u_{k+1}\}_{2\alpha}=-v_ku_{k+1}\;,
\end{equation}
define a Poisson bracket on $\cR\cV$.  The systems ${\rm RVL}_{\pm}(\alpha)$ are 
Hamiltonian flows on $\Big(\cR\cV,\{\cdot,\cdot\}_{2\alpha}\Big)$, with 
the Hamilton functions 
\begin{equation}\label{RVL1+ H1}
\rH_{1\alpha}^{(+)}(u,v)=\sum_{k=1}^N(u_k+v_k+\alpha u_kv_k)
\end{equation}
and 
\begin{equation}\label{RVL1- H1}
\rH_{1\alpha}^{(-)}(u,v)=\alpha^{-1}\,\sum_{k=1}^N\log
\Big(1+\alpha(u_k+v_{k-1})\Big)\;,
\end{equation} 
respectively. The functions $\rH_{1\alpha}^{(+)}$ and $\rH_{1\alpha}^{(-)}$
are in involution in the bracket $\{\cdot,\cdot\}_{2\alpha}$.

{\rm b)} The relations 
\begin{eqnarray}\label{RVL c br}
\{u_k,v_k\}_{3\alpha} & = & -u_kv_k(u_k+v_k+\alpha u_kv_k)\;,
\nonumber\\ \nonumber\\
\{v_k,u_{k+1}\}_{3\alpha} & = & -v_ku_{k+1}(v_k+u_{k+1})\;,
\nonumber\\ \nonumber\\
\{u_k,u_{k+1}\}_{3\alpha} & = & -u_kv_ku_{k+1}(1+\alpha u_k)\;,\\
\nonumber\\
\{v_k,v_{k+1}\}_{3\alpha} & = & -v_ku_{k+1}v_{k+1}(1+\alpha v_{k+1})\;,
\nonumber\\ \nonumber\\
\{v_k,u_{k+2}\}_{3\alpha} & = & -\alpha v_ku_{k+1}v_{k+1}u_{k+2}\nonumber
\end{eqnarray}
define a Poisson bracket on $\cR\cV$ compatible with 
$\{\cdot,\cdot\}_{2\alpha}$.  The systems ${\rm RVL}_{\pm}(\alpha)$ are 
Hamiltonian flows on $\Big(\cR\cV,\{\cdot,\cdot\}_{3\alpha}\Big)$, with the 
Hamilton functions 
\begin{equation}\label{RVL1+ H0}
\rH_{0\alpha}^{(+)}(u,v)=\frac{1}{2}\sum_{k=1}^N\log(u_kv_k)
\end{equation}
and
\begin{equation}\label{RVL1- H0}
\rH_{0\alpha}^{(-)}(u,v)=\frac{1}{2}\sum_{k=1}^N\log(u_kv_k)-
\sum_{k=1}^N\log\Big(1+\alpha(u_k+v_{k-1})\Big)\;,
\end{equation}
respectively. The functions $\rH_{0\alpha}^{(+)}$ and $\rH_{0\alpha}^{(-)}$
are in involution in the bracket $\{\cdot,\cdot\}_{3\alpha}$.
\end{proposition}

\subsection{Miura relations}
The relativistic Volterra hierarchies are related to the relativistic
Toda hierarchy by means of {\it literally the same} Miura transformations as
the nonrelativistic one, i.e.
\begin{equation}\label{RVL Miuras}
M_1:\left\{\begin{array}{l}
b_k=u_k+v_{k-1}\;,\\ \\ a_k=u_kv_k\;,\end{array}\right. \qquad
M_2:\left\{\begin{array}{l}
b_k=u_k+v_k\;,\\ \\ a_k=u_{k+1}v_k\;.\end{array}\right.
\end{equation}
It turns out that the flows (\ref{RVL1+}), (\ref{RVL1-}) are connected with
$M_1$, while the flows (\ref{RVL2+}), (\ref{RVL2-}) are connected with $M_2$.
We restrict ourselves with the $M_1$ case.
\begin{theorem}
Both Miura maps $M_{1,2}:\cR\cV(u,v)\mapsto\cR\cT(a,b)$
given in {\rm(\ref{RVL Miuras})} are Poisson, provided both spaces
$\cR\cV$ and $\cR\cT$ carry the corresponding brackets 
$\{\cdot,\cdot\}_{2\alpha}$. The Miura map $M_1$ is also Poisson, provided 
both spaces $\cR\cV$ and $\cR\cT$ carry the corresponding brackets 
$\{\cdot,\cdot\}_{3\alpha}$. The pull--back of the flows 
${\rm RTL}_{\pm}(\alpha)$ under the Miura map $M_1$ coincides with
${\rm RVL}_{\pm}(\alpha)$.
\end{theorem}

The bracket on $\cR\cV$ analogous to $\{\cdot,\cdot\}_{3\alpha}$ and
assuring the Poisson property of $M_2$ case can be obtained via the shift
transformation (\ref{RVL shift}).

\subsection{Lax representation}
\label{Sect RVL Lax}

Lax matrix $(U,W^{-1},V):\cR\cV\mapsto\g\otimes\g\otimes\g$:
\begin{eqnarray}
U(u,v,\lambda) & = & \sum_{k=1}^N u_k E_{kk}+\lambda\sum_{k=1}^N E_{k+1,k}\;, 
\label{RVL U}\\
V(u,v,\lambda) & = & I+\lambda^{-1} \sum_{k=1}^N v_kE_{k,k+1}\;, 
\label{RVL V}\\
W(u,v,\lambda) & = & I-\alpha\lambda^{-1}\sum_{k=1}^N u_kv_kE_{k,k+1}\;.
\label{RVL W}
\end{eqnarray}
Notice that the formulas 
\[
b_k=u_k+v_{k-1}\;,\qquad a_k=u_kv_k
\] 
for the Miura map $M_1$ are equivalent to the following factorization:
\[
L(a,b,\lambda)-W(a,b,\lambda)=\alpha U(u,v,\lambda)V(u,v,\lambda)\;,
\]
while the formulas 
\[
b_k=u_k+v_k\;,\qquad a_k=u_{k+1}v_k
\] 
for the Miura map $M_2$ are equivalent to the following factorization:
\[
L(a,b,\lambda)-W(a,b,\lambda)=\alpha V(u,v,\lambda)U(u,v,\lambda)\;.
\]
\vspace{2mm}

Lax representation of the flow ${\rm RVL}_+(\alpha)$:
\begin{equation}\label{RVL+ Lax triads}
\dot{U}=UA_3-A_1U\;,\qquad \dot{V}=VA_2-A_3V\;,\qquad
\dot{W}=WA_2-A_1W\;,
\end{equation}
with 
\begin{eqnarray}
A_1=\pi_+(UVW^{-1}) & = & 
\sum_{k=1}^N (u_k+v_{k-1}+\alpha u_{k-1}v_{k-1})E_{kk}+
\lambda\sum_{k=1}^N E_{k+1,k}\;, \label{RVL+ A1}\\
A_2=\pi_+(W^{-1}UV) & = & \sum_{k=1}^N (u_k+v_{k-1}+\alpha u_kv_k)E_{kk}+
\lambda\sum_{k=1}^N E_{k+1,k}\;, \label{RVL+ A2}\\
A_3=\pi_+(VW^{-1}U) & = & \sum_{k=1}^N (u_k+v_k+\alpha u_kv_k) E_{kk}+
\lambda\sum_{k=1}^N E_{k+1,k}\;. \label{RVL+ A3} 
\end{eqnarray}
Lax representation of the flow ${\rm RVL}_-(\alpha)$: 
\begin{equation}\label{RVL- Lax triads}
\dot{U}=C_1U-UC_3\;,\qquad \dot{V}=C_3V-VC_2\;,\qquad
\dot{W}=C_1W-WC_2\;,
\end{equation}
with 
\begin{eqnarray}
C_1=\pi_-\Big(f(UVW^{-1})\Big) & = & \lambda^{-1}\sum_{k=1}^N 
\frac{u_kv_k}{1+\alpha (u_{k+1}+v_k)}\,E_{k,k+1}\;, \label{RVL- C1}\\
C_2=\pi_-\Big(f(W^{-1}UV)\Big) & = & \lambda^{-1}\sum_{k=1}^N 
\frac{u_kv_k}{1+\alpha (u_k+v_{k-1})}\,E_{k,k+1}\;. \label{RVL- C2}\\
C_3=\pi_-\Big(f(VW^{-1}U)\Big) & = & \lambda^{-1}\sum_{k=1}^N 
\frac{u_{k+1}v_k}{1+\alpha (u_{k+1}+v_k)}\,E_{k,k+1}\;,  \label{RVL- C3}
\end{eqnarray}
where $f(T)=\Big(I-(I+\alpha T)^{-1}\Big)/\alpha$.

\subsection{Discretization}
\label{Sect RVL+ discretization}
The discrete Lax equations for the map  ${\rm dRVL}_+(\alpha)$:
\begin{equation}\label{dRVL Lax}
\wU=\mbA_1^{-1}U\mbA_3\;,\qquad \wV=\mbA_3^{-1}V\mbA_2\;, \qquad
\wW=\mbA_1^{-1}W\mbA_2\;,
\end{equation}
where
\begin{eqnarray}
\mbA_1 & = & \Pi_+\Big(I+hUVW^{-1}\Big)\;=\;
\sum_{k=1}^N \alpha_kE_{kk}+h\lambda\sum_{k=1}^N E_{k+1,k}\;,
\label{dRVL A1}\\
\mbA_2 & = & \Pi_+\Big(I+hW^{-1}UV\Big)\;=\;
\sum_{k=1}^N \beta_kE_{kk}+h\lambda\sum_{k=1}^N E_{k+1,k}\;,
\label{dRVL A2}\\
\mbA_3 & = & \Pi_+\Big(I+hVW^{-1}U\Big)=
\sum_{k=1}^N \gamma_kE_{kk}+h\lambda\sum_{k=1}^N E_{k+1,k}\;.
\label{dRVL A3}
\end{eqnarray}
\vspace{2mm}

The discrete Lax equations for the map ${\rm dRVL}_-(\alpha)$:
\begin{equation}\label{dRVL- Lax}
\wU=\mbC_1U\mbC_3^{-1}\;,\qquad \wV=\mbC_3V\mbC_2^{-1}\;, \qquad
\wW=\mbC_1W\mbC_2^{-1}\;,
\end{equation}
where
\begin{eqnarray}
\mbC_1 & = & \Pi_-\Big(I+hf(UVW^{-1})\Big)\;=\;
I+h\lambda^{-1}\sum_{k=1}^N \varepsilon_kE_{k,k+1}\;,
\label{dRVL C1}\\
\mbC_2 & = & \Pi_-\Big(I+hf(W^{-1}UV)\Big)=
I+h\lambda^{-1}\sum_{k=1}^N \delta_kE_{k,k+1}\;,
\label{dRVL C2}\\
\mbC_3 & = & \Pi_-\Big(I+hf(VW^{-1}U)\Big)=
I+h\lambda^{-1}\sum_{k=1}^N \zeta_kE_{k,k+1}\;.
\label{dRVL C3}
\end{eqnarray}

\setcounter{equation}{0}
\section{Modified relativistic Volterra lattice}
\label{Sect MRVL}

\subsection{Equations of motion}

The phase space of the {\it modified relativistic Volterra hierarchy}
${\rm MRVL}(\alpha;\epsilon)$ will be denoted by 
$\cM\cR\cV={\Bbb R}^{2N}(y,z)$. The modification parameter $\epsilon$ plays
a role somewhat different from the relativistic parameter $\alpha$. The
equations of motion of the flow ${\rm MRVL}_+(\alpha;\epsilon)$ read:
\begin{eqnarray}\label{MRVL+}
\dot{y}_k & = & y_k(1+\epsilon y_k)\left(\displaystyle
\frac{z_k+\alpha y_kz_k}{1-\epsilon\alpha y_kz_k}-
\displaystyle\frac{z_{k-1}+\alpha y_{k-1}z_{k-1}}
{1-\epsilon\alpha y_{k-1}z_{k-1}}\right)\;,\nonumber\\ \\
\dot{z}_k & = & z_k(1+\epsilon z_k)\left(\displaystyle\frac
{y_{k+1}+\alpha y_{k+1}z_{k+1}}{1-\epsilon\alpha y_{k+1}z_{k+1}}-
\displaystyle\frac{y_k+\alpha y_kz_k}{1-\epsilon\alpha y_kz_k}
\right)\;.\nonumber
\end{eqnarray}
The equations of motion of the flow ${\rm MRVL}_-(\alpha;\epsilon)$ read:
\begin{eqnarray}\label{MRVL-}
\dot{y}_k & = & y_k(1+\epsilon y_k)\left(\displaystyle
\frac{z_k}{1+\alpha(y_{k+1}+z_k)+\epsilon\alpha y_{k+1}z_k}-
\displaystyle\frac{z_{k-1}}{1+\alpha(y_k+z_{k-1})+\epsilon\alpha y_kz_{k-1}}
\right)\;,\nonumber\\ \\
\dot{z}_k & = & z_k(1+\epsilon z_k)\left(\displaystyle\frac
{y_{k+1}}{1+\alpha(y_{k+1}+z_k)+\epsilon\alpha y_{k+1}z_k}-
\displaystyle\frac{y_k}{1+\alpha(y_k+z_{k-1})+\epsilon\alpha y_kz_{k-1}}
\right)\;.\nonumber
\end{eqnarray}

\subsection{Hamiltonian structure}
\begin{proposition}  
The relations
\begin{equation}\label{MRVL br}
\begin{array}{rcl}
\{y_k,z_k\}_{23\alpha} & = & 
-y_kz_k(1+\epsilon y_k)(1+\epsilon z_k)\;,\\ \\
\{z_k,y_{k+1}\}_{23\alpha} & = & 
-z_ky_{k+1}(1+\epsilon z_k)(1+\epsilon y_{k+1})
\end{array}
\end{equation}
define a Poisson bracket on $\cM\cR\cV(y,z)$. The flows 
${\rm MRVL}_{\pm}(\alpha;\epsilon)$ are Hamiltonian with respect to this 
bracket with the Hamilton functions
\begin{equation}\label{MRVL+ H}
\rG_{0\alpha}^{(+)}(y,z)=\epsilon^{-1}\sum_{k=1}^N\log(1+\epsilon y_k)+
\epsilon^{-1}\sum_{k=1}^N\log(1+\epsilon z_k)-
\epsilon^{-1}\sum_{k=1}^N\log(1-\epsilon\alpha y_kz_k)
\end{equation}
and
\begin{eqnarray}\label{MRVL- H}
\rG_{0\alpha}^{(-)}(y,z) & = & (\epsilon-\alpha)^{-1}\sum_{k=1}^N\log
(1+\epsilon y_k)+
(\epsilon-\alpha)^{-1}\sum_{k=1}^N\log(1+\epsilon z_k) \nonumber\\
& & -(\epsilon-\alpha)^{-1}\sum_{k=1}^N\log
\Big(1+\alpha(y_k+z_{k-1})+\epsilon \alpha y_kz_{k-1}\Big)\;,
\end{eqnarray}
respectively. The functions $\rG_{0\alpha}^{(+)}$ and $\rG_{0\alpha}^{(-)}$
are in involution in the bracket $\{\cdot,\cdot\}_{23\alpha}$.
\end{proposition}

Notice that the bracket $\{\cdot,\cdot\}_{23\alpha}$ actually does not depend
on $\alpha$.

\subsection{Miura relations}
\begin{theorem}\label{MRVL Miuras permutability}
{\rm a)} The flows ${\rm MRVL}_{\pm}(\alpha;\epsilon)$ are the 
pull--backs of the flows ${\rm MRTL}_{\pm}^{(+)}(\alpha;\epsilon)$  
under the Miura transformation $M_1^{(+)}(\alpha;\epsilon):\,
\cM\cR\cV(y,z)\mapsto\cM\cR\cT(q,p)$ defined by 
\begin{equation}\label{MRVL Miura to MRTL}
M_1^{(+)}(\alpha;\epsilon)\;:\left\{\begin{array}{l}
p_k=y_k+z_{k-1}+\epsilon y_kz_{k-1}\;,\\ \\ 
q_k=\displaystyle\frac{y_kz_k}{1-\epsilon\alpha y_kz_k}\;.\end{array}
\right.
\end{equation}
This Miura transformation is Poisson, if both $\cM\cR\cV(y,z)$ and 
$\cM\cR\cT(q,r)$ carry the corresponding brackets $\{\cdot,\cdot\}_{23\alpha}$.

{\rm b)}  The flows ${\rm MRVL}_{\pm}(\alpha;\epsilon)$ are the 
pull--backs of the flows ${\rm RVL}_{\rm}(\alpha)$ under 
either of the Miura transformations $\mbM_{1,2}^{(+)}(\alpha;\epsilon):\,
\cM\cR\cV(y,z)\mapsto\cR\cV(u,v)$ defined by
\begin{equation}\label{MRVL Miura 1 to RVL}
\mbM_1^{(+)}(\alpha;\epsilon):\left\{\begin{array}{l}
u_k=\displaystyle\frac{y_k(1+\epsilon z_{k-1})}
{1-\epsilon\alpha y_{k-1}z_{k-1}}\;,\\ \\ 
v_k=\displaystyle\frac{z_k(1+\epsilon y_k)}{1-\epsilon\alpha y_kz_k}\;,
\end{array}
\right.
\end{equation}
\[
\]
\begin{equation}\label{MRVL Miura 2 to RVL}
\mbM_2^{(+)}(\alpha;\epsilon):\left\{\begin{array}{l}
u_k=\displaystyle\frac{y_k(1+\epsilon z_k)}{1-\epsilon\alpha y_kz_k}\;,\\ \\ 
v_k=\displaystyle\frac{z_k(1+\epsilon y_{k+1})}
{1-\epsilon\alpha y_{k+1}z_{k+1}}\;.\end{array}
\right.
\end{equation}
Both Miura transformations $\mbM_{1,2}^{(+)}(\alpha;\epsilon)$ are Poisson, if 
$\cM\cR\cV(y,z)$ is equipped with the bracket $\{\cdot,\cdot\}_{23\alpha}$ 
and $\cR\cV(u,v)$ is equipped with the bracket $\{\cdot,\cdot\}_{2\alpha}
+\epsilon\{\cdot,\cdot\}_{3\alpha}$.

{\rm c)} the following diagram is commutative for $j=1,2$:
\begin{center}
\unitlength1cm
\begin{picture}(9,6.5)
\put(3.5,1.1){\vector(1,0){2}}
\put(3.5,5.1){\vector(1,0){2}}
\put(2,4.1){\vector(0,-1){2}}
\put(7,4.1){\vector(0,-1){2}}
\put(1,0.6){\makebox(2,1){$\cR\cV$}} 
\put(1,4.6){\makebox(2,1){$\cM\cR\cV$}}
\put(6,4.6){\makebox(2,1){$\cM\cR\cT$}}
\put(6,0.6){\makebox(2,1){$\cR\cT$}}
\put(0,2.6){\makebox(1.8,1){$\mbM_j^{(+)}(\alpha;\epsilon)$}}
\put(7,2.6){\makebox(2.2,1){$\rM_j^{(+)}(\alpha;\epsilon)$}}
\put(3.8,-0.2){\makebox(1.4,1.4){$M_1$}}
\put(3.8,5.0){\makebox(1.4,1.4){$M_1^{(+)}(\alpha;\epsilon)$}}
\end{picture}
\end{center} 
\end{theorem}

There is only {\it one} Miura transformation 
$M_1^{(+)}(\alpha;\epsilon)$ relating the modified relativistic Volterra
hierarchy ${\rm MRVL}(\alpha;\epsilon)$ to the modified relativistic Toda 
hierarchy ${\rm MRTL}^{(+)}(\alpha;\epsilon)$. Of course, this is no surprise:
the situation with the unmodified systems was exactly the same: the
hierarchies ${\rm RVL}(\alpha)$ and ${\rm RTL}(\alpha)$ are related 
by only one Miura transformation $M_1$. Actually, there exists
a ``twin'' ${\rm MRVL}(\alpha;\epsilon)$ hierarchy related to 
${\rm MRTL}^{(+)}(\alpha;\epsilon)$ via the ``twin'' Miura transformation
$M_2^{(+)}(\alpha;\epsilon)$.

\subsection{Application: localizing change of variables for 
${\rm dRVL}_+(\alpha)$}
The Miura map $\mbM_1^{(+)}(\alpha;h)$ plays the role of the localizing
change of variables for the map ${\rm dRVL}_+(\alpha)$. Indeed, consider
the change of variables $\cR\cV(\bu,\bv)\mapsto\cR\cV(u,v)$:
\begin{equation}\label{dRVL loc map}
\mbM_1^{(+)}(\alpha;h)\,:\quad
u_k=\bu_k\,\frac{1+h\bv_{k-1}}{1-h\alpha \bu_{k-1}\bv_{k-1}}\;,\quad
v_k=\bv_k\,\frac{1+h\bu_k}{1-h\alpha \bu_k\bv_k}\;.
\end{equation}
The the entries of the factors $\mbA_j$, $j=1,2,3$ (see
(\ref{dRVL A1})--(\ref{dRVL A3})) admit local expressions in the 
coordinates $(\bu,\bv)$:
\begin{equation}\label{dRVL loc alpha}
\alpha_k=\frac{(1+h\bu_k)(1+h\bv_{k-1})}{1-h\alpha \bu_{k-1}\bv_{k-1}}\;,
\end{equation}
\begin{equation}\label{dRVL loc beta}
\beta_k=\frac{(1+h\bu_k)(1+h\bv_{k-1})}{1-h\alpha \bu_k\bv_k}\;,
\end{equation}
\begin{equation}\label{dRVL loc gamma}
\gamma_k=\frac{(1+h\bu_k)(1+h\bv_k)}{1-h\alpha \bu_k\bv_k}\;.
\end{equation}
\begin{theorem}
The change of variables {\rm(\ref{dRVL loc map})} conjugates 
${\rm dRVL}_+(\alpha)$ with the map on $\cR\cV(\bu,\bv)$ described by
the following equations:
\begin{eqnarray}
\widetilde{\bu}_k\,\frac{1+h\widetilde{\bv}_{k-1}}
{1-h\alpha \widetilde{\bu}_{k-1}\widetilde{\bv}_{k-1}} 
& = &
\bu_k\,\frac{1+h\bv_k}{1-h\alpha \bu_k\bv_k} \;, \nonumber\\
\label{dRVL loc} \\
\widetilde{\bv}_k\,\frac{1+h\widetilde{\bu}_k}{1-h\alpha \widetilde{\bu}_k
\widetilde{\bv}_k} &  = &
\bv_k\,\frac{1+h\bu_{k+1}}{1-h\alpha \bu_{k+1}\bv_{k+1}} \;.\nonumber
\end{eqnarray}
\end{theorem}
We see that the local form of ${\rm dRVL}_+(\alpha)$ lives in the hierarchy
${\rm MRVL}(\alpha;h)$.

{\bf Corollary.} {\it The local form of ${\rm dRVL}_+(\alpha)$ is a Poisson
map with respect to the following bracket on $\cR\cV(\bu,\bv)$:
\begin{eqnarray}\label{dRVL loc PB loc}
\{\bu_k,\bv_k\} & = & -\bu_k\bv_k(1+h\bu_k)(1+h\bv_k)\;,\nonumber\\ \\
\{\bv_k,\bu_{k+1}\} & = & -\bv_k\bu_{k+1}(1+h\bv_k)(1+h\bu_{k+1})\;,\nonumber
\end{eqnarray}
which is the pull--back of the bracket
\begin{equation}
\{\cdot,\cdot\}_{2\alpha}+h\{\cdot,\cdot\}_{3\alpha}
\end{equation}
on the space $\cR\cV(u,v)$ under the change of variables 
{\rm(\ref{dRVL loc map})}.}

Finally, we give the translations of the Miura relations between the maps 
${\rm dRVL}_+(\alpha)$ and ${\rm dRTL}_+(\alpha)$ into the language of 
localizing variables. This is achieved by reformulating the part c) of
Theorem \ref{MRVL Miuras permutability}.
\begin{theorem} 
{\rm a)} The following diagram is commutative:
\begin{center}
\unitlength1cm
\begin{picture}(9,6.5)
\put(3.5,1.1){\vector(1,0){2}}
\put(3.5,5.1){\vector(1,0){2}}
\put(2,4.1){\vector(0,-1){2}}
\put(7,4.1){\vector(0,-1){2}}
\put(1,0.6){\makebox(2,1){$\cR\cV(u,v)$}} 
\put(1,4.6){\makebox(2,1){$\cR\cV(\bu,\bv)$}}
\put(6,4.6){\makebox(2,1){$\cR\cT(\ba,\bb)$}}
\put(6,0.6){\makebox(2,1){$\cR\cT(a,b)$}}
\put(0,2.6){\makebox(1.8,1){$\mbM_1^{(+)}(\alpha;h)$}}
\put(7,2.6){\makebox(2.2,1){$\rM_1^{(+)}(\alpha;h)$}}
\put(3.8,-0.2){\makebox(1.4,1.4){$M_1$}}
\put(3.8,5.0){\makebox(1.4,1.4){$M_1^{(+)}(\alpha;h)$}}
\end{picture}
\end{center}
where the map $M_1^{(+)}(\alpha;h):\,\cR\cV(\bu,\bv)\mapsto\cR\cT(\ba,\bb)$ 
is given by the formulas
\begin{equation}\label{Miura 1 RVL+ RTL+ loc}
M_1^{(+)}(\alpha;h)\,: \quad \left\{\begin{array}{l}
\ba_k=\displaystyle\frac{\bu_k\bv_k}{1-h\alpha\bu_k\bv_k}
\;,\\ \\
\bb_k=\bu_k+\bv_{k-1}+h\bu_k\bv_{k-1}\;.
\end{array}\right.
\end{equation}

{\rm b)} The map $M_1^{(+)}(\alpha;h)$ conjugates the local form of 
${\rm dRVL}_+(\alpha)$ {\rm (\ref{dRVL loc})} with the local form of 
${\rm dRTL}_+(\alpha)$ {\rm (\ref{dRTL+ param loc})}. 

{\rm c)} The map $M_1^{(+)}(\alpha;h)$ is Poisson, provided $\cR\cV(\bu,\bv)$ 
is equipped with the bracket {\rm (\ref{dRVL loc PB loc})}, and 
$\cR\cT(\ba,\bb)$ is equipped with the bracket {\rm(\ref{dRTL+ param loc PB2})}.
\end{theorem}

\section{Bibliographical remarks.}
{\bf a)} Our account of the Miura maps is very close in spirit to that of 
{\it Kupershmidt}, see has works
\begin{itemize}
\item[${\rm [K1]}$] B.A. Kupershmidt. On the nature of the Gardner transformation. 
{\em J. Math. Phys.} {\bf  22} (1981) 449-452. 
\item[${\rm [K2]}$] B.A. Kupershmidt. Deformations of integrable systems. 
{\em Proc. R. Ir. Acad.}, {\bf A83} (1983) 45-74.
\item[${\rm [K3]}$] B.A. Kupershmidt.  Discrete Lax equations and 
differential--difference calculus. {\em Asterisque} {\bf 123} (1985) 212pp.
\end{itemize}
In particular, our presentation of the original Miura (Gardner) transformation
in the introduction follows [K1]. The Hamiltonian formalism for lattice
systems is developed in [K3], in particular, the tri--Hamiltonian structure of 
the Toda lattice and the bi--Hamiltonian structure of the Volterra lattice was
found there for the first time. The papers [K2], [K3] treat also the modified
and the double modified Toda lattice, as well as the modified Volterra lattice.
Accordingly, the corresponding Miura maps and their Poisson and permutability
properties are given there. However, the role of Miura maps as localizing
changes of variables for integrable discretizations is a novel aspect, not
discussed in the above mentioned references.
\vspace{2mm}

{\bf b)} The whole tower of modifications for the nonrelativistic Toda--like 
systems, including the triple modified Toda and the double modified Volterra 
lattices, was discovered by {\it Yamilov} in the course of classifying 
integrable lattice systems with nearest neighbors interactions:
\begin{itemize}
\item[${\rm [Y1]}$] R. Yamilov. On the classification of discrete equations. 
In: {\it Integrable systems}, Ufa 1982, 95-114 (in Russian).
\item[${\rm [Y1]}$] R. Yamilov. Classification of discrete evolution
equations. {\em Uspekhi Matem. Nauk} {\bf 38} (1983) 155--156 (in Russian).
\item[${\rm [Y3]}$] R. Yamilov. Construction scheme for discrete Miura 
transformations. {\em J. Phys. A: Math. Gen.} {\bf 27} (1994) 6839-6851.
\end{itemize}
The Hamiltonian properties of these systems and of Miura maps were not his 
main concern.
\vspace{2mm}

{\bf c)} The local integrable discretizations of the Toda lattice
(\ref{dTL loc}), the Volterra lattice (\ref{dVL loc}), and the modified Toda 
lattice (\ref{dMTL loc}) were found for the first time by {\it Hirota},
including the Miura relations between them:
\begin{itemize}
\item[${\rm [H1]}$] R. Hirota, S. Tsujimoto, T. Imai. Difference scheme of
soliton equations. In: {\it ''Future directions of nonlinear dynamics in
physical and biological systems''}, Eds. P.L.Christiansen, J.C.Eilbeck,
and R.D.Parmentier (Plenum, 1993), 7--15.
\item[${\rm [H2]}$] R. Hirota, S. Tsujimoto. Conserved quantities of discrete 
Lotka-Volterra equations. {\em RIMS Kokyuroku} {\bf 868} (1994) 31--38 
(in Japanes); Conserved quantities of a class of nonlinear 
difference--difference equations. {\em J. Phys. Soc. Japan} {\bf 64} (1995) 
3125--3127.
\end{itemize} 
However, he did not identify these discretizations as the members of the 
modified hierarchies, and did not study their Hamiltonian properties.
\vspace{2mm}

{\rm d)} The present paper is a companion for 
\begin{itemize}
\item[${\rm [S]}$] Yu.B. Suris. Integrable discretizations for lattice systems:
local equations of motion and their Hamiltonian properties. {\em 
Rev. Math. Phys.} (1999, to appear).
\end{itemize}
The reader should consult this paper for general concepts from the theory
of $r$--matrix hierarchies, relevant for the problem of integrable 
discretizations, as well as for an extensive list of references.
The concept of localizing changes of variables appeared there for the 
first time. The present paper contains more examples, in particular, 
almost all results of the ``relativistic'' part are new.
\end{document}